     \documentclass[oneside,12pt]{book}

  \usepackage[T1]{fontenc}
\usepackage{color}
\usepackage[breaklinks=true,colorlinks=true,anchorcolor=blue,citecolor=blue,filecolor=blue,menucolor=blue,pagecolor=blue,urlcolor=blue,linkcolor=blue]{hyperref}
 \usepackage{natbib,url}
     \usepackage{graphicx}
     \usepackage{epsfig}
    \usepackage{amsmath}
\begin{document}

     \chapter[
     %
     %
             Svozil (rev. 2011 02 28)
     %
     %
     ]{\huge{
                             Physical Unknowables
     }\bigskip
     \\ \large{
                             Karl Svozil
     }\\
     \bigskip
     }

\begin{quote}
\begin{flushright}
{\footnotesize
{
As we know, there are known knowns; \\
there are things we know we know. \\
We also know there are known unknowns; \\
that is to say we know there are some things we do not know. \\
But there are also unknown unknowns --\\
the ones we don't know we don't know.} \\
{ --~United States Secretary of Defense Donald H. Rumsfeld \\
at a Department of Defense news briefing on February 12, 2002}
\\ $\;$
\\ $\;$
{ Ei mihi, qui nescio saltem quid nesciam!}\\
{ (Alas for me, that I do not at least know the extent of my own ignorance!)}  \\
{  --~Aurelius Augustinus, 354--430, ``Confessiones'' (Book XI, chapter 25)}
}
\end{flushright}
\end{quote}

\section{Rise and fall of determinism}

In what follows, a variety of physical unknowables will be discussed.
Provable lack of physical omniscience, omnipredictability and omnipotence is derived by reduction to problems that are known to be recursively unsolvable.
``Chaotic'' symbolic dynamical systems are unstable with respect to variations of initial states.
Quantum unknowables include the random occurrence of single events, complementarity, and value indefiniteness.

From antiquity onward, various waves of (in)determinism have influenced human thought.
Regardless of whether they were shaped by some {\it Zeitgeist,}
or whether, as Goethe's {\it Faust} puts it,
{``what you the Spirit of the Ages call,
is nothing but the spirit of you all,
wherein the Ages are reflected,''}
their proponents have sometimes vigorously defended their stance in irrational,  unscientific, and ideologic ways.
Indeed, from an emotional point of view, may it not appear frightening
to be ``imprisoned'' by remorseless, relentless predetermination,
even in a dualistic setup \citep{descartes-meditation}; and, equally frightening,
to accept that one's fate depends on total arbitrariness and chance?
Does determinism expose freedom, self-determination and human dignity as an idealistic illusion?
On the other extreme, what kind of morale, merits and efforts appear worthy in a universe governed by pure chance?
Is there some reasonable in-between straddling those extreme positions that may also be consistent with science?

We shall, for the sake of separating the scientific debate
from emotional overtones and possible bias, adopt a contemplative strategy of {\em evenly-suspended attention}
outlined by  \citet*{Freud-1912}, who admonishes analysts to be aware of the dangers
caused by {``temptations to project,
what  [the analyst]  in dull self-perception recognizes as the peculiarities of his own personality,
as generally valid theory into science.''}
Nature is thereby treated as a  client-patient,  and whatever findings come up are accepted  as is  without any
immediate emphasis or judgment.

\subsection{Toward explanation and feasibility}

Throughout history,
the human desire to foresee and manipulate the physical world for survival and prosperity,
and in accord with personal wishes and  fantasies,
has been confronted with the inability to predict and manipulate
large portions of the habitat.
As time passed, people have figured out various ways to tune
ever increasing fragments of the world according to their needs.
From a purely behavioral perspective, this is brought about in the way of
pragmatic quasi-causal conditional rules  of the following kind,
``if one does this, one obtains that.''
A typical example of such a rule is ``if I rub my hands, they get warmer.''

How does one arrive at those kinds of rules?
Guided by suspicions, thoughts, formalisms and by pure chance,
inquiries start by  roaming around,  inspecting portions of the world
and examining their behavior.
Repeating phenomena or  patterns of behavior are observed and  pinned down  by reproducing and evoking them.
A physical behavior is anything that can be observed and thus operationally obtained and measured;
for example, the rise and fall of the sun, the ignition of fire, the formation and  melting of ice (in principle even time series of financial entities traded
at stock exchanges or over-the-counter).


As physical behaviors are observed,
people attempt to  understand  them by trying to figure out some
 cause \citep{schlick,frank} or reason for their occurrences.
Researchers invent virtual parallel worlds of thoughts
and intellectual concepts such as ``electric field'' or ``mechanical force''
to  explain and manipulate the physical behaviors,
calling these creations of their minds ``physical theories.''
Contemporary physical theories are heavily formalized and spelled out in the language of mathematics.
A good theory provides people with the feeling of a key unlocking new ways of world comprehension and manipulation.
Ideally, an explanation should be as compact as possible
and should apply to as many behavioral patterns as possible.

Ultimately, theories of everything \citep{schlick-35,barrow-TOE,Kragh-qg} should be able to predict and manipulate all phenomena.
In the extreme form, science becomes omniscient and omnipotent,
and we envision ourselves almost as becoming empowered with  magic:
we presume that our ability to manipulate and tune the world is limited by our fantasies alone,
and any constraints whatsoever can be bypassed or overcome one way or another.
Indeed, some of what in the past has been called ``supernatural,'' ``mystery,'' and ``the beyond''
has been realized in everyday life.
Many wonders of witchcraft have been transferred into the realm of the physical sciences.
Take, for example, our abilities to fly,
to transmute mercury into gold \citep{PhysRev.60.473},
to listen and speak to far away friends,
or to cure bacterial diseases with a few pills of antibiotics.

Until about 1900, the fast-growing natural sciences, guided  by
rational \citep{Descartes-Discourse} and empirical \citep{lock-thu,hume-thu} thinking,
and seconded by the European Enlightenment, prospered under the assumption of physical determinism.
Under the {\em aegis} of physical determinism, all incapacities to predict and manipulate physical behavior were
interpreted to be merely {\em epistemic} in nature, purporting that, with growing precision of measurements and
improvements of theory, all physical unknowables will eventually be overcome and turned into knowables;
that is,   everything should in principle be knowable.
Even  statistical  quantities would describe underlying deterministic behaviors.
Consequently, there could not exist any physical behavior or entity
without a cause  stimulating or pushing  it into existence.

The uprise of determinism culminated in the following statement by \citet[chap.~2]{laplace-prob}:
\begin{quote}
{
Present events are connected with preceding ones
by a tie based upon the evident principle that a thing
cannot occur without a cause which produces it. This
axiom, known by the name of the principle of sufficient
reason, extends even to actions which are considered
indifferent $\ldots$

We ought then to regard the present state of the
universe as the effect of its anterior state and as the
cause of the one which is to follow. Given for one
instant an intelligence which could comprehend all the
forces by which nature is animated and the respective
situation of the beings who compose it an intelligence
sufficiently vast to submit these data to analysis it
would embrace in the same formula the movements of
the greatest bodies of the universe and those of the
lightest atom; for it, nothing would be uncertain and
the future, as the past, would be present to its eyes.
}
\end{quote}
The invention of (analytic) functions reflects this paradigm quite nicely:
some dispersionless point coordinate $x(t)$ of infinite precision
serves as the representation \citep{hertz-94} of a physical state
as a (unique) function of physical time~$t$.

Indeed, the possibility to formulate theories {\it per se},
and in particular, the applicability of formal, mathematical models, comes as a mind-boggling surprise and
cannot be taken for granted;
there appears to be what \citet{wigner}  called an
``unreasonable effectiveness of mathematics in the natural sciences.''
Even today, there is a Pythagorean consensus that there is no limit to
dealing with physical entities in terms of mathematical formalism.
And, as mathematics increasingly served as a proper representation of reality,
and computational deduction systems were increasingly  introduced to delineate formalizable truth,
algorithmics started to become a metaphor for physics.
In algorithmic terms, nature computes, and can be (re)programmed to perform certain tasks.

The natural sciences continued to be uninhibited by any sense of limits until
about {\it fin-de-si\`ecle}, around 1900.
In parallel, the formalization of mathematics progressed in an equally uninhibited way.
\citet[170]{hilbert-26} argued that
nobody should ever expel mathematicians from the paradise created by Cantor's set theory
and posed a challenge \citep{hilbert-1900e}
to search for a consistent, finite system of formal axioms which would be able to render all
mathematical and physical truths; just like quasi-finitistic ways to cope with infinitesimal calculus had been found.

This type of belief system that claims omniscience could be called ``deterministic conjecture''
because no proof for its validity can be given,
nor is there any way of falsification \citep{popper-en}.
Alas, from a pragmatic point of view, omniscience  can be  effectively  disproved
on a daily basis by tuning in to local weather forecasts.

Furthermore, it seems to be an enduring desire of human nature to be able not merely to trust
the rules and theories syntactically and operationally \citep{bridgman}
but also to be able to semantically interpret them as implying and carrying some
ontological significance or truth -- as if
reality would communicate with us, mediated through our senses, thereby revealing the laws governing nature.
Stated pointedly, we not only wish to accept physical theories as pure abstractions and constructions of our own mind \citep{berkeley}
but we associate meaning and truth to them
so much so that only very reluctantly do we admit their preliminary, transient, and changing character \citep{lakatosch}.

\subsection{Rise of indeterminism}

Almost unnoticed,  the tide of indeterminism started to build
 toward the end of the nineteenth century \citep{purrington,Kragh-qg}.
At that time, mechanistic theories faced an increasing number of anomalies:
Poincar\'e's discovery of  instabilities  of trajectories of celestial bodies
(which made them extremely sensible to initial conditions),
radioactivity \citep{Kragh-1997AHESradioact,Kragh-2009_RePoss5},
X-rays,
specific heats of gases and solids,
emission and absorption of light
(in particular, blackbody radiation),
the (ir)reversibility dichotomy
between classical reversible mechanics and
Boltzmann's statistical-mechanical theory of entropy {\it versus} the second law of thermodynamics,
and the experimental refutation of classical  constructions  of the ether as a medium for the propagation of light waves.

After the year 1900 followed a short period of revolutionary new physics, in particular,
quantum theory and relativity theory,
without any strong inclination toward (in)determinism.
Then indeterminism erupted with Born's claim that quantum mechanics has it both ways:
the quantum state evolves strictly deterministically,
whereas the individual event or measurement outcome occurs indeterministically.
Born also stated that he {\em believed} that there is no cause for an individual quantum event;
that is, such an outcome occurs irreducibly at random.

There followed a fierce controversy, with many researchers such as Born, Bohr, Heisenberg, and Pauli
taking the indeterministic stance,
whereas others,
like Planck \citep{born-55}, Einstein \citep{epr,ein-reply}, Schr\"odinger, and De Brogli, leaning toward determinism.
This latter position was pointedly put forward by Einstein's {\it dictum} in a letter to Born,
dated December~12, 1926 \cite[113]{born-69}:
{``In any case I am convinced that he [the Old One] does not throw dice.''}
At present, indeterminism is clearly favored, the canonical position being expressed by \citet{zeil-05_nature_ofQuantum}:
{``The discovery that individual events are
irreducibly random is probably one of the
most significant findings of the twentieth
century. $\ldots$~For the individual event in quantum physics,
not only do we not know the cause, there is no cause.''}

The last quarter of the twentieth century saw the rise of yet another form of physical indeterminism,
originating in Poincar\'e's aforementioned
discovery of instabilities of the motion of classical bodies
against variations of initial conditions \citep{Campbell-1882,poincare14,Diacu96-ce}.
This scenario of {\em deterministic chaos} resulted in a plethora of claims regarding indeterminism
that resonated with a general public susceptible to fables and fairy tales \citep{bricmont}.

In parallel, G\"odel's incompleteness theorems \citep{godel1,tarski:32,davis-58,davis,smullyan-92},
as well as related findings in the computer sciences \citep{turing-36,chaitin3,calude:02,gruenwald-vitanyi},
put an end to Hilbert's program of finding a finite axiom system for all mathematics.
G\"odel's incompleteness theorems also established formal
bounds on provability, predictability, and induction.
(The incompleteness theorems also put an end to philosophical contentions
expressed by \citet[101]{schlick-35} that, beyond epistemic unknowables and
the ``essential incompetence  of human knowledge,'' there is ``not a single real
question for which it would be {\em logically} impossible to find a solution.'')

Alas, just like determinism, physical indeterminism cannot be proved, nor can there be given any reasonable criterion for its falsification.
After all, how can one check against all laws and find none applicable?
Unless one is willing to denote any system whose laws are currently unknown
or whose behavior is hard to predict with present techniques as indeterministic,
there is no scientific substance to such absolute claims,
especially  if one takes into account the bounds imposed by the theory of recursive functions discussed later.
So, just as in the deterministic case, this position should be considered conjectural.

In discussing the present status of physical (in)determinism,
we shall first consider provable unknowables through reduction to incompleteness theorems of recursion theory,
then discuss classical deterministic chaos, and finally deal with the three types of quantum indeterminism:
the occurrence of certain single events, complementarity, and value indefiniteness.
The latter quantum unknowables are not commonly accepted by the entire community of physicists;
a minority is still hoping for a more complete quantum theory than the present statistical theory.

\section{Provable physical unknowables}

In the past century,
unknowability has been formally defined and {\em derived} in terms of a precise,
formal notion of unprovability \citep{godel1,tarski:32,tarski:56,turing-36,rogers1,davis-58,odi:89,smullyan-92}.
This is a remarkable departure from informal suspicions and observations regarding the limitations
of our worldview.
No longer is one reduced to informal, heuristic contemplations and comparisons about what one knows and can do {\it versus}
one's ignorance and incapability.
Formal unknowability is about formal proofs of unpredictability and impossibility.

There are several pathways to formal undecidability.
For contemporaries accustomed to computer programs (and their respective codes),
a straight route may be algorithmic.
What is an algorithm? In Turing's \citeyearpar[34]{Turing-Intelligent_Machinery} own words,
 \begin{quote}
{
a man provided with paper, pencil and rubber, and subject to strict discipline [carrying out a set of rules of procedure written down] is in effect a universal computer.
}
\end{quote}
From a purely syntactic point of view,
formal systems in mathematics  can be identified with computations
and {\it vice versa}.
Indeed, as stated by G\"odel \citeyearpar[369-370]{godel-ges1} in a {\em postscript,} dated from  June 3, 1964:
 \begin{quote}
 {
 due to A. M. Turing's work,
 a precise and unquestionably
 adequate definition of the general concept of formal system can now be
 given, the existence of undecidable arithmetical propositions and the
 non-demonstrability of the consistency of a system in the same system
 can now be proved rigorously for {\em every} consistent formal system
 containing a certain amount of finitary number theory.

 Turing's work gives an analysis of the
 concept of ``mechanical
 procedure'' (alias ``algorithm'' or ``computation procedure'' or
 ``finite combinatorial procedure''). This concept is shown to be
 equivalent with that of a ``Turing machine.'' A formal system can
 simply be defined to be any mechanical procedure for producing
 formulas, called provable formulas.
}
 \end{quote}

Almost since its discovery, attempts \citep{popper-50i,popper-50ii} have been made to translate
formal incompleteness into physics,
mostly by reduction to some provable undecidable problem
of recursion theory  such as the halting
problem \citep{wolfram84,kanter,moore,wolfram85b,dc-d91a,dc-d91b,suppes-1993,svozil-93,1994IJTP...33.1085H,casti:94-onlimits_book,casti:96-onlimits,barrow-impossibilities}.
Here the term {\em reduction} indicates that physical undecidability is linked or reduced to logical undecidability.
A typical example is the embedding of a Turing machine or any type of computer capable of
universal computation into a physical system.
As a consequence, the physical system inherits
any type of unsolvability derivable for universal computers such as the
unsolvability of the halting problem:
because the computer is part of the physical system, so are its behavioral patterns
[and {\em vice versa} \citep{bridgman,landauer:86,landauer}].

Note that these logical and recursion-theoretical
types of physical unknowables are only derivable within deterministic systems that are
strong enough to express {\em self-reference}, {\em substitution} \cite[chap.~1]{smullyan-92},
and {\em universal computation}.
Indeterministic systems are not deterministic by definition,
and too-weak forms of expressibility are trivially incomplete \citep{bruk-08},
as they are incapable of expressing universal computation or self-reference and substitution.

G\"odel himself did not believe that his incompleteness theorems had any relevance for physics,
especially not for quantum mechanics.
The author was told by professor Wheeler that G\"odel's resentments
[also mentioned in \citet[140--141]{bernstein}]
may have been due to Einstein's negative opinion about quantum theory,
because Einstein may have brainwashed G\"odel
into believing that all efforts in this direction were in vain.

\subsection{Intrinsic self-referential observers}

Embedded \citep{toffoli:79}, intrinsic observers \citep{svozil-94} cannot leave their Cartesian prison \cite[Meditation~1.12]{descartes-meditation}
and step outside the universe examining it from some  Archimedean point \cite[sect.~11,  405--409]{bos1}.
Thus every physical observation is reflexive \citep{nagel-ViewFromNowhere,sosa-rk2} and circular \citep{Kauffman198753}.
The self-referential and substitution capability of observers results in very diverse,
unpredictable forms of behavior  and in provable unknowables.

For the sake of the further analysis, suppose that there exist observers measuring objects  and that
observers and objects are distinct from one another, separated by a cut.
Through that cut, information is exchanged.
Symbolically, we may regard the object as an agent contained in a black box,
whose only relevant emanations are representable by finite strings of zeroes and ones
appearing on the cut, which can be modeled by any kind of screen or display.
According to this purely syntactic point of view,
a physical theory should be able to render identical symbols like the ones appearing through the cut;
that is, a physical theory should be able to mimic or emulate the black box to which it purports to apply.
This view is often adapted in quantum mechanics \citep{fuchs-peres},
where the question regarding any  meaning of the quantum formalism is notorious \cite[129]{feynman-law}.

A sharp distinction between a physical object and an extrinsic
outside observer is a rarely affordable abstraction.
Mostly the observer is part of the system to be observed.
In such cases,
the measurement process is modeled symmetrically,
and information is exchanged between observer and object bidirectionally.
This symmetrical configuration makes a distinction between observer and object
purely conventional \citep{svozil-2001-convention}.
The cut is constituted by the information exchanged.
We tend to associate with the  measurement apparatus
one of the two subsystems that, in comparison, is  larger, more classical,
and up-linked with some conscious observer \citep{wigner:mb}.
The rest of the system can then be called the measured object.

Intrinsic observers face all kinds of paradoxical self-referential situations.
These have been expressed informally as puzzling amusement and artistic perplexity,
and as a formalized, scientifically valuable resource.
The {\em liar paradox}, for instance, is already mentioned in the Bible's Epistle to Titus 1:12, stating that
``one of Crete's own prophets has said it: `Cretans are always liars, evil brutes, lazy gluttons.'
He has surely told the truth.''
In what follows, paradoxical self-referentiality will be applied to argue
against the solvability of the general induction problem
as well as for a pandemonium of undecidabilities related to physical systems
and their behaviors. All are based on intrinsic observers embedded
in the systems they observe.

It is not totally unreasonable to speculate that the
limits of intrinsic self-expression seems to be
what G\"odel himself
considered the gist of his incompleteness theorems.
In a reply to a letter by Burks
[reprinted in \citet[55]{v-neumann-66}; see also \citet[554]{fef-84}],
G\"odel states:
 \begin{quote}
 {
that a complete epistemological description
 of a language $A$ cannot be given in the same language $A$, because
 the concept of truth of sentences of $A$ cannot be defined in $A$. It
 is this theorem which is the true reason for the existence of
 undecidable propositions in the formal systems containing arithmetic.
}
 \end{quote}

One of the first researchers to become interested in the application
of paradoxical self-reference to physics
was the philosopher \citet{popper-50i,popper-50ii}
who published two almost forgotten papers
discussing, among other issues, Russell's paradox of
Tristram Shandy \citep{sterne}:
In volume 1, chapter 14, Shandy finds that he could publish
two volumes of his life every year,
covering a time span far shorter than the time it took him to write
these volumes. This de-synchronization, Shandy concedes,
will rather increase than diminish as he advances; one may thus have serious doubts about
whether he will ever complete his autobiography.
This relates to a question of whether there can be a physical computer that can be assured
of correctly {\em processing information faster than the universe does.}
\citet[016128-1]{PhysRevE.65.016128} states that [see also \citet[sect.~5]{CalCamSvo-Stef-1995}]
{``In a certain sense, the universe is more powerful
than any information-processing system constructed within it
could be. This result can alternatively be viewed as a restriction
on the computational power of the universe -- the universe
cannot support the existence within it of a computer
that can process information as fast as it can.''}

\subsection{Unpredictability}
\label{2010-unknowable-s2}

For any deterministic system strong enough to support
universal computation,  the general forecast or prediction
problem is provable unsolvable.
This proposition will be argued by reduction to the halting problem, which is provable unsolvable.
A straightforward embedding of a universal computer
into a physical system results in the fact that,
owing to the reduction to the recursive undecidability of the halting problem,
certain future events cannot be predicted
and are thus provable indeterministic.
Here reduction again means that physical undecidability is linked or reduced
to logical undecidability.

A clear distinction should be made between {\em determinism}
(such as {\em computable evolution laws}) and {\em predictability} \citep{suppes-1993}.
Determinism does not exclude unpredictability in the long run.
The local (temporal), step-by-step evolution of the system can be perfectly deterministic and computable,
whereas recursion-theoretic unknowables correspond to global observables at unbounded time scales.
Indeed, (nontrivial) provable unpredictability requires determinism,
because formalized proofs require formal systems or algorithmic behavior.

Unpredictability in indeterministic systems is tautological and trivial.
At the other extreme, one should also keep in mind that there exist rather straightforward
pre-G\"odelian impossibilities \citep{bruk-08}  to
express certain mathematical truths
 in weak systems that are incapable of
representing universal computation or Peano arithmetic.

For the sake of exploring (algorithmically)
what paradoxical self-reference is like,
one can consider the sketch of a proof by contradiction
of the unsolvability of the halting problem.
The halting problem is about whether or not a computer will eventually halt on a given input,
that is, will evolve into a state indicating the completion of a computation task or will stop altogether.
Stated differently, a solution of the halting problem will be an algorithm that
decides whether another arbitrary algorithm on arbitrary input will finish running or will run forever.

The scheme of the proof by contradiction is as follows:
the existence of a hypothetical halting algorithm
capable of solving the halting problem will be {\em assumed.}
This could, for instance, be a subprogram of some suspicious supermacro library
that takes the code of an arbitrary program as input and outputs 1 or 0,
depending on whether or not the program halts.
One may also think of it as a sort of oracle or black box analyzing an arbitrary
program in terms of its symbolic code and outputting one of two symbolic states, say, 1 or 0,
referring to termination or nontermination of the input program, respectively.

On the basis of this {\em hypothetical halting algorithm}
one constructs another {\em diagonalization program} as follows:
on receiving some arbitrary {\em input program} code as input, the {diagonalization program}
consults the {\em hypothetical halting algorithm} to find out whether or not this
{input program} halts; on receiving the answer, it does the {\em opposite:}
If  the   hypothetical halting algorithm  decides that the   input program  {\em halts,}
the   diagonalization program  does {\em not halt} (it may do so easily by entering an infinite loop).
Alternatively, if  the   hypothetical halting algorithm  decides that the  input program  does {\em not halt,}
the {diagonalization program} will {\em halt} immediately.

The {diagonalization program} can be forced to execute a paradoxical task by
receiving {\em its own program code} as input.
This is so because, by considering the {diagonalization program,}
the {hypothetical halting algorithm} steers the {diagonalization program} into
{\em halting} if it discovers that it {\em does not halt;}
conversely,  the {hypothetical halting algorithm} steers the {diagonalization program} into
{\em not halting} if it discovers that it {\em halts.}

The contradiction obtained in applying the {\em  diagonalization program} to its own code proves that this program
and, in particular, the {hypothetical halting algorithm} cannot exist.
A slightly revised form of the proof (using quantum diagonalizaton operators that are equivalent to a classical {\em derangement} or {\em subfactorial})
holds for quantum diagonalization \citep{1612095},
as quantum information could be in a fifty-fifty fixed-point halting state.
Procedurally, in the absence of any fixed-point halting state,
the aforemetioned task might turn into a nonterminating
alteration of oscillations between halting and nonhalting states \citep{Kauffman198753}.

A universal computer
can in principle be embedded into, or realized by, certain physical systems designed to universally compute.
An example of such a physical system is the computer on which I am currently typing this chapter.
Assuming unbounded space [i.e., memory \citep{calude-staiger-09}] and time,
it follows by
reduction \citep{wolfram84,kanter,moore,wolfram85b,dc-d91a,dc-d91b,suppes-1993,svozil-93,1994IJTP...33.1085H,casti:94-onlimits_book,CalCamSvo-Stef-1995,casti:96-onlimits,barrow-impossibilities}
that there exist physical observables,
in particular, forecasts about whether or not an embedded computer will ever
halt in the sense sketched earlier,
that are provably undecidable.

\subsection{The busy beaver function as the maximal recurrence time}

The busy beaver function \citep{rado,chaitin-ACM,dewdney,brady}
addresses the following
question: suppose one considers all  programs (on a particular computer)
up to length (in terms of the number of symbols) $n$.
What is the {\em largest number} producible by such a program before halting?
(Note that non-halting programs, possibly producing an infinite number, e.g., by a non-terminating loop, do not apply.)
This number may be called the {\em  busy beaver function} of $n$.
The first values of a certain universal computer's busy beaver function with two states and n symbols
 are, for $n=$ 2, 3, 4, 5, 7 and 8, known to be, or estimated by \citep{dewdney,brady},
 4, 6, 13, greater than $10^{3}$, greater than  $10^{4}$, and greater than
$10^{44}$.

Consider a related question: what is the upper bound of running time  --  or,
alternatively, recurrence time  --  of a program of length $n$ bits before
terminating  or, alternatively, recurring?
An answer to this question will explain just how long we have to
wait for the most time-consuming program of length $n$ bits to
halt. That, of course, is a worst-case scenario. Many programs of
length $n$ bits will have halted long  before the maximal halting time.
We mention without proof \citep{chaitin-ACM,chaitin-bb}  that
this bound can be represented by the busy beaver function.

Knowledge of the maximal halting time would solve the halting
problem quantitatively
because if the maximal halting time were known
and bounded by any computable function of the program size of $n$ bits,
one would have to wait
just a little longer than the maximal halting time to make sure
that every program of length $n$  --  also this particular program, if it is destined for termination  --
has terminated.
Otherwise, the program would run forever.
Hence, because of the recursive unsolvability of the halting problem
the maximal halting time cannot be a computable function.
Indeed, for large values of $n$, the maximal halting time explodes and
grows faster than any computable function  of $n$.

By reduction, upper bounds for the recurrence of any kind of physical behavior can be obtained;
for deterministic systems representable by $n$ bits,
the maximal recurrence time grows faster than any computable number
of $n$.
This bound from below for possible behaviors may be interpreted quite generally
as a measure
of the impossibility to predict and forecast such behaviors by algorithmic means.

\subsection{Undecidability of the induction problem}

Induction, in physics, is the inference of general rules
dominating and generating physical behaviors from these behaviors alone.
For any deterministic system strong enough to support
universal computation, the general induction problem
is provable unsolvable.
Induction is thereby reduced to the unsolvability of
the rule inference problem \citep{go-67,blum75blum,angluin:83,ad-91,li:92}
of identifying a rule or law reproducing the behavior of a deterministic system
by observing its input-output performance by purely algorithmic means
(not by intuition).

Informally, the algorithmic idea of the proof is to take any sufficiently powerful
rule or method of induction and, by using it, to define some
functional behavior that is not identified by it.
This amounts
to constructing an algorithm which
(passively)
fakes the guesser by simulating some particular function
until the guesser
pretends to be able to guess the function correctly.
In a second,  diagonalization step, the faking algorithm then switches to a different
 function to invalidate the guesser's guess.

%
%
%

One can also interpret this result in terms of the recursive
unsolvability of the halting problem, which in turn is related to the busy beaver function;
there is no recursive bound on the
time the guesser has to wait  to make sure that the guess is
correct.

\subsection{Impossibility}

Physical tasks which would result in paradoxical
behavior \citep{hilbert-26} are impossible to perform.
One such task is the solution of the general halting problem, as discussed earlier.
Thus omnipotence appears infeasible, at least as long as one sticks to the usual
formal rules opposing inconsistencies \cite[163]{hilbert-26}.

Another such paradoxical task (requiring substitution and self-reference) can be forced upon  {\it La Bocca della Verit\'a} (Mouth of Truth),
located in the {\it portico} of the church of {\it Santa Maria in Cosmedin} in Rome.
It is believed that if one tells a lie with one's hand in the mouth of the sculpture,
the hand will be bitten off; another less violent legend has it that anyone sticking a hand in the mouth while
uttering a false statement will never be able to pull the hand back out.
\citet[178]{rucker} once allegedly put in his hand in the sculpture's mouth uttering, {``I will not be able to pull my hand back out.''}
The author leaves it to the reader to imagine {\it La Bocca della Verit\'a}'s confusion when confronted with such as statement!

There is a {\em pandemonium} of conceivable physical tasks \citep{barrow-impossibilities},
some quite entertaining \citep{smullyan-78}, which would result in paradoxical behavior and
are thus impossible to perform.
Some of these tasks are pre-G\"odelian and merely require {\em substitution}.

For the sake of demonstrating paradoxical substitution and the resulting impossibility,
consider the following  {\em printing task} discussed by \citet[2--4]{smullyan-92}.
Let the expressions  (not), (printable), (self-substitute),
have a standard interpretation in terms of negation, printing,
 and  self-reference by substitution
[i.e., if $X$ is some expression formed by the earlier three expressions and brackets, then
{(self-substitute)}$(X)=X(X)$], respectively,
and define {(not)(printable)}$(X)$  for arbitray expressions $X$
to be true if and only if $X$ cannot be printed.
Likewise,   {(not)(printable)}{(self-substitute)}$(X)$  is defined
to be true if and only if {(self-substitute)}$X$ cannot be printed.
Whatever the rules deriving expressions (subject to the notion of truth defined earlier) may be,
as long as the system is consistent and produces only true propositions (and no false ones),
within this small system,
the following proposition is {\em true but unprintable:}
{(not)(printable)(self-substitute)}$[${(not)(printable)(self-substitute)}$]$.
By definition, this proposition is true if and only if  (self-substitute)$[${(not)(printable)(self-substitute)}$]$
cannot be printed.
As per definition, (self-substitute)$[${(not)(printable)(self-substitute)}$]$
is just {(not)(printable)(self-substitute)}$[${(not)(printable)(self-substitute)}$]$,
the proposition is true if and only if it is not printable.
Thus the proposition  is either true and cannot be printed,
or it is printable and thus false.
The latter alternative is excluded by the assumption of consistency.
Thus one is left with the only consistent alternative that the proposition
{(not)(printable)(self-substitute)}$[${(not)(printable)(self-substitute)}$]$
is true but unprintable.
Note also that, since its negation {(printable)(self-substitute)}$[${(not)(printable)(self-substitute)}$]$
is false, it is also not printable (by the consistency assumption),
and hence {(printable)(self-substitute)}$[${(not)(printable)(self-substitute)}$]$
is an example of a proposition which is undecidable within the system --
neither it nor its negation will ever be printed in a consistent formalized system with the
notion of truth defined earlier.

\subsection{Results in classical recursion theory with implications for theoretical physics}

The following theorems of recursive  analysis \citep{aberth-80,Weihrauch} have some
implications for theoretical physics \citep{kreisel}:
(1)
There exist recursive monotone bounded sequences of rational numbers
whose limit is no computable number
\citep{Specker49}.
A concrete example of such a number is Chaitin's Omega number \citep{chaitin3,calude:02,calude-dinneen06},
the halting probability for a computer (using prefix-free code),
which can be defined by a sequence of rational numbers
with no computable rate of convergence.
(2)
There exist a recursive real function which has its maximum in the unit interval
at no recursive real number \citep{Specker57}.
This has implications for the principle of least action.
(3)
There exists a real number $r$ such that $G(r) = 0$ is recursively undecidable for $G(x)$
in a class of functions which involves polynomials and the sine function
\citep{wang}.
This, again, has some bearing on  the principle of least action.
(4)
There exist incomputable solutions of the wave equations for computable initial values
\citep{pr1,bridges1}.
(5)
On the basis of theorems of recursive analysis \citep{Scarpellini-63,richardson68},
many questions in dynamical systems theory are provable undecidable \citep{1985cfd..book.....F,dc-d93,Stewart-91,calude:037103}.

\section{Deterministic chaos}

The wording {\em deterministic chaos} appears to be a {\it contradictio in adjecto},
indicating a hybrid form of chaotic behavior in deterministic systems \citep{li-83,nld-chaos}.
Operationally, it is characterized by the practical impossibility of forecasting the future
because the system is unstable \citep{Lyapunov-92} and very sensitive
to tiny variations of the initial state.
Because the initial state can only be determined with finite accuracy,
its evolution will soon become totally unpredictable.

\subsection{Instabilities in classical motion}

In 1885 King Oscar II of Sweden and Norway, stimulated by Weierstrass, Hermite, and Mittag-Leffler,
offered a prize to anybody contributing toward the solution of the so-called {\em $n$-body problem} \cite[2]{weierstrass-1885}:
\begin{quote}
{
Given a system of arbitrarily many mass points that attract each
according to Newton's law, try to find, under the assumption that no two points ever collide,
 a representation of the coordinates of each point
as a series in a variable that is some known function of time and for
all of whose values the series converges uniformly.
}
\end{quote}

The prize-winning work was expected to render systematic techniques toward a solution to {\em stable} motion
such that systems whose states start out close together will stay close together forever \cite[69]{Diacu96-ce}.
To everyone's surprise, the exciting course of events \citep{peterson-NC,Diacu96,Diacu96-ce}
resulted in Poincar{\'e}'s prize-winning centennial revised contribution \citep{poincare-1890},
which predicted unexpected and irreducible {\em instabilities} in the mechanical motion of bodies.
Poincar{\'e} was led to the conclusion that sometimes small
variations in the initial state could lead to huge variations in the
evolution of a physical system at later times.
In Poincar{\'e}'s own words \cite[chapt.~4, sect.~2,  56--57]{poincare14}:
\begin{quote}
{
If we would know the laws of nature and the state of the Universe precisely
for a certain time,
we would be able to predict with certainty
the state of the Universe for any later time.
But
$\ldots$
it can be the case that small differences in the initial values
produce great differences in the later phenomena;
a small error in the former may result in a large error in the latter.
The prediction becomes impossible and we have a ``random phenomenon.''}
\end{quote}

Note that Poincar{\'e} adheres to a Laplacian-type determinism
but recognizes the possibility that systems whose states start out close together
will stay close together {\em for a while} \cite[69]{Diacu96-ce}
and then diverge into totally different behaviors.
Today such behaviors are subsumed under the name {\em deterministic chaos.}
In chaotic systems, it is practically impossible to specify
the initial value precise enough to allow long-term predictions.

Already in 1873, Maxwell mentioned \cite[211-212]{Campbell-1882}
\begin{quote}
{
When an infinitely small variation in the present state may bring about a finite difference in the state of the
system in a finite time, the condition of the system is said to be unstable.
It is manifest that the existence of unstable conditions renders impossible the prediction of future events, if our
knowledge of the present state is only approximate, and not accurate.
}
\end{quote}
Maxwell also discussed unstable states of high potential energy whose
spontaneous \citep{frank}  decay or change \cite[212]{Campbell-1882}
{``requires an expenditure of work, which in certain cases may be
infinitesimally small, and in general bears no definite proportion to the energy developed in consequence thereof.''}

Today, after more than a century of research into unstable chaotic motion, {\em symbolic dynamics}
identified the
 {\em Poincar{\'e} map near a homocyclic orbit},
the {\em horseshoe map} \citep{smale-hm},
and the {\em shift map}
as equivalent origins of classical deterministic chaotic motion,
which is characterized by a {\em computable evolution law}
and the {\em sensitivity}  and instability with respect to variations of the
{\em initial value} \citep{shaw,li-83,nld-chaos}.

This scenario can be demonstrated by considering the shift map $\sigma$ as it
pushes up dormant information residing in the successive bits of the initial state represented by the sequence
$s=0.\text{(bit~1)}\text{(bit~2)}\text{(bit~3)}\cdots$,
thereby truncating the bits before the comma;
that is, $\sigma (s)= 0.\text{(bit~2)}\text{(bit~3)}\text{(bit~4)}\cdots$,
$\sigma (\sigma (s))= 0.\text{(bit~3)}\text{(bit~4)}\text{(bit~5)}\cdots$, and so on.
Suppose a measurement device operates with a precision of, say, two bits after the comma,
indicated by a two bit window of measurability;  thus intially
all information beyond the second bit after the comma is hidden to the experimenter.
Consider two initial states
$s=[0.\text{(bit~1)}\text{(bit~2)}] \text{(bit~3)}\cdots$ and
$s'=[0.\text{(bit~1)}\text{(bit~2)}] \text{(bit~3)}'\cdots$,
where the square brackets
indicate the boundaries of the window of measurability (two bits in this case).
Initially, as the representations of both states start with the same two bits after the comma
$[0.\text{(bit~1)}\text{(bit~2)}]$,
these states appear operationally identical and cannot be discriminated experimentally.
Suppose further that, after the second bit, when compared,
the successive bits
$\text{(bit }i\text{)}$ and $\text{(bit }i\text{)}'$
in both state representations at identical positions $i=3,4,\ldots$ are totally
independent and uncorrelated.
After just two iterations of the shift map $\sigma$, $s$ and
$s'$
may result in totally  different, diverging observables
$\sigma (\sigma (s))= [0.\text{(bit~3)}\text{(bit~4)}]\text{(bit~5)}\cdots$
and
$\sigma (\sigma (s'))= [0.\text{(bit~3)}'\text{(bit~4)}']\text{(bit~5)}'\cdots$.

If the initial values are {\em defined} to be elements of
a continuum, then almost all (of measure one) of them
are not representable by any algorithmically compressible number;
in short,  they are random \citep{MartinLöf1966602,calude:02}.
Classical deterministic chaos results from the assumption of such a random initial value
 --  drawn somehow [one needs the {\em axiom of choice} \citep{wagon1,svozil-set} for doing this]
from the continuum urn  --  and the unfolding of the information contained therein by a recursively enumerable (computable),
deterministic (temporal evolution) function.
Of course, if one restricts the initial values to finite sets,
or, say, to the rationals, then the behavior will be periodic.
The randomness of classical, deterministic chaos resides
in the {\em assumption of the continuum};
an assumption which might be considered a convenience (for the sake of applying the infinitesimal calculus),
as it is difficult to conceive of any convincing
physical operational evidence supporting the full structure of continua.
If the continuum assumption is dropped, then what remains is Maxwell's
and Poincar{\'e}'s observation of the unpredictability
of the behavior of a deterministic system
due to instabilities and diverging evolutions from almost identical initial states \citep{Lyapunov-92}.

\subsection{Rate of convergence}

The connections between symbolic dynamical systems and universal computation
result in provable unknowables \citep{dc-d93,Stewart-91}.
These symbolic dynamic unknowables are different in type from the dynamical instabilities,
and should be interpreted recursion theoretically, as outlined in Section~\ref{2010-unknowable-s2}.

Let us come back to the original $n$-body problem.
About one hundred years after its formulation, as quoted earlier,
the $n$-body problem has been solved \citep{1969VeLen...7..121B,Babadzanjanz-1979,Wang91,Diacu96,Wang01,Babadzanjanz-1993,Babadzanjanz-2006}.
The three-body problem was already solved by \citet{Sundman12}.
The solutions are given in terms of  convergent power series.

Yet, to be practically applicable,
the rate of convergence of the series must be computable and even reasonably good.
One might already expect from symbolic dynamics,
in particular, from chaotic motion,
that these series solutions could converge very slowly.
Even the short-term prediction of future behaviors
may require the summation of a huge number of terms,
making these series unusable for all practical purposes
\citep{Diacu96,rousseau-2004}.

Alas, the complications regarding convergence may be more serious.
Consider a universal computer based on the $n$-body problem.
This can, for instance, be achieved by ballistic computation, such as the
``Billiard Ball'' model of computation
\citep{fred-tof-82,margolus-02}
that effectively embeds a universal computer into an  $n$-body system \citep{svozil-2007-cestial}.
It follows by reduction that certain predictions,
say, for instance, the general halting problem, are impossible.

What are the consequences of this reduction for the convergence of the series solutions?
It can be expected that not only do the series converge very slowly,
like in deterministic chaos,
but that, in general, there does not exist any computable rate of convergence
for the series solutions of particular observables.
This is very similar to the busy beaver function or to Chaitin's Omega number \citep{chaitin3,calude:02},
representing the halting probability of a universal computer.
The Omega number can be enumerated
by series solutions from quasi-algorithms
computing its very first digits \citep{calude-dinneen06}.
Yet, because of the incomputable growth of the time
required to determine whether certain summation terms corresponding to halting programs possibly contribute,
the series lack any computable rate of convergence.

Though it may be possible to evaluate
the state of the $n$ bodies by Wang's power series solution
for any finite time with a computable rate of convergence,
global observables, referring to (recursively) unbounded times, may be incomputable.
Examples of global observables correspond to solutions of certain decision problems
such as the stability of some solar system (we do not claim that this is provable incomputable), or the
halting problem.

This, of course,
stems from the metaphor and robustness of universal computation
and the capacity of the $n$-bodies to implement universality.
It is no particularity or peculiarity of Wang's power series solution.
Indeed, the troubles reside in the capacity to implement substitution, self-reference, universal computation,
and Peano arithmetic
by $n$-body problems.
Because of this capacity, there cannot exist other formalizable methods,
analytic solutions, or approximations capable of deciding and computing certain decision problems
or observables for the $n$-body problem.

\section{Quantum unknowables}

In addition to provable physical unknowables by reduction to recursion-theoretic ones,
and chaotic symbolic dynamic systems, a third group of physical unknowables resides in the quantum domain.
Although it has turned out to be a highly successful theory,
quantum mechanics, in particular, its interpretation and meaning,
has been controversially received within the physics community.
Some of its founding fathers, like Schr\"odinger and, in particular,  Einstein,
considered quantum mechanics to be an unsatisfactory theory:
Einstein, Podolsky and Rosen \citeyearpar{epr,ein-reply} argued
that there exist counterfactual \citep{svozil-2006-omni,vaidman:2009}
ways to infer observables from experiment that, according to quantum mechanics, cannot coexist simultaneously;
hence quantum mechanics cannot predict what experiment can (counterfactually) measure.
Thus quantum mechanics is {\em incomplete} and should eventually be substituted by a more complete theory.
Others, among them Born, Bohr, and Heisenberg,
claimed that unknowability in quantum mechanics is irreducible,  is ontic, and will remain so forever.
Over the years, the latter view seems to have prevailed
\citep{fuchs-peres,Bub-1999}, although not totally unchallenged
\citep{jammer:66,jammer1,jammer-92}.
Already Sommerfeld warned his students not to get
into the meaning behind quantum mechanics,
and as mentioned by \citet{clauser-talkvie},
not long ago, scientists working in that field
had to be very careful not to become discredited as {\em quacks.}
Richard Feynman \cite[129]{feynman-law}
once mentioned the
\begin{quote}
{perpetual torment that results
from [the question], ``But how can it be like that?'' which
is a reflection of uncontrolled but utterly vain desire to see
[quantum mechanics] in terms of an analogy with something familiar.
$\ldots$
Do not keep saying to yourself, if you can possibly avoid it,
``But how can it be like that?''
because you will get ``down the drain,'' into a blind alley from which nobody has yet
escaped.}
\end{quote}
This antirationalistic postulate of  irreducible indeterminism and meaninglessness came
after a period of fierce debate on the quantum foundations,
followed by decades of vain attempts to complete quantum mechanics in any operationally testable way,
and after the discovery of proofs of the incompatibility of local, realistic, context-independent ways to complete
quantum mechanics \citep{clauser,mermin-93}.

In what follows, we shall discuss three realms of quantum unknowables:
(1) randomness of single events,
(2) complementarity, and
(3) value indefiniteness.

\subsection{Random individual events}

In 1926, \citet[866]{born-26-1} [see an English translation in \citet[54]{wheeler-Zurek:83}] postulated that
\begin{quote}
{``from the standpoint of our quantum mechanics, there is no quantity
which in any individual case causally fixes the consequence of the collision;
but also experimentally we have so far no reason to believe that there are some inner properties of the atom
which condition a definite outcome for the collision.
Ought we to hope later to discover such properties $\ldots$  and determine them in individual cases?
Or ought we to  believe that the agreement of theory and experiment  --  as to the impossibility
of prescribing conditions? I myself am inclined  to give up determinism in the world of atoms.''
}
\end{quote}

Furthermore, Born suggested that, though {\em individual particles behave irreducibly indeterministic},
the {\em quantum state evolves deterministically}  in a strictly Laplacian causal way.
Indeed, between (supposedly irreversible) measurements the (unitary) quantum state evolution
is even reversible, that is, one-to-one, and amounts to a  generalized (distance preserving) rotation in complex Hilbert space.
In Born's \citeyearpar[804]{born-26-2} [see an English translation in \citet[302]{jammer:89}] own words,
\begin{quote}
{the motion of particles conforms to the laws of probability, but the probability itself
is propagated in accordance with the law of causality.
[This means that knowledge of a state in all points in a given time determines the distribution of
the state at all later times.]
}
\end{quote}

This distinction between a reversible, deterministic evolution of the quantum state, on one hand,
and the irreversible measurement, on the other hand, has left some physicists with an uneasy feeling;
in particular, because of the possibility to erase
\citep{PhysRevD.22.879,PhysRevA.25.2208,greenberger2,Nature351,Zajonc-91,PhysRevA.45.7729,PhysRevLett.73.1223,PhysRevLett.75.3783,hkwz}
measurements by reconstructing the quantum state,
accompanied by a complete loss of the information obtained from the quantum state before the (undone) measurement --
unlike in classical reversible computation \citep{bennett-73,bennett-82,maxwell-demon},
which still allows copying, that is, one-to-many operations, the quantum state evolution is strictly one-to-one.
Indeed, the possibility to undo measurements on quantum states appears to be  not bound by any fundamental principle,
and limited merely by the experimenter's  technological capacities.
Stated pointedly, it would in principle be possible to undo all measurements,
yet this cannot be accomplished  most of the time (for almost all measurements)
for all practical purposes~\cite{bell:a1}.
But then, one could speculate, Born's statement seems to suggest that the deterministic state evolution uniformly prevails.
Pointedly stated, if, at least in principle, there is no such thing as an irreversible measurement,
and the quantum state evolves uniformly deterministically, why should there exist indeterministic individual events?
In this view, the insistence in irreversible measurements as well as in an irreducible indeterminism associated with individual quantum events
appears to be an idealistic, subjective illusion -- in fact, this kind of indeterminism depends
on measurement irreversibility and decays into thin air if the latter is denied.

Similar arguments have been brought forth by \citet{everett} and \citet{schroedinger-interpretation}.
Note that it is not entirely clear [and indeed remains conventional \citep{svozil-2001-convention}]
where exactly the measurement cut \citep{wigner:mb,roessler-98}
between the observer and the object is located.
By assuming the universal applicability of quantum mechanics,
the object and the measurement apparatus could be uniformly {\em combined} into a
larger system whose quantum mechanical evolution should be deterministic;
otherwise quantum mechanics would not be universally valid.
Such frameworks hardly offer objective opportunities for indeterminism besides subjective ones --
in the {\em many worlds} resolution \citep{everett}, every one
of many simultaneous observers branching off to different universes
subjectively experiences the arbitrariness of the occurrence of events as indeterminism.
(This resembles the perception of a particular sequence of bits as compared to all possible ones.)

Alas, the deterministic evolution of the quantum state could result in the
{\em superposition} of classically contradictory states.
One of the mind-boggling, perplexing and counterintuitive consequences associated with this coexistence of classical contradictions is
Schr\"odinger's \citeyearpar[812]{schrodinger} cat paradox
implying the simultaneous coexistence of death and life of a macroscopic object such as a mammal.
Another one is Everett's \citeyearpar{everett} aforementioned many-worlds interpretation suggesting that
our universe perpetually branches
off into zillions of consistent alternatives.

Thus one is faced with a {\em dilemma:}
either to accept a somehow spurious nonuniformity in the evolution of the quantum state
during (irreversible) measurement processes -- an {\it ad hoc} assumption challenged by quantum erasure experiments --
or being confronted with the counterintuitive decay of quantum states into superpositions of
classically mutually exclusive states -- a sort of jelly -- not backed by our everday
experience as conscious beings (although often ambivalent we usually dont reside in mental ambiguity for too long).
\citet[19--20]{schroedinger-interpretation} sharply addressed the difficulties of a quantum theorist
coping with this aspect of the quantum formalism:
\begin{quote}
{
The idea that  [the alternate measurement outcomes] be not alternatives but {\em all} really happening simultaneously
seems lunatic to [the quantum theorist], just {\em impossible.}
He thinks that if the laws of nature took {\em this} form for,
let me say,
a quarter of an hour, we should find our surroundings rapidly turning into a quagmire, a sort of a featureless jelly or plasma,
all contours becoming blurred, we ourselves probably becoming jelly fish.
It is strange that he should believe this.
For I understand he grants that unobserved nature does behave this way -- namely according to the wave equation.
$\ldots$ according to the quantum theorist, nature is prevented from rapid
jellification only by our perceiving or observing it.
}
\end{quote}

If, however, an additional irreducible irreversible evolution or some other,
possibly environmental \citep{PhysRevD.22.879,RevModPhys.75.715},
effect associated with measurements (and  the collapse of the quantum wave function) is  postulated
or somehow emerges,
individual events may occur indeterministically.
The considerations might appear to be sophistries,  but
they have direct consequences for the supposedly most advanced random number generators of our time.
These devices operate with beam splitters
\citep{svozil-qct,rarity-94,zeilinger:qct,stefanov-2000,wang:056107,PhysRevA.82.022102},
which are strictly reversible
\citep{Mandel-Ou1987118,green-horn-zei,zeilinger:882,svozil-2004-analog}
-- one could demonstrate reversibility on beam splitters by  forming a Mach-Zehnder interforemeter with two
serially connected ones --
or parametric down-conversions  and entanglement \citep{0256-307X-21-10-027,fiorentino:032334,10.1038/nature09008}.

Born did not address these questions, nor did he specify the formal notion of indeterminism to which he was relating.
So far, no mathematical characterization of quantum randomness has been proved \citep{2008-cal-svo}.
In the absence of any indication to the contrary, it is mostly implicitly assumed
that quantum randomness is of the strongest possible kind,
which amounts to postulating that the symbolic sequences
associated with measurement outcomes are uncomputable
or even algorithmically incompressible.

Indeed, the quantum formalism does not predict the outcome of single events
when there is a mismatch between the context in which a state was prepared,
and the context in which it is measured.
Here, the term {\em context} \citep{svozil-2006-omni,svozil-2008-ql}
denotes a maximal collection of comeasurable observables,
or, more technically,
the maximal operator from which all commuting operators can be functionally derived
\cite[sect.~84]{halmos-vs}.
Ideally, a quantized system can be prepared
to yield exactly one answer in exactly one context \citep{zeil-99,DonSvo01,svozil-2002-statepart-prl}.
Other outcomes associated with other contexts occur indeterministically \citep{2008-cal-svo}.

Furthermore, the quantum formalism is incapable of predicting deterministically the radioactive decay of individual particles.
Attempts to find causal laws lost steam \citep{Kragh-1997AHESradioact,Kragh-2009_RePoss5} at the time of Born's suggestion
of the indeterministic interpretation of individual measurement outcomes,
and nobody has come up with a operationally satisfactory deterministic prediction since then.

In the absence of other explanations,
it is not too unreasonable to pragmatically presume that these single events occur without any causation
and thus at random.
Presently, this appears to be the prevalent opinion among physicists.
Such random quantum coin tosses
\citep{svozil-qct,rarity-94,zeilinger:qct,stefanov-2000,0256-307X-21-10-027,wang:056107,fiorentino:032334,svozil-2009-howto,10.1038/nature09008}
have been used for various purposes, such as delayed choice experiments
\citep{wjswz-98,zeilinger:qct}.

Note that randomness of this type \citep{Cris04,calude-dinneen05}
is postulated rather than proved and thus, unless disproved, remains conjectural.
This is necessarily so, for any claim of randomness can only be corroborated  {\em relative} to, and
{\em with respect} to, a more or less large class of laws or behaviors;
it is impossible to inspect the hypothesis against an infinity of  --  and even less so all  --  conceivable laws.
To rephrase a statement about computability \cite[11]{davis-58}, how can we ever exclude the possibility of our
presented, some day (perhaps by some extraterrestrial visitors), with a (perhaps
extremely complex) device  that computes and predicts
a certain type of hitherto random physical phenomenon?

\subsection{Complementarity}

{\em Complementarity} is the impossibility of measuring
two or more complementary observables
with arbitrary precision simultaneously.
In 1933, \citet[7]{pauli:58} gave the first explicit definition of {\em complementarity} stating that
[see the partial English translation in \cite[369]{jammer:89}]
\begin{quote}
{
in the case of  an indeterminacy of a property of a system at a certain configuration
(at a certain state of a system), any attempt to measure the respective property (at least partially)
annihilates the influence of the previous knowledge of the system on the (possibly statistical) propositions
about possible later measurement results.
$\ldots$
The impact
on the system by the  measurement apparatus for momentum (position) is such that
within the limits of the uncertainty relations
the value of the knowledge of the previous position (momentum) for the
prediction of later measurements of position and momentum is lost.
}
\end{quote}

Einstein, Podolsky, and Rosen \citeyearpar{epr} challenged quantum complementarity (and doubted the completeness of quantum theory)
by utilizing a configuration
of two entangled \citep{schrodinger,CambridgeJournals:1737068,CambridgeJournals:2027212} particles.
They
claimed to be able to empirically infer two different complementary contexts counterfactually simultaneously,
thus circumventing quantum complementarity.
Thereby, one context is measured on one side of the setup, whereas the other context is measured on the other side of it.
By the uniqueness property \citep{svozil-2006-uniquenessprinciple}  of certain two-particle states,
knowledge of a property of one particle entails the certainty
that, if this property were measured on the other particle as well, the outcome of the measurement would be
a unique function of the outcome of the measurement performed.

This makes possible the measurement of one context {\em as well as} the {\em simultaneous counterfactual inference}
of a different complementary context.
Because, one could argue, although one has actually measured on one side a different,
incompatible context compared to the context measured on the other side,
if, on both sides, the same  context {\em would be measured},
the outcomes on both sides {\em would be uniquely correlated}.
(This can indeed be verified in another experiment.)
Hence, the Einstein, Podolsky, and Rosen argument continues,
measurement of one context per side is sufficient,
for the outcome could be counterfactually inferred on the other side.
Thus, effectively two complementary contexts are knowable.
Based on this argument, Einstein, Podolsky, and Rosen suggested that quantum mechanics must be considered incomplete,
because it cannot predict what can be measured; thus a more complete theory is needed.

Complementarity was first encountered in quantum mechanics,
but it is a phenomenon also observable in the classical world.
To get  better intuition of complementarity, we shall consider generalized urn models \citep{wright,wright:pent} or,
equivalently \citep{svozil-2001-eua},
finite deterministic automata \citep{e-f-moore,svozil-93,schaller-96,dvur-pul-svo,cal-sv-yu} in an unknown initial state.
Both quasi-classic examples mimic complementarity to the extent that even quasi-quantum cryptography
can be performed with them \citep{svozil-2005-ln1e} as long as value indefiniteness is not a feature of the protocol
\citep{PhysRevLett.85.3313,2010-qchocolate},
that is, for instance,
the Bennett and  Brassard \citeyearpar{benn-84} protocol \citep{benn-92} can be implemented with generalized urn models,
whereas the Ekert protocol
\citep{ekert91} cannot.

A generalized urn model is
characterized by an ensemble of balls with black background color.
Printed on these balls are some color symbols.
Every ball contains just one  symbol per color.
Further assume some filters or eyeglasses that are
perfect because they totally absorb light of all other colors
but a particular  one.
In that way, every color can be associated with a particular pair of eyeglasses and {\it vice versa.}

When a spectator looks at a ball through such a particular pair of eyeglasses,
the only operationally recognizable symbol will be the one in the particular
color that is transmitted through the eyeglasses.
All other colors are absorbed, and the symbols printed on them will appear black
and therefore will not be differentiable from the black background.
Hence the ball will appear to carry a different message or symbol,
depending on the color with which it is viewed.

For the sake of demonstration, let us consider a generalized urn model with four
ball types, two colors, say red and green, and two symbols, say ``0'' and ``1,'' per color, that is,
ball type~1:~({\color{red}red~0}~{\color{green}green~0}),
ball type~2:~({\color{red}red~0}~{\color{green}green~1}),
ball type~3:~({\color{red}red~1}~{\color{green}green~0}),  and
ball type~4:~({\color{red}red~1}~{\color{green}green~1}).
The {green} pair of eyeglasses associated with the green observable
allows the observer to differentiate between
ball types 1 or 3 (associated with the green symbol ``0''),
and
ball types 2 or 4 (associated with the green symbol ``1'').
The {red} pair of eyeglasses  associated with the red observable
allows the observer to differentiate between
ball types 1 or 2 (associated with the green symbol ``0''),
and
ball types 3 or 4 (associated with the green symbol ``1'').
[Without going into details in general this yields sets of partitions of the set of ball types
resulting in partition logics \cite[chapt.~10]{svozil-93}.]

The difference between the balls and the quanta is the possibility
of viewing all the different symbols on the balls
in all different colors by taking off the eyeglasses;
also, one can consecutively look at one and the same ball with differently colored pair of eyeglasses,
thereby identifying the ball completely.
Quantum mechanics does not provide us with  a possibility to look across the quantum veil,
as it allows neither a global, simultaneous measurement of all complementary observables
nor a measurement of one observable without disturbing the measurement of another complimentary observable
(with the exception of Einstein, Podolsky, and Rosen counterfactual measurements discussed earlier).
On the contrary, there are strong formal arguments suggesting
that the assumption of a simultaneous
physical coexistence of such complementary observables yields a complete contradiction.
These issues will be discussed next.

\subsection{Value indefiniteness {\it versus} omniscience}

Still another quantum unknowable results from the fact that no global
(in the sense of all or at least certain finite sets of  complementary observables)
classical truth
assignment exists which is consistent with even a finite number of local (in the sense of comeasurable) ones,
that is, no consistent classical truth table can be given by pasting together the possible outcomes of measurements of certain complementary observables.
This phenomenon is also known as {\em value indefiniteness} or, by an option to interpret this result, {\em contextuality}  (see later).
Here the term {\em local} refers to a particular context \citep{svozil-2008-ql}
that, operationally, should be thought of as the collection of all comeasurable or copreparable \citep{zeil-99} observables.
The structure of quantum propositions \citep{birkhoff-36,kochen3,kalmbach-83,kalmbach-86,pulmannova-91,nav:91,svozil-ql}
can be obtained by pasting contexts together.

As by definition, only {\em one} such context is directly measurable,
arguments based on more than one context must necessarily involve counterfactuals \citep{svozil-2006-omni,vaidman:2009}.
A {\em counterfactual} is a would-be-observable or
{\em contrary-to-fact conditional}
\citep{chisholm-46}
which has not been measured but potentially could have been measured
if an observer would have decided to do so; alas the observer decided to measure a different, presumably complementary, observable.

Already scholastic philosophy,
for instance, Thomas Aquinas,
considered similar questions such as whether God has knowledge of
non-existing  things \cite[part one, question 14, article 9]{Aquinas} or things
that are not yet \cite[part one, question 14, article 13]{Aquinas};
see also Specker's \citeyearpar[243]{specker-60}  reference to {\it infuturabilities}.
Classical omniscience, at least its naive expression that,
if a proposition is true, then an omniscient agent (such as God) knows that it is true,
is plagued by controversies and paradoxes.
Even without evoking quantum mechanics, there exist bounds on omniscience because of the self-referential
perception of intrinsic observers endowed with free will:
if such an observer is omniscient and has absolute predictive power,
then free will could counteract omniscience and, in particular, the observer's own predictions.
Within a consistent formal framework, the only alternative is to either abandon free will,
stating that it is an idealistic illusion,
or  accept that omniscience and absolute predictive power is bound by paradoxical self-reference.

The empirical sciences implement classical omniscience by assuming that
in principle, all observables of classical physics are comeasurable without any restrictions,
regardless of whether they are actually measured.
No ontological distinction is made between an observable obtained by an actual and a potential or counterfactual measurement.
[In contrast, compare Schr\"odinger's \citeyearpar[sect.~7]{schrodinger} own epistemological interpretation of the wave function as a
{\em catalog of expectations.}]
Classically, precision and comeasurability are limited only by the technical capacities of the experimenter.
The principle of empirical classical omniscience has given rise to the realistic believe that
all observables exist regardless of their observation, that is, regardless and independent of
any particular measurement.

Physical (co-)existence is thereby related to the realistic assumption
[sometimes referred to as the ``ontic'' \citep{atman:05} viewpoint] that \citep{stace}
``some entities sometimes exist without being experienced by any finite mind.''
With regards to such unexperienced counterfactual entities,
\citet[364, 365, 368]{stace}  questions their existence (compare also Schr\"odinger's remark quoted earlier):
\begin{quote}
In front of me is a piece of paper. I assume that the realist believes
that this paper will continue to exist when it is put away in my desk for the night,
and when no finite mind is experiencing it.
$\ldots$
I will state clearly at the outset that I cannot prove that no entities exist without being experienced
by minds. For all I know completely unexperienced entities may exist,
but what I shall assert is that
$\ldots$
there is absolutely no reason for asserting that these non-mental, or physical,
entities ever exist except when they are being experienced,
and the proposition that they do so exist is utterly groundless and gratuitous,
and one which ought not to be believed.
$\ldots$
As regards [a] unicorn on Mars, the correct position,
as far as logic is concerned,
is obviously that if anyone asserts that there is a unicorn there,
the onus is on him to prove it;
and that until they do prove it,
we ought not to believe that they exist.
\end{quote}

One might criticize Stace's idealistic position by responding that
suppose an experimenter can choose which observable among a collection of  different, complementary,
observables is actually measured.
Regardless of this choice, a measurement of any  observable that {\em could} be measured
{\em would} produce some result.
This contrary-to-fact conditional
could be interpreted as an existing {\em element of physical reality.}
Furthermore, according to the argument of Einstein, Podolsky and Rosen \citeyearpar[777]{epr},
even certain sets of {\em complementary} counterfactual elements of physical reality
coexist
``if, without in any way disturbing a system, we can predict with certainty (i.e.,
with probability equal to unity) the value of [these] physical quantit[ies].''
The idealist  might repond that these arguments are unconvincing
because they are merely based on conterfactual inference and are thus empirically ``utterly groundless and gratuitous.''

The formal expression of classical omniscience is the Boolean algebra of observable propositions \citep{Boole},
in particular the abundance of two-valued states interpretable
as omniscience about the system.
Thereby, any such dispersionless quasi-classical two-valued state  --  associated with a truth assignment
 --  can be defined for all observables,
regardless of whether they have been actually observed.


After the discovery of complementarity, a further indication against quantum omniscience came from Boole's \citeyearpar{Boole-62}
{\em conditions of possible (classical) experience} which
are bounds for the occurrence of (classical) events that are derivable within classical probability theory
\citep{pitowsky-89a,pitowsky,Pit-94,2000-poly} for quantum probabilities and
quantum expectation functions.
\citet{bell-66} pointed out that experiments
based on counterfactually inferred observables
discussed by Einstein, Podolsky and Rosen \citeyearpar{epr} discussed earlier violate these  conditions of possible (classical) experience and thus
seem to indicate the impossibility
of a faithful embedding (i.e., preserving the logical structure)
of quantum observables into classical Boolean algebras.
Stated pointedly, under some (presumably mild) side assumptions,
{\em unperformed experiments have no results} \citep{peres222};
that is,
there cannot exist a table enumerating all
actual and hypothetical context independent (see later) experimental
outcomes consistent with the observed quantum frequencies \citep{zeilinger-epr-98,svozil_2010-pc09}.
As any such table could be interpreted as omniscience with respect to
the observables in the Boole-Bell-Einstein-Podolsky-Rosen-type experiments,
the impossibility to consistently enumerate such tables (under the noncontextual assumption)
appears to be a very serious indication against
omniscience in the quantum domain.

The quantum nonlocal (i.e., the particles are spatially separated)
correlations among observables in the Boole-Bell-Einstein-Podolsky-Rosen-type experiments
are stronger than classical in the sense that {\it ex post facto,} when the two outcomes are communicated and compared,
in the case of dichotomic observables, say ``0'' and ``1,''
for some measurement parameter regions,
there appear to be
{\em more equal} occurrences ``00'' or ``11'' and thus
{\em fewer unequal} occurrences ``01'' or ``10'' than could be classically accounted for;
likewise,
for other  measurement parameter regions,
there appear to be
{\em fewer equal} occurrences ``00'' or ``11'' and thus
{\em more unequal} occurrences ``01'' or ``10'' than could be classically accounted for.
These conclusions can only be drawn {\it in retrospect}, that is, after bringing together and comparing the outcomes.
Individual outcomes occur indeterministically and, in particular, independently of the  measurement parameter regions
[but not of outcomes \citep{shimony2}]
of other distant, measurements.
No faster-than-light signaling can occur.
Indeed, even stronger-than-quantum correlations would, in this scenario,
not violate relativistic causality \citep{pop-rohr,popescu-97b,svozil-krenn,svozil-2004-brainteaser}.

The reason that it is impossible to describe all quantum observables
simultaneously by classical tables of experimental outcomes
can be understood in terms of a stronger conclusion that,
for quantum systems whose Hilbert space is of dimension greater than two,
there does not exist any dispersionless quasi-classical, two-valued state
interpretable as truth assignment.
This conclusion, which is known as the
Kochen-Specker theorem
\citep{specker-60,kochen1,ZirlSchl-65,Alda,Alda2,kamber64,kamber65,mermin-93,svozil-ql,svozil-tkadlec,cabello-96,svozil-2008-ql},
has a finitistic proof by contradiction.
Proofs of the Kochen-Specker theorem  amount to  brain teasers in graph coloring
resulting in the fact that, for the geometric configurations considered,
there does not exist any possibility to consistently and context independently
enumerate and tabulate the values of all the observables occurring in a  Kochen-Specker-type argument \citep{cabello-96}.

The violations of conditions of possible classical experience in Boole-Bell-type experiments or
the Kochen-Specker theorem do not exclude realism restricted to a single context
but (noncontextual) realistic omniscience beyond it.
It may thus not be totally unreasonable to suspect that the assumption of \mbox{(pre-)}determined observables outside
a single context may be unjustified \citep{svozil-2003-garda}.

If one nevertheless {\em insists} in the simultaneous physical coexistence
of counterfactual observables,
any {\em forced} tabulation \citep{peres222,svozil_2010-pc09} of truth values for Boole-Bell-type or Kochen-Specker-type configurations
would  either result in a complete contradiction or
in {\em context dependence}, also termed {\em contextuality}, that is,
the outcome of a measurement of an observable would depend on what other comeasurable observables are measured
alongside  it \citep{bohr-1949,bell-66,hey-red,redhead,svozil-2008-ql}.

Indeed, the current mainstream interpretation
of the Boole-Bell-type or Kochen-Specker-type theorems is in terms of contextuality, that is, by assuming a dependence of the outcome
of a single observable {\em on what other observables} are actually measured
or at least what could have been consistently known alongside it.
This insistence in the coexistence of complementary observables could be interpreted as an attempt to rescue
classical omniscience accompanied by ontological realism at the price of accepting contextuality.
The realist \citet[451]{bell-66}
suggested that
``the result of an observation may reasonably depend $\ldots$
on the complete disposition of the apparatus.''
(Already \citet{bohr-1949} mentioned
``the impossibility of any sharp separation between the behaviour of atomic
objects and the interaction with the measuring instruments which serve to define the conditions
under which the phenomena appear.'')

For the sake of demonstrating contextuality \citep{svozil_2010-pc09} consider a dichotomic observable
(with outcomes ``0'' or ``1'').
Contextuality predicts that, when measured together with some particular set of  observables,
this observable yields a certain outcome, say ``0,''
whereas when measured together with another, complementary,
set of other observables, the observable may yield a different outcome, say ``1.''

However, statistically the quantum probability and expectation value of this observable is
noncontextual and thus  {\em independent} of the set of co-observables.
Thus contextuality is a hypothetical (counterfactual)
phenomenon regarding complementary measurements on an individual particle, making it inaccessible for direct tests.
Alas, as far as Einstein-Podolsky-Rosen-type measurements might reproduce such contextual behavior
for individual particles,
quantum mechanics predicts noncontextuality \citep{svozil:040102} and thus contradicts the assumption of quantum contextuality.
(Often claims of experimental evidence of quantum contextuality
do not deal with its individual particle character but deal with statistical violations of Boole-Bell-type or Kochen-Specker-type configurations.
The terms which contribute to (in)equalities are not measured on one and the same particle;
operationally they even originate in very different measurement setups.)
One may argue that contextuality occurs only when absolutely necessary, that is,
when the set of observables allows only an insufficient number of two-valued states for a
homeomorphic embedding into (classical) Boolean algebras; but
in view of the fact that quantum noncontextuality for single events occurs for configurations
which can be pasted together to construct a Kochen-Specker-type scheme,
any such argument might appear {\it ad hoc.}.

On the basis of the aforementioned lack of quantum omniscience, it is possible to postulate
the existence of absolute sources of indeterminism;
if there are no (preexisting) observables,
and no causal laws yielding individual outcomes,
the occurrence of any such outcome can only be unpredictable and incomputable \citep{2008-cal-svo}.
This quantum dice approach has first been proposed \citep{svozil-qct,rarity-94,zeil-99}
and realized \citep{zeilinger:qct,stefanov-2000,0256-307X-21-10-027,wang:056107}
in setups which utilize {\em complementarity}, yet still allow omniscience.
More recently, it was suggested \citep{svozil-2009-howto,10.1038/nature09008}
to utilize quantum systems with more than two exclusive outcomes  that are  are subject to {\em value indefiniteness}
(two-dimensional systems cannot be proven to be value indefinite).
The additional advantage over devices utilizing merely complementarity is that these new type of quantum oracles \citep{fiorentino:032334,1367-2630-12-1-013019,10.1038/nature09008}
are ``quantum mechanically certified''
by Boole-Bell-type, Kochen-Specker-type, and Greenberger-Horne-Zeilinger-type \citep{ghsz} theorems not to allow omniscience.
Of course, all these devices operate under the assumption that there are no hidden variables that could complete the quantum mechanical description of nature,
especially no contextual ones,
as well as no quasi-indeterminism caused by environmental influences
[such as in the context translation principle \citep{svozil-2003-garda}].
Thus, ultimately, these sources of quantum randomness are grounded in our belief
that quantum mechanics is the most complete representation of physical phenomenology.

\section{Miracles due to gaps in causal description}

A different issue, discussed by \citet{frank},
is the possible occurrence of miracles in the presence of {\em gaps} of physical determinism.
Already Maxwell has considered {\em singular points} \cite[212--213]{Campbell-1882}, {``where prediction,
except from absolutely perfect data, and guided by the omniscience of contingency, becomes impossible.''}
One might perceive individual events occurring
outside the validity of classical and quantum physics without any apparent cause as miracles.
For if there is no cause to an event,
why should such an event occur altogether rather than not occur?

Although such thoughts remain highly speculative, miracles
could be the basis for an operator-directed evolution in otherwise deterministic physical systems.
Similar models have  been applied to dualistic models of the mind \citep{popper-eccles,Eccles22051986,eccles:papal}.
The objection that this scenario is unnecessarily complicating an otherwise monistic model
should be carefully reevaluated in view of computer-generated {\em virtual realities} \citep{descartes-meditation,putnam:81,svozil-nat-acad}.
In such algorithmic universes, there are computable evolution laws as well as inputs from interfaces.
From the intrinsic perspective \citep{svozil-94}, the inputs cannot be causally accounted for,
and hence they remain irreducibly transcendental with respect to the otherwise algorithmic universe.

\section{Concluding thoughts}

\subsection{Metaphysical status of (in)determinism}

Hilbert's \citeyearpar{hilbert-1900e} sixth problem is about the axiomatization of physics.
Regardless of whether this goal is achievable,
omniscience cannot be gained
via the formalized, syntactic route,
which will remain blocked forever by the paradoxical self-reference
to which intrinsic observers and operational methods are bound.
Even if the universe were a computer \citep{zuse-70,fredkin,wolfram-2002,svozil-2005-cu},
we would intrinsically experience unpredictability and complementarity.

With regard to conjectures about the (in)deterministic evolution of physical events,
the situation is unsettled and can be expected to remain unsettled forever.
The reason for this is the provable impossibility to formally prove (in)determinism:
it is not possible to ensure that physical behaviors are causal and will remain so forever,
nor is it possible to exclude all causal behaviors.

The postulate of indeterministic behavior in physics or elsewhere is impossible to {\em prove} by
considering a finite operationally obtained encoded phenotype
such as a finite sequence of (supposedly random) bits from physical experiments
alone.
Furthermore, recursion theory and algorithmic information theory \citep{chaitin3,calude:02,gruenwald-vitanyi} imply that
an unbounded system of axioms is required to prove the unbounded
algorithmic information content of an unbounded symbolic sequence.
There also exist irreducible complexities in pure mathematics \citep{chaitin-04,s00032-006-0064-2}.

The opportunistic approach that (as historically,  many ingenious scientists have failed to come up with a causal description)
indeterminism will prevail appears to be anecdotal, at best, and  misleading, at worst.
Likewise, the advice of authoritative researchers to
avoid asking questions related
to completing a theory,
or to avoid thinking about the meaning of quantum mechanics or any  kind of rational interpretation,
and to avoid searching for causal laws for phenomena which are, at the same time,
postulated to occur indeterministically by the same authorities --  even wisely and benevolently posted  --
hardly qualify as proof.

Any kind of lawlessness can thus be claimed only
{\em with reference to,} and {\em relative to,} certain criteria, laws, or quantitative statistical or algorithmic tests.
For instance, randomness could be established merely {\em with respect to} certain tests,
such as some batteries of tests of randomness, for instance, {\em diehard} \citep{diehard}, {\it NIST} \citep{Rukhin-nist},  {\it TestU01} \citep{1268777},
or algorithmic \citep{calude-dinneen05,PhysRevA.82.022102} tests.
Note, however, that even the decimal expansion of $\pi$, the ratio between the circumference and the diameter of an ideal circle \citep{bailey97,bailey05},
behaves reasonably random \citep{PhysRevA.82.022102};
$\pi$ might even be a good source of randomness for many Monte Carlo calculations.

Thus, both from a  formal as well as from an operational point of view,
any rational investigation into, or claim of, absolute (in)determinism is metaphysical and can only be proved
{\em relative to} a limited number of statistical or algorithmic tests
which some specialists happen to choose;
with very limited validity for the formal and the natural sciences.

\subsection{Harnessing unknowables and indeterminism}

Physical indeterminism need not necessarily be perceived negatively as the absence of causal laws
but rather as a {\em valuable resource.}
Indeed, ingenious quasi-programs to compute the {\em halting probability} \citep{chaitin3,calude-dinneen06,rtx100200236p}
through summation of series without any computable rate of convergence could,
at least in principle, and in the limit of unbounded computational resources,
be interpreted as generating  provable random sequences.
However, as has already been expressed by  \citet[768]{von-neumann1},
{``anyone who considers arithmetical methods of producing random digits is, of course, in a state of sin.''}

Besides recursion-theoretic undecidability,
there appear to be at least two principal sources of indeterminism and randomness in physics:
(1) one scenario is associated with instabilities of classical physical systems
and with a strong dependence of future behaviors on the initial value, and
(2) quantum indeterminism, which can be subdivided into three subcategories, including  random outcomes of individual events,
 complementarity, and
value indefiniteness.

The production of random numbers by  physical generators has a long history \citep{rand-55}.
The similarities and differences between classical and quantum randomness can be conceptualized
in terms of two  black boxes: the first  of them,  called the {\em ``Poincar{\'e} box,''}
containing a classical, deterministic, chaotic source of randomness and
the second,  called the {\em ``Born box,''}
containing a quantum source of randomness.

A Poincar{\'e} box could be realized by operating a classical dynamical system in the shift map region.
Major principles for  Born boxes utilizing beam
splitters or parametric down conversion
include the following:
(1) there should be at least three mutually exclusive outcomes to ensure value indefiniteness
\citep{PhysRevLett.85.3313,2008-cal-svo,svozil-2009-howto,1367-2630-12-1-013019,10.1038/nature09008};
(2) the states prepared and measured should be pure and in mutually  [possibly interlinked \citep{svozil:040102}]
unbiased bases or contexts; and
(3) events should be independent  to be able to apply proper normalization procedures \citep{von-neumann1,Samuelson-1968}.

Suppose an agent is being presented with both boxes without any label on, or hint about, them;
that is, the origin of indeterminism
is unknown to the agent.
In a modified Turing test, an agent's task would be to find out which is the Born and which is
the Poincar{\'e} box solely by observing their output.
In the absence of any criteria, there should not exist any operational  method or procedure
capable of discriminating among these boxes.
Moreover, both types of indeterminism appear to be based on speculative assumptions:
in the classical case, it is the existence of continua and the possibility to randomly choose
elements thereof, representing the initial values;
in the quantum case, it is the irreducible indeterminism of single events.

\subsection{Personal remarks}

It is perpetually amazing, perplexing and  mind-boggling
how many laws and mathematical form\ae~can be found to express and program or induce physical behavior
with high precision.
There definitely is substance to the Pythagorean belief that, at least in a restricted manner,
nature is numbers and God computes; maybe also throwing dice sometimes.

The apparent impossibility to explain certain phenomena by any causal law should be perceived carefully and cautiously in a historic, transient perspective.
The author has the impression that in their attempts to canonize beliefs in the irreducible randomness of (quantum) mechanics,
many physicists, philosophers, and communicators may have prematurely thrown out a thorough rationalistic worldview with
the provably unfounded claims of total omniscience and omnipotence.

Let me sketch some very speculative attempts to undo the {\it Goridan Knot}
that haunts the perception of randomness in the classical and quantum domains in recent times.
(1) G\"odel-Turing-Tarski-type undecidability will remain with us forever, at least as long
one allows substitution, self-reference,  and universal computation.
(2) Most classical as well quantum unknowables might  be epistemic and not ontic.
(3) The classical continua might be convenient abstractions
that will have to be abandoned in favor of
granular, course-graining structures eventually.
As a consequence, classical randomness originating from deterministic chaos
might turn out to be formally computable but for all practical purposes impossible to predict.
(4) Space and time might turn out to be intrinsic constructions to represent dichotomic events
 in a world dominated by  one-to-one state evolution.
(5) There might only exist pure quantum states that can be associated with a unique (measurement and preparation) context.
Mixed quantum states might turn out to be purely epistemic, that is,
based on our ignorance of the pure state we are dealing with.
(6) Kochen-Specker and Boole-Bell-type arguments should be interpreted to indicate value indefiniteness
beyond a single context. The idea that there is physical existence beyond a single context at a time (and, associated with it, contextuality)
might  be misleading.
(7) Quantum randomness originate in the process of  context translation between
different, mismatching preparation and measurement contexts.
It might thus be induced by the environment of the measurement apparatus and our technologic inability to
maintain universal coherence.
(8) Dualistic operator controlled scenarios might present an option that are consistent
or at least in {\em peaceful coexistence}  with a certain type of
determinism (leaving room for miracles or gaps of causality). The information flow from and through the interface might either
be experienced as miracle, or, within the statistical bounds, as incomputable event or input.
Whether these specutations and feelings are justified only generations to come will know.

\section*{Acknowledgements}

The author gratefully acknowledges discussions with
Matthias Baaz,
Norbert Brunner,
Cristian S. Calude,
Elena Calude,
John Casti,
Gregory Chaitin,
Robert K. Clifton,
Michael~J. Dinneen,
Paul Adrien Maurice Dirac
Anatolij Dvure{\v{c}}enskij,
Klaus Ehrenberger,
Kurt Rudolf Fischer,
Daniel Greenberger,
Hans Havlicek,
Gudrun Kalmbach,
G\"unther Krenn,
Eckehart K{\"o}hler,
Alexander Leitsch,
David Mermin,
Mirko Navara,
Pavel Pt{\'{a}}k,
Sylvia Pulmannov{\'{a}},
Werner DePauli Schimanovich,
Ernst Specker,
Friedrich Stadler
Johann Summhammer,
Josef Tkadlec,
John Archibald Wheeler,
Ron Wright,
and
Anton Zeilinger.
None of these persons should be blamed for my ignorance.



\begin{thebibliography}{}

\bibitem[\protect\citeauthoryear{Aberth}{Aberth}{1980}]{aberth-80}
Aberth, O. (1980).
\newblock {\em Computable Analysis}.
\newblock New York: McGraw-Hill.

\bibitem[\protect\citeauthoryear{Adamatzky}{Adamatzky}{2002}]{adama02}
Adamatzky, A. (2002).
\newblock {\em Collision-based Computing}.
\newblock London: Springer.

\bibitem[\protect\citeauthoryear{Adleman \& Blum}{Adleman \&
  Blum}{1991}]{ad-91}
Adleman, L.~M. \& Blum, M. (1991).
\newblock Inductive inference and unsolvability.
\newblock {\em Journal of Symbolic Logic}, {\em 56}, 891--900.
\newblock Available from: \url{http://dx.doi.org/10.2307/2275058}, \href
  {http://dx.doi.org/10.2307/2275058} {\path{doi:10.2307/2275058}}.

\bibitem[\protect\citeauthoryear{Alda}{Alda}{1980}]{Alda}
Alda, V. (1980).
\newblock On\/ {\rm 0-1} measures for projectors {I}.
\newblock {\em Aplikace matematiky (Applications of Mathematics)}, {\em 25},
  373--374.
\newblock Available from: \url{http://dml.cz/dmlcz/103871}.

\bibitem[\protect\citeauthoryear{Alda}{Alda}{1981}]{Alda2}
Alda, V. (1981).
\newblock On\/ {\rm 0-1} measures for projectors {II}.
\newblock {\em Aplikace matematiky (Applications of Mathematics)}, {\em 26},
  57--58.
\newblock Available from: \url{http://dml.cz/dmlcz/103894}.

\bibitem[\protect\citeauthoryear{Angluin \& Smith}{Angluin \&
  Smith}{1983}]{angluin:83}
Angluin, D. \& Smith, C.~H. (1983).
\newblock A survey of inductive inference: Theory and methods.
\newblock {\em Computing Surveys}, {\em 15}, 237--269.

\bibitem[\protect\citeauthoryear{Anishchenko, Astakhov, Neiman, Vadivasova \&
  Schimansky-Geier}{Anishchenko et~al.}{2007}]{nld-chaos}
Anishchenko, V.~S., Astakhov, V., Neiman, A., Vadivasova, T., \&
  Schimansky-Geier, L. (2007).
\newblock {\em Nonlinear Dynamics of Chaotic and Stochastic Systems Tutorial
  and Modern Developments. {T}utorial and Modern Developments\/} (second ed.).
\newblock Berlin, Heidelberg: Springer.

\bibitem[\protect\citeauthoryear{Aquinas}{Aquinas}{1981}]{Aquinas}
Aquinas, T. (1981).
\newblock {\em Summa Theologica. {T}ranslated by {F}athers of the {E}nglish
  {D}ominican {P}rovince}.
\newblock Grand Rapids, MI: Christian Classics Ethereal Library.
\newblock Available from: \url{http://www.ccel.org/ccel/aquinas/summa.html}.

\bibitem[\protect\citeauthoryear{Atmanspacher \& Primas}{Atmanspacher \&
  Primas}{2005}]{atman:05}
Atmanspacher, H. \& Primas, H. (2005).
\newblock Epistemic and ontic quantum realities.
\newblock In Khrennikov, A. (Ed.), {\em Foundations of Probability and Physics
  -- 3, AIP Conference Proceedings Volume 750}, (pp.\ 49--62)., New York,
  Berlin. Springer.
\newblock Available from: \url{http://dx.doi.org/10.1063/1.1874557}, \href
  {http://dx.doi.org/10.1063/1.1874557} {\path{doi:10.1063/1.1874557}}.

\bibitem[\protect\citeauthoryear{Babadzanjanz}{Babadzanjanz}{1969}]{1969VeLen...7..121B}
Babadzanjanz, L.~K. (1969).
\newblock Analytical methods of computing perturbations of coordinates of the
  planets.
\newblock {\em Leningradskii Universitet Vestnik Matematika Mekhanika
  Astronomiia}, {\em 7}, 121--132.

\bibitem[\protect\citeauthoryear{Babadzanjanz}{Babadzanjanz}{1979}]{Babadzanjanz-1979}
Babadzanjanz, L.~K. (1979).
\newblock Existence of the continuations in the $n$-body problem.
\newblock {\em Celestial Mechanics and Dynamical Astronomy}, {\em 20\/}(1),
  43--57.
\newblock Available from: \url{http://dx.doi.org/10.1007/BF01236607}, \href
  {http://dx.doi.org/10.1007/BF01236607} {\path{doi:10.1007/BF01236607}}.

\bibitem[\protect\citeauthoryear{Babadzanjanz}{Babadzanjanz}{1993}]{Babadzanjanz-1993}
Babadzanjanz, L.~K. (1993).
\newblock On the global solution of the $n$-body problem.
\newblock {\em Celestial Mechanics and Dynamical Astronomy}, {\em 56},
  427--449.
\newblock Available from: \url{http://dx.doi.org/10.1007/BF00691812}, \href
  {http://dx.doi.org/10.1007/BF00691812} {\path{doi:10.1007/BF00691812}}.

\bibitem[\protect\citeauthoryear{Babadzanjanz \& Sarkissian}{Babadzanjanz \&
  Sarkissian}{2006}]{Babadzanjanz-2006}
Babadzanjanz, L.~K. \& Sarkissian, D.~R. (2006).
\newblock Taylor series method for dynamical systems with control: Convergence
  and error estimates.
\newblock {\em Journal of Mathematical Sciences}, {\em 139}, 7025--7046.
\newblock Available from: \url{http://dx.doi.org/10.1007/s10958-006-0404-3},
  \href {http://dx.doi.org/10.1007/s10958-006-0404-3}
  {\path{doi:10.1007/s10958-006-0404-3}}.

\bibitem[\protect\citeauthoryear{Bailey, Borwein \& Plouffe}{Bailey
  et~al.}{1997}]{bailey97}
Bailey, D., Borwein, P., \& Plouffe, S. (1997).
\newblock On the rapid computation of various polylogarithmic constants.
\newblock {\em Mathematics of Computation}, {\em 66}, 903--913.
\newblock Available from:
  \url{http://dx.doi.org/10.1090/S0025-5718-97-00856-9}, \href
  {http://dx.doi.org/10.1090/S0025-5718-97-00856-9}
  {\path{doi:10.1090/S0025-5718-97-00856-9}}.

\bibitem[\protect\citeauthoryear{Bailey \& Borwein}{Bailey \&
  Borwein}{2005}]{bailey05}
Bailey, D.~H. \& Borwein, J.~M. (2005).
\newblock Experimental mathematics: Examples, methods and implications.
\newblock {\em Notices of the American Mathematical Society}, {\em 52},
  502--514.
\newblock Available from:
  \url{http://www.ams.org/notices/200505/fea-borwein.pdf}.

\bibitem[\protect\citeauthoryear{Barrow}{Barrow}{1991}]{barrow-TOE}
Barrow, J.~D. (1991).
\newblock {\em Theories of Everything}.
\newblock Oxford: Oxford University Press.

\bibitem[\protect\citeauthoryear{Barrow}{Barrow}{1998}]{barrow-impossibilities}
Barrow, J.~D. (1998).
\newblock {\em Impossibility. {T}he Limits of Science and the Science of
  Limits}.
\newblock Oxford: Oxford University Press.

\bibitem[\protect\citeauthoryear{Bechmann-Pasquinucci \&
  Peres}{Bechmann-Pasquinucci \& Peres}{2000}]{PhysRevLett.85.3313}
Bechmann-Pasquinucci, H. \& Peres, A. (2000).
\newblock Quantum cryptography with 3-state systems.
\newblock {\em Physical Review Letters}, {\em 85\/}(15), 3313--3316.
\newblock Available from: \url{http://dx.doi.org/10.1103/PhysRevLett.85.3313},
  \href {http://dx.doi.org/10.1103/PhysRevLett.85.3313}
  {\path{doi:10.1103/PhysRevLett.85.3313}}.

\bibitem[\protect\citeauthoryear{Bell}{Bell}{1966}]{bell-66}
Bell, J.~S. (1966).
\newblock On the problem of hidden variables in quantum mechanics.
\newblock {\em Reviews of Modern Physics}, {\em 38}, 447--452.
\newblock Reprinted in Ref.~\cite[pp. 1-13]{bell-87}.
\newblock Available from: \url{http://dx.doi.org/10.1103/RevModPhys.38.447},
  \href {http://dx.doi.org/10.1103/RevModPhys.38.447}
  {\path{doi:10.1103/RevModPhys.38.447}}.

\bibitem[\protect\citeauthoryear{Bell}{Bell}{1987}]{bell-87}
Bell, J.~S. (1987).
\newblock {\em Speakable and Unspeakable in Quantum Mechanics}.
\newblock Cambridge: Cambridge University Press.

\bibitem[\protect\citeauthoryear{Bell}{Bell}{1992}]{bell:a1}
Bell, J.~S. (1992).
\newblock Against `measurement'.
\newblock {\em Physikalische Bl{\"{a}}tter}, {\em 48\/}(4), 267.

\bibitem[\protect\citeauthoryear{Bennett}{Bennett}{1973}]{bennett-73}
Bennett, C.~H. (1973).
\newblock Logical reversibility of computation.
\newblock {\em IBM Journal of Research and Development}, {\em 17}, 525--532.
\newblock Reprinted in Ref.~\cite[pp. 197-204]{maxwell-demon}.

\bibitem[\protect\citeauthoryear{Bennett}{Bennett}{1982}]{bennett-82}
Bennett, C.~H. (1982).
\newblock The thermodynamics of computation---a review.
\newblock {\em International Journal of Theoretical Physics}, {\em 21},
  905--940.
\newblock Reprinted in Ref.~\cite[pp. 283-318]{maxwell-demon2}.
\newblock Available from: \url{http://dx.doi.org/10.1007/BF02084158}, \href
  {http://dx.doi.org/10.1007/BF02084158} {\path{doi:10.1007/BF02084158}}.

\bibitem[\protect\citeauthoryear{Bennett, Bessette, Brassard, Salvail \&
  Smolin}{Bennett et~al.}{1992}]{benn-92}
Bennett, C.~H., Bessette, F., Brassard, G., Salvail, L., \& Smolin, J. (1992).
\newblock Experimental quantum cryptography.
\newblock {\em Journal of Cryptology}, {\em 5}, 3--28.
\newblock Available from: \url{http://dx.doi.org/10.1007/BF00191318}, \href
  {http://dx.doi.org/10.1007/BF00191318} {\path{doi:10.1007/BF00191318}}.

\bibitem[\protect\citeauthoryear{Bennett \& Brassard}{Bennett \&
  Brassard}{1984}]{benn-84}
Bennett, C.~H. \& Brassard, G. (1984).
\newblock Quantum cryptography: Public key distribution and coin tossing.
\newblock In {\em Proceedings of the IEEE International Conference on
  Computers, Systems, and Signal Processing, Bangalore, India}, (pp.\
  175--179). IEEE Computer Society Press.
\newblock Available from:
  \url{http://www.research.ibm.com/people/b/bennetc/bennettc198469790513.pdf}.

\bibitem[\protect\citeauthoryear{Berkeley}{Berkeley}{1710}]{berkeley}
Berkeley, G. (1710).
\newblock {\em A Treatise Concerning the Principles of Human Knowledge}.
\newblock Available from: \url{http://www.gutenberg.org/etext/4723}.

\bibitem[\protect\citeauthoryear{Bernstein}{Bernstein}{1991}]{bernstein}
Bernstein, J. (1991).
\newblock {\em Quantum Profiles}.
\newblock Princeton, NJ: Princeton University Press.

\bibitem[\protect\citeauthoryear{Birkhoff \& {von Neumann}}{Birkhoff \& {von
  Neumann}}{1936}]{birkhoff-36}
Birkhoff, G. \& {von Neumann}, J. (1936).
\newblock The logic of quantum mechanics.
\newblock {\em Annals of Mathematics}, {\em 37\/}(4), 823--843.
\newblock Available from: \url{http://dx.doi.org/10.2307/1968621}, \href
  {http://dx.doi.org/10.2307/1968621} {\path{doi:10.2307/1968621}}.

\bibitem[\protect\citeauthoryear{Blum \& Blum}{Blum \& Blum}{1975}]{blum75blum}
Blum, L. \& Blum, M. (1975).
\newblock Toward a mathematical theory of inductive inference.
\newblock {\em Information and Control}, {\em 28\/}(2), 125--155.
\newblock Available from:
  \url{http://dx.doi.org/10.1016/S0019-9958(75)90261-2}, \href
  {http://dx.doi.org/10.1016/S0019-9958(75)90261-2}
  {\path{doi:10.1016/S0019-9958(75)90261-2}}.

\bibitem[\protect\citeauthoryear{Bohr}{Bohr}{1949}]{bohr-1949}
Bohr, N. (1949).
\newblock Discussion with {E}instein on epistemological problems in atomic
  physics.
\newblock In P.~A. Schilpp (Ed.), {\em {A}lbert {E}instein:
  Philosopher-Scientist}  (pp.\ 200--241). Evanston, Ill.: The Library of
  Living Philosophers.
\newblock Available from:
  \url{http://dx.doi.org/10.1016/S1876-0503(08)70379-7}, \href
  {http://dx.doi.org/10.1016/S1876-0503(08)70379-7}
  {\path{doi:10.1016/S1876-0503(08)70379-7}}.

\bibitem[\protect\citeauthoryear{Boole}{Boole}{1862}]{Boole-62}
Boole, G. (1862).
\newblock On the theory of probabilities.
\newblock {\em Philosophical Transactions of the Royal Society of London}, {\em
  152}, 225--252.
\newblock Available from: \url{http://www.jstor.org/stable/108830}.

\bibitem[\protect\citeauthoryear{Boole}{Boole}{1958}]{Boole}
Boole, G. (1958).
\newblock {\em An Investigation of the Laws of Thought}.
\newblock New York: Dover.

\bibitem[\protect\citeauthoryear{Born}{Born}{1926a}]{born-26-2}
Born, M. (1926a).
\newblock {Q}uantenmechanik der {S}to{\ss}vorg{\"{a}}nge.
\newblock {\em Zeitschrift f{\"{u}}r Physik}, {\em 38}, 803--827.
\newblock Available from: \url{http://dx.doi.org/10.1007/BF01397184}, \href
  {http://dx.doi.org/10.1007/BF01397184} {\path{doi:10.1007/BF01397184}}.

\bibitem[\protect\citeauthoryear{Born}{Born}{1926b}]{born-26-1}
Born, M. (1926b).
\newblock Zur {Q}uantenmechanik der {S}to{\ss}vorg{\"{a}}nge.
\newblock {\em Zeitschrift f{\"{u}}r Physik}, {\em 37}, 863--867.
\newblock Available from: \url{http://dx.doi.org/10.1007/BF01397477}, \href
  {http://dx.doi.org/10.1007/BF01397477} {\path{doi:10.1007/BF01397477}}.

\bibitem[\protect\citeauthoryear{Born}{Born}{1955}]{born-55}
Born, M. (1955).
\newblock {I}st die klassische {M}echanik tats{\"{a}}chlich deterministisch?
\newblock {\em Physikalische Bl{\"{a}}tter}, {\em 11}, 49--54.
\newblock {E}nglish translation ``Is classical mechanics in fact
  deterministic?'' Reprinted in Ref.~\cite[p.~78-83]{born-69}.

\bibitem[\protect\citeauthoryear{Born}{Born}{1969}]{born-69}
Born, M. (1969).
\newblock {\em Physics in my generation\/} (second ed.).
\newblock New York: Springer.

\bibitem[\protect\citeauthoryear{Boskovich}{Boskovich}{1966}]{bos1}
Boskovich, R.~J. (1966).
\newblock De spacio et tempore, ut a nobis cognoscuntur.
\newblock In J.~M. Child (Ed.), {\em A Theory of Natural Philosophy}  (pp.\
  203--205). Cambridge, MA: Open Court (1922) and MIT Press.
\newblock Available from:
  \url{ttp://www.archive.org/details/theoryofnaturalp00boscrich}.

\bibitem[\protect\citeauthoryear{Brady}{Brady}{1988}]{brady}
Brady, A.~H. (1988).
\newblock The busy beaver game and the meaning of life.
\newblock In R.~Herken (Ed.), {\em The Universal {T}uring Machine. A
  Half-Century Survey}  (pp.\ 259). Hamburg: Kammerer und Unverzagt.

\bibitem[\protect\citeauthoryear{Bricmont}{Bricmont}{1996}]{bricmont}
Bricmont, J. (1996).
\newblock Science of chaos or chaos in science?
\newblock {\em Annals of the New York Academy of Sciences}, {\em 775},
  131--175.
\newblock Available from:
  \url{http://dx.doi.org/10.1111/j.1749-6632.1996.tb23135.x}, \href
  {http://arxiv.org/abs/chao-dyn/9603009} {\path{arXiv:chao-dyn/9603009}},
  \href {http://dx.doi.org/10.1111/j.1749-6632.1996.tb23135.x}
  {\path{doi:10.1111/j.1749-6632.1996.tb23135.x}}.

\bibitem[\protect\citeauthoryear{Bridges}{Bridges}{1999}]{bridges1}
Bridges, D.~S. (1999).
\newblock Can constructive mathematics be applied in physics?
\newblock {\em Journal of Philosophical Logic}, {\em 28\/}(5), 439--453.
\newblock Available from: \url{http://dx.doi.org/10.1023/A:1004420413391},
  \href {http://dx.doi.org/10.1023/A:1004420413391}
  {\path{doi:10.1023/A:1004420413391}}.

\bibitem[\protect\citeauthoryear{Bridgman}{Bridgman}{1934}]{bridgman}
Bridgman, P.~W. (1934).
\newblock A physicist's second reaction to {M}engenlehre.
\newblock {\em Scripta Mathematica}, {\em 2}, 101--117, 224--234.
\newblock Discussed in \cite{landauer-95}.

\bibitem[\protect\citeauthoryear{Brukner}{Brukner}{2003}]{bruk-08}
Brukner, {\v{C}}. (2003).
\newblock Quantum experiments can test mathematical undecidability.
\newblock In C.~S. Calude, J.~F.~G. Costa, R.~Freund, M.~Oswald, \&
  G.~Rozenberg (Eds.), {\em Unconventional Computing. Unconventional
  Computation. {P}roceedings of the 7th International Conference, {UC} 2008,
  {V}ienna, {A}ustria, {A}ugust 25-28, 2008. {L}ecture Notes in Computer
  Science. {V}olume 5204/2008}  (pp.\ 1--5). Berlin: Springer.
\newblock Available from: \url{http://dx.doi.org/10.1007/978-3-540-85194-3_1},
  \href {http://dx.doi.org/10.1007/978-3-540-85194-3_1}
  {\path{doi:10.1007/978-3-540-85194-3_1}}.

\bibitem[\protect\citeauthoryear{Bub}{Bub}{1999}]{Bub-1999}
Bub, J. (1999).
\newblock {\em Interpreting the Quantum World}.
\newblock Cambridge: Cambridge University Press.

\bibitem[\protect\citeauthoryear{Cabello, Estebaranz \&
  Garc{\'{i}}a-Alcaine}{Cabello et~al.}{1996}]{cabello-96}
Cabello, A., Estebaranz, J.~M., \& Garc{\'{i}}a-Alcaine, G. (1996).
\newblock {B}ell-{K}ochen-{S}pecker theorem: A proof with 18 vectors.
\newblock {\em Physics Letters A}, {\em 212\/}(4), 183--187.
\newblock Available from: \url{http://dx.doi.org/10.1016/0375-9601(96)00134-X},
  \href {http://dx.doi.org/10.1016/0375-9601(96)00134-X}
  {\path{doi:10.1016/0375-9601(96)00134-X}}.

\bibitem[\protect\citeauthoryear{Calude}{Calude}{2002}]{calude:02}
Calude, C. (2002).
\newblock {\em Information and Randomness---An Algorithmic Perspective\/}
  (second ed.).
\newblock Berlin: Springer.

\bibitem[\protect\citeauthoryear{Calude, Calude, Svozil \& Yu}{Calude
  et~al.}{1997}]{cal-sv-yu}
Calude, C., Calude, E., Svozil, K., \& Yu, S. (1997).
\newblock Physical versus computational complementarity {I}.
\newblock {\em International Journal of Theoretical Physics}, {\em 36\/}(7),
  1495--1523.
\newblock Available from: \url{http://dx.doi.org/10.1007/BF02435752}, \href
  {http://arxiv.org/abs/arXiv:quant-ph/9412004}
  {\path{arXiv:arXiv:quant-ph/9412004}}, \href
  {http://dx.doi.org/10.1007/BF02435752} {\path{doi:10.1007/BF02435752}}.

\bibitem[\protect\citeauthoryear{Calude, Campbell, Svozil \& \c{S}tef\u
  anescu}{Calude et~al.}{1995}]{CalCamSvo-Stef-1995}
Calude, C., Campbell, D.~I., Svozil, K., \& \c{S}tef\u anescu, D. (1995).
\newblock Strong determinism vs. computability.
\newblock In Schimanovich, W.~D., K{\"o}hler, E., \& Stadler, F. (Eds.), {\em
  The Foundational Debate. {C}omplexity and Constructivity in Mathematics and
  Physics. Vienna Circle Institute Yearbook, {V}ol. 3}, (pp.\ 115--131).,
  Dordrecht, Boston, London. Kluwer.
\newblock Available from: \url{http://arxiv.org/abs/quant-ph/9412004}, \href
  {http://arxiv.org/abs/arXiv:quant-ph/9412004}
  {\path{arXiv:arXiv:quant-ph/9412004}}.

\bibitem[\protect\citeauthoryear{Calude}{Calude}{2005}]{Cris04}
Calude, C.~S. (2005).
\newblock {A}lgorithmic randomness, quantum physics, and incompleteness.
\newblock In M.~Margenstern (Ed.), {\em Proceedings of the Conference
  ``Machines, Computations and Universality'' (MCU'2004)}  (pp.\ 1--17).
  Berlin: Lectures Notes in Comput. Sci. 3354, Springer.

\bibitem[\protect\citeauthoryear{Calude, Calude \& Svozil}{Calude
  et~al.}{2010}]{calude:037103}
Calude, C.~S., Calude, E., \& Svozil, K. (2010).
\newblock The complexity of proving chaoticity and the {C}hurch--{T}uring
  thesis.
\newblock {\em Chaos: An Interdisciplinary Journal of Nonlinear Science}, {\em
  20\/}(3), 037103.
\newblock Available from: \url{http://dx.doi.org/10.1063/1.3489096}, \href
  {http://dx.doi.org/10.1063/1.3489096} {\path{doi:10.1063/1.3489096}}.

\bibitem[\protect\citeauthoryear{Calude \& Chaitin}{Calude \&
  Chaitin}{2007}]{rtx100200236p}
Calude, C.~S. \& Chaitin, G.~J. (2007).
\newblock What is $\ldots$ a halting probability?
\newblock {\em Notices of the AMS}, {\em 57\/}(2), 236--237.
\newblock Available from:
  \url{http://www.ams.org/notices/201002/rtx100200236p.pdf}.

\bibitem[\protect\citeauthoryear{Calude \& Dinneen}{Calude \&
  Dinneen}{2005}]{calude-dinneen05}
Calude, C.~S. \& Dinneen, M.~J. (2005).
\newblock Is quantum randomness algorithmic random? a preliminary attack.
\newblock In Bozapalidis, S., Kalampakas, A., \& Rahonis, G. (Eds.), {\em
  Proceedings of the 1st International Conference on Algebraic Informatics},
  (pp.\ 195--196)., Thessaloniki, Greece. Aristotle University of Thessaloniki.

\bibitem[\protect\citeauthoryear{Calude \& Dinneen}{Calude \&
  Dinneen}{2007}]{calude-dinneen06}
Calude, C.~S. \& Dinneen, M.~J. (2007).
\newblock Exact approximations of omega numbers.
\newblock {\em International Journal of Bifurcation and Chaos}, {\em 17},
  1937--1954.
\newblock {CDMTCS} report series 293.
\newblock Available from: \url{http://dx.doi.org/10.1142/S0218127407018130},
  \href {http://dx.doi.org/10.1142/S0218127407018130}
  {\path{doi:10.1142/S0218127407018130}}.

\bibitem[\protect\citeauthoryear{Calude, Dinneen, Dumitrescu \& Svozil}{Calude
  et~al.}{2010}]{PhysRevA.82.022102}
Calude, C.~S., Dinneen, M.~J., Dumitrescu, M., \& Svozil, K. (2010).
\newblock Experimental evidence of quantum randomness incomputability.
\newblock {\em Phys. Rev. A}, {\em 82\/}(2), 022102.
\newblock Available from: \url{http://dx.doi.org/10.1103/PhysRevA.82.022102},
  \href {http://dx.doi.org/10.1103/PhysRevA.82.022102}
  {\path{doi:10.1103/PhysRevA.82.022102}}.

\bibitem[\protect\citeauthoryear{Calude \& Staiger}{Calude \&
  Staiger}{2010}]{calude-staiger-09}
Calude, C.~S. \& Staiger, L. (2010).
\newblock A note on accelerated turing machines.
\newblock {\em Mathematical Structures in Computer Science}, {\em 20\/}(Special
  Issue 06), 1011--1017.
\newblock Available from: \url{http://dx.doi.org/10.1017/S0960129510000344},
  \href {http://dx.doi.org/10.1017/S0960129510000344}
  {\path{doi:10.1017/S0960129510000344}}.

\bibitem[\protect\citeauthoryear{Calude \& Svozil}{Calude \&
  Svozil}{2008}]{2008-cal-svo}
Calude, C.~S. \& Svozil, K. (2008).
\newblock Quantum randomness and value indefiniteness.
\newblock {\em Advanced Science Letters}, {\em 1\/}(2), 165--168.
\newblock Available from:
  \url{http://www.ingentaconnect.com/content/asp/asl/2008/00000001/00000002/art00004},
  \href {http://arxiv.org/abs/arXiv:quant-ph/0611029}
  {\path{arXiv:arXiv:quant-ph/0611029}}, \href
  {http://dx.doi.org/10.1166/asl.2008.016} {\path{doi:10.1166/asl.2008.016}}.

\bibitem[\protect\citeauthoryear{Campbell \& Garnett}{Campbell \&
  Garnett}{1882}]{Campbell-1882}
Campbell, L. \& Garnett, W. (1882).
\newblock {\em The life of {J}ames {C}lerk {M}axwell. {W}ith a selection from
  his correspondence and occasional writings and a sketch of his contributions
  to science}.
\newblock London: MacMillan.
\newblock Available from: \url{http://www.sonnetsoftware.com/bio/maxbio.pdf}.

\bibitem[\protect\citeauthoryear{Casti \& Karlquist}{Casti \&
  Karlquist}{1996}]{casti:96-onlimits}
Casti, J.~L. \& Karlquist, A. (1996).
\newblock {\em Boundaries and Barriers. On the Limits to Scientific Knowledge}.
\newblock Reading, MA: Addison-Wesley.

\bibitem[\protect\citeauthoryear{Casti \& Traub}{Casti \&
  Traub}{1994}]{casti:94-onlimits_book}
Casti, J.~L. \& Traub, J.~F. (1994).
\newblock {\em On Limits}.
\newblock Santa Fe, NM: Santa Fe Institute.
\newblock Report 94-10-056.
\newblock Available from:
  \url{http://www.santafe.edu/research/publications/workingpapers/94-10-056.pdf}.

\bibitem[\protect\citeauthoryear{Chaitin}{Chaitin}{1974}]{chaitin-ACM}
Chaitin, G.~J. (1974).
\newblock Information-theoretic limitations of formal systems.
\newblock {\em Journal of the Association of Computing Machinery}, {\em 21},
  403--424.
\newblock Reprinted in Ref.~\cite{chaitin2}.
\newblock Available from:
  \url{http://www.cs.auckland.ac.nz/CDMTCS/chaitin/acm74.pdf}.

\bibitem[\protect\citeauthoryear{Chaitin}{Chaitin}{1987a}]{chaitin3}
Chaitin, G.~J. (1987a).
\newblock {\em Algorithmic Information Theory}.
\newblock Cambridge: Cambridge University Press.
\newblock Available from: \url{http://www.cs.auckland.ac.nz/~chaitin/cup.pdf}.

\bibitem[\protect\citeauthoryear{Chaitin}{Chaitin}{1987b}]{chaitin-bb}
Chaitin, G.~J. (1987b).
\newblock Computing the busy beaver function.
\newblock In T.~M. Cover \& B.~Gopinath (Eds.), {\em Open Problems in
  Communication and Computation}  (pp.\ 108). New York: Springer.
\newblock Reprinted in Ref.~\cite{chaitin2}.

\bibitem[\protect\citeauthoryear{Chaitin}{Chaitin}{1990}]{chaitin2}
Chaitin, G.~J. (1990).
\newblock {\em Information, Randomness and Incompleteness\/} (second ed.).
\newblock Singapore: World Scientific.
\newblock This is a collection of G. Chaitin's early publications.

\bibitem[\protect\citeauthoryear{Chaitin}{Chaitin}{2004}]{chaitin-04}
Chaitin, G.~J. (2004).
\newblock Irreducible complexity in pure mathematics.
\newblock eprint {math/0411091}.
\newblock Available from: \url{http://arxiv.org/abs/math/0411091}, \href
  {http://arxiv.org/abs/math/0411091} {\path{arXiv:math/0411091}}.

\bibitem[\protect\citeauthoryear{Chaitin}{Chaitin}{2007}]{s00032-006-0064-2}
Chaitin, G.~J. (2007).
\newblock The halting probability {O}mega: {I}rreducible complexity in pure
  mathematics.
\newblock {\em Milan Journal of Mathematics}, {\em 75\/}(1), 291--304.
\newblock Available from: \url{http://dx.doi.org/10.1007/s00032-006-0064-2},
  \href {http://dx.doi.org/10.1007/s00032-006-0064-2}
  {\path{doi:10.1007/s00032-006-0064-2}}.

\bibitem[\protect\citeauthoryear{Chapman, Hammond, Lenef, Schmiedmayer,
  Rubenstein, Smith \& Pritchard}{Chapman et~al.}{1995}]{PhysRevLett.75.3783}
Chapman, M.~S., Hammond, T.~D., Lenef, A., Schmiedmayer, J., Rubenstein, R.~A.,
  Smith, E., \& Pritchard, D.~E. (1995).
\newblock Photon scattering from atoms in an atom interferometer: Coherence
  lost and regained.
\newblock {\em Physical Review Letters}, {\em 75\/}(21), 3783--3787.
\newblock Available from: \url{http://dx.doi.org/10.1103/PhysRevLett.75.3783},
  \href {http://dx.doi.org/10.1103/PhysRevLett.75.3783}
  {\path{doi:10.1103/PhysRevLett.75.3783}}.

\bibitem[\protect\citeauthoryear{Chisholm}{Chisholm}{1946}]{chisholm-46}
Chisholm, R.~M. (1946).
\newblock The contrary-to-fact conditional.
\newblock {\em Mind}, {\em 55\/}(220), 289--307.
\newblock reprinted in \cite[pp.~482-497]{chisholm}.
\newblock Available from: \url{http://www.jstor.org/stable/2250757}.

\bibitem[\protect\citeauthoryear{Chisholm}{Chisholm}{1949}]{chisholm}
Chisholm, R.~M. (1949).
\newblock The contrary-to-fact conditional.
\newblock In H.~Feigl \& W.~Sellars (Eds.), {\em Readings in Philosophical
  Analysis}  (pp.\ 482--497). New York: Appleton-Century-Crofts.

\bibitem[\protect\citeauthoryear{Clauser}{Clauser}{2002}]{clauser-talkvie}
Clauser, J. (2002).
\newblock Early history of {B}ell's theorem.
\newblock In R.~Bertlmann \& A.~Zeilinger (Eds.), {\em Quantum (Un)speakables.
  {F}rom {B}ell to Quantum Information}  (pp.\ 61--96). Berlin: Springer.

\bibitem[\protect\citeauthoryear{Clauser \& Shimony}{Clauser \&
  Shimony}{1978}]{clauser}
Clauser, J.~F. \& Shimony, A. (1978).
\newblock {B}ell's theorem: experimental tests and implications.
\newblock {\em Reports on Progress in Physics}, {\em 41}, 1881--1926.
\newblock Available from: \url{http://dx.doi.org/10.1088/0034-4885/41/12/002},
  \href {http://dx.doi.org/10.1088/0034-4885/41/12/002}
  {\path{doi:10.1088/0034-4885/41/12/002}}.

\bibitem[\protect\citeauthoryear{{Costa} \& Doria}{{Costa} \&
  Doria}{1991}]{dc-d91a}
{Costa}, N.~C.~A. \& Doria, F.~A. (1991).
\newblock Classical physics and {P}enrose's thesis.
\newblock {\em Foundations of Physics Letters}, {\em 4}, 363--373.
\newblock Available from: \url{http://dx.doi.org/10.1007/BF00671484}, \href
  {http://dx.doi.org/10.1007/BF00665895} {\path{doi:10.1007/BF00665895}}.

\bibitem[\protect\citeauthoryear{{da Costa} \& Doria}{{da Costa} \&
  Doria}{1991}]{dc-d91b}
{da Costa}, N.~C.~A. \& Doria, F.~A. (1991).
\newblock Undecidability and incompleteness in classical mechanics.
\newblock {\em International Journal of Theoretical Physics}, {\em 30},
  1041--1073.
\newblock Available from: \url{http://dx.doi.org/10.1007/BF00665895}, \href
  {http://dx.doi.org/10.1007/BF00671484} {\path{doi:10.1007/BF00671484}}.

\bibitem[\protect\citeauthoryear{{da Costa}, Doria \& do~Amaral}{{da Costa}
  et~al.}{1993}]{dc-d93}
{da Costa}, N. C.~A., Doria, F.~A., \& do~Amaral, A. F.~F. (1993).
\newblock Dynamical system where proving chaos is equivalent to proving
  {F}ermat's conjecture.
\newblock {\em International Journal of Theoretical Physics}, {\em 32\/}(11),
  2187--2206.
\newblock Available from: \url{http://dx.doi.org/10.1007/BF00675030}, \href
  {http://dx.doi.org/10.1007/BF00675030} {\path{doi:10.1007/BF00675030}}.

\bibitem[\protect\citeauthoryear{Davis}{Davis}{1958}]{davis-58}
Davis, M. (1958).
\newblock {\em Computability and Unsolvability}.
\newblock New York: McGraw-Hill.

\bibitem[\protect\citeauthoryear{Davis}{Davis}{1965}]{davis}
Davis, M. (1965).
\newblock {\em The Undecidable. Basic Papers on Undecidable, Unsolvable
  Problems and Computable Functions}.
\newblock Hewlett, N.Y.: Raven Press.

\bibitem[\protect\citeauthoryear{Descartes}{Descartes}{1637}]{Descartes-Discourse}
Descartes, R. (1637).
\newblock {\em Discours de la m{\'e}thode pour bien conduire sa raison et
  chercher la verit{\'e} dans les sciences (Discourse on the Method of Rightly
  Conducting One's Reason and of Seeking Truth)}.
\newblock Available from: \url{http://www.gutenberg.org/etext/59}.

\bibitem[\protect\citeauthoryear{Descartes}{Descartes}{1641}]{descartes-meditation}
Descartes, R. (1641).
\newblock {\em Meditation on First Philosophy}.
\newblock Available from:
  \url{http://oregonstate.edu/instruct/phl302/texts/descartes/meditations/meditations.html}.

\bibitem[\protect\citeauthoryear{Dewdney}{Dewdney}{1984}]{dewdney}
Dewdney, A.~K. (1984).
\newblock Computer recreations: A computer trap for the busy beaver, the
  hardest-working {T}uring machine.
\newblock {\em Scientific American}, {\em 251\/}(2), 19--23.

\bibitem[\protect\citeauthoryear{Diacu}{Diacu}{1996}]{Diacu96}
Diacu, F. (1996).
\newblock The solution of the n-body problem.
\newblock {\em The Mathematical Intelligencer}, {\em 18\/}(3), 66--70.
\newblock Available from: \url{http://dx.doi.org/10.1007/BF03024313}, \href
  {http://dx.doi.org/10.1007/BF03024313} {\path{doi:10.1007/BF03024313}}.

\bibitem[\protect\citeauthoryear{Diacu \& Holmes}{Diacu \&
  Holmes}{1996}]{Diacu96-ce}
Diacu, F. \& Holmes, P. (1996).
\newblock {\em Celestial Encounters - the Origins of Chaos and Stability}.
\newblock Princeton, NJ: Princeton University Press.

\bibitem[\protect\citeauthoryear{Donath \& Svozil}{Donath \&
  Svozil}{2002}]{DonSvo01}
Donath, N. \& Svozil, K. (2002).
\newblock Finding a state among a complete set of orthogonal ones.
\newblock {\em Physical Review A}, {\em 65}, 044302.
\newblock Available from: \url{http://dx.doi.org/10.1103/PhysRevA.65.044302},
  \href {http://arxiv.org/abs/quant-ph/0105046}
  {\path{arXiv:quant-ph/0105046}}, \href
  {http://dx.doi.org/10.1103/PhysRevA.65.044302}
  {\path{doi:10.1103/PhysRevA.65.044302}}.

\bibitem[\protect\citeauthoryear{Dvure{\v{c}}enskij, Pulmannov{\'{a}} \&
  Svozil}{Dvure{\v{c}}enskij et~al.}{1995}]{dvur-pul-svo}
Dvure{\v{c}}enskij, A., Pulmannov{\'{a}}, S., \& Svozil, K. (1995).
\newblock Partition logics, orthoalgebras and automata.
\newblock {\em Helvetica Physica Acta}, {\em 68}, 407--428.

\bibitem[\protect\citeauthoryear{Eccles}{Eccles}{1986}]{Eccles22051986}
Eccles, J.~C. (1986).
\newblock Do mental events cause neural events analogously to the probability
  fields of quantum mechanics?
\newblock {\em Proceedings of the Royal Society of London. Series B. Biological
  Sciences}, {\em 227\/}(1249), 411--428.
\newblock Available from: \url{http://dx.doi.org/10.1098/rspb.1986.0031}, \href
  {http://dx.doi.org/10.1098/rspb.1986.0031}
  {\path{doi:10.1098/rspb.1986.0031}}.

\bibitem[\protect\citeauthoryear{Eccles}{Eccles}{1990}]{eccles:papal}
Eccles, J.~C. (1990).
\newblock The mind-brain problem revisited: The microsite hypothesis.
\newblock In J.~C. Eccles \& O.~Creutzfeldt (Eds.), {\em The Principles of
  Design and Operation of the Brain}  (pp.\ 549--572). Berlin: Springer.
\newblock Available from: \url{http://dx.doi.org/10.1017/S0960129510000344},
  \href {http://dx.doi.org/10.1017/S0960129510000344}
  {\path{doi:10.1017/S0960129510000344}}.

\bibitem[\protect\citeauthoryear{Einstein}{Einstein}{1938}]{ein-reply}
Einstein, A. (1938).
\newblock Reply to criticism: Remarks concerning the essays brought together in
  this co-operative volume.
\newblock In P.~A. Schilpp (Ed.), {\em {A}lbert {E}instein:
  {P}hilosopher-Scientist (New York: Library of Living Philosophers, 1949), on
  p. 668.}  (pp.\ 665--688). New York, NY: Harper and Brothers Publishers.

\bibitem[\protect\citeauthoryear{Einstein, Podolsky \& Rosen}{Einstein
  et~al.}{1935}]{epr}
Einstein, A., Podolsky, B., \& Rosen, N. (1935).
\newblock Can quantum-mechanical description of physical reality be considered
  complete?
\newblock {\em Physical Review}, {\em 47\/}(10), 777--780.
\newblock Available from: \url{http://dx.doi.org/10.1103/PhysRev.47.777}, \href
  {http://dx.doi.org/10.1103/PhysRev.47.777}
  {\path{doi:10.1103/PhysRev.47.777}}.

\bibitem[\protect\citeauthoryear{Ekert}{Ekert}{1991}]{ekert91}
Ekert, A.~K. (1991).
\newblock Quantum cryptography based on {B}ell's theorem.
\newblock {\em Physical Review Letters}, {\em 67}, 661--663.
\newblock Available from: \url{http://dx.doi.org/10.1103/PhysRevLett.67.661},
  \href {http://dx.doi.org/10.1103/PhysRevLett.67.661}
  {\path{doi:10.1103/PhysRevLett.67.661}}.

\bibitem[\protect\citeauthoryear{Everett}{Everett}{1957}]{everett}
Everett, H. (1957).
\newblock `relative state' formulation of quantum mechanics.
\newblock {\em Reviews of Modern Physics}, {\em 29}, 454--462.
\newblock Reprinted in Ref.~\cite[pp. 315-323]{wheeler-Zurek:83}.
\newblock Available from: \url{http://dx.doi.org/10.1103/RevModPhys.29.454},
  \href {http://dx.doi.org/10.1103/RevModPhys.29.454}
  {\path{doi:10.1103/RevModPhys.29.454}}.

\bibitem[\protect\citeauthoryear{Feferman}{Feferman}{1984}]{fef-84}
Feferman, S. (1984).
\newblock {K}urt {G}\"odel: conviction and caution.
\newblock {\em Philosophia Naturalis}, {\em 21}, 546--562.

\bibitem[\protect\citeauthoryear{Feynman}{Feynman}{1965}]{feynman-law}
Feynman, R.~P. (1965).
\newblock {\em The Character of Physical Law}.
\newblock Cambridge, MA: MIT Press.

\bibitem[\protect\citeauthoryear{Fiorentino, Santori, Spillane, Beausoleil \&
  Munro}{Fiorentino et~al.}{2007}]{fiorentino:032334}
Fiorentino, M., Santori, C., Spillane, S.~M., Beausoleil, R.~G., \& Munro,
  W.~J. (2007).
\newblock Secure self-calibrating quantum random-bit generator.
\newblock {\em Physical Review A}, {\em 75\/}(3), 032334.
\newblock Available from: \url{http://dx.doi.org/10.1103/PhysRevA.75.032334},
  \href {http://dx.doi.org/10.1103/PhysRevA.75.032334}
  {\path{doi:10.1103/PhysRevA.75.032334}}.

\bibitem[\protect\citeauthoryear{Frank}{Frank}{1932}]{frank}
Frank, P. (1932).
\newblock {\em Das Kausalgesetz und seine Grenzen}.
\newblock Vienna: Springer.
\newblock {E}nglish translation in Ref.~\cite{franke}.

\bibitem[\protect\citeauthoryear{Frank \& {R. S. Cohen (Editor)}}{Frank \& {R.
  S. Cohen (Editor)}}{1997}]{franke}
Frank, P. \& {R. S. Cohen (Editor)} (1997).
\newblock {\em The Law of Causality and its Limits (Vienna Circle Collection)}.
\newblock Vienna: Springer.

\bibitem[\protect\citeauthoryear{Fredkin}{Fredkin}{1990}]{fredkin}
Fredkin, E. (1990).
\newblock Digital mechanics. an informational process based on reversible
  universal cellular automata.
\newblock {\em Physica}, {\em D45}, 254--270.
\newblock Available from: \url{http://dx.doi.org/10.1016/0167-2789(90)90186-S},
  \href {http://dx.doi.org/10.1016/0167-2789(90)90186-S}
  {\path{doi:10.1016/0167-2789(90)90186-S}}.

\bibitem[\protect\citeauthoryear{Fredkin \& Toffoli}{Fredkin \&
  Toffoli}{1982}]{fred-tof-82}
Fredkin, E. \& Toffoli, T. (1982).
\newblock Conservative logic.
\newblock {\em International Journal of Theoretical Physics}, {\em 21\/}(3-4),
  219--253.
\newblock Reprinted in Ref.~\cite[Part I, Chapter 3]{adama02}.
\newblock Available from: \url{http://dx.doi.org/10.1007/BF01857727}, \href
  {http://dx.doi.org/10.1007/BF01857727} {\path{doi:10.1007/BF01857727}}.

\bibitem[\protect\citeauthoryear{Freud}{Freud}{1999}]{Freud-1912}
Freud, S. (1999).
\newblock {R}atschl{\"{a}}ge f{\"{u}}r den {A}rzt bei der psychoanalytischen
  {B}ehandlung.
\newblock In Freud, A., Bibring, E., Hoffer, W., Kris, E., \& Isakower, O.
  (Eds.), {\em {G}esammelte {W}erke. {C}hronologisch geordnet. {A}chter {B}and.
  {W}erke aus den {J}ahren 1909--1913}, (pp.\ 376--387)., Frankfurt am Main.
  Fischer.

\bibitem[\protect\citeauthoryear{Fuchs \& Peres}{Fuchs \&
  Peres}{2000}]{fuchs-peres}
Fuchs, C.~A. \& Peres, A. (2000).
\newblock Quantum theory needs no `interpretation´.
\newblock {\em Physics Today}, {\em 53\/}(4), 70--71.
\newblock Further discussions of and reactions to the article can be found in
  the September issue of Physics Today, {\it 53}, 11-14 (2000).
\newblock Available from:
  \url{http://www.aip.org/web2/aiphome/pt/vol-53/iss-9/p11.html and
  http://www.aip.org/web2/aiphome/pt/vol-53/iss-9/p14.html}.

\bibitem[\protect\citeauthoryear{G{\"{o}}del}{G{\"{o}}del}{1931}]{godel1}
G{\"{o}}del, K. (1931).
\newblock {\"{U}}ber formal unentscheidbare {S}\"{a}tze der {P}rincipia
  {M}athematica und verwandter {S}ysteme.
\newblock {\em Monatshefte f{\"{u}}r Mathematik und Physik}, {\em 38\/}(1),
  173--198.
\newblock Reprint and {E}nglish translation in
  Ref.~\cite[p.~144-195]{godel-ges1}, and in Ref.~\cite[p.~5-40]{davis}.
\newblock Available from: \url{http://dx.doi.org/10.1007/s00605-006-0423-7},
  \href {http://dx.doi.org/10.1007/s00605-006-0423-7}
  {\path{doi:10.1007/s00605-006-0423-7}}.

\bibitem[\protect\citeauthoryear{G{\"{o}}del}{G{\"{o}}del}{1986}]{godel-ges1}
G{\"{o}}del, K. (1986).
\newblock In S.~Feferman, J.~W. Dawson, S.~C. Kleene, G.~H. Moore, R.~M.
  Solovay, \& J.~van Heijenoort (Eds.), {\em Collected Works. Publications
  1929-1936. Volume {I}}. Oxford: Oxford University Press.

\bibitem[\protect\citeauthoryear{Gold}{Gold}{1967}]{go-67}
Gold, M.~E. (1967).
\newblock Language identification in the limit.
\newblock {\em Information and Control}, {\em 10}, 447--474.
\newblock Available from:
  \url{http://dx.doi.org/10.1016/S0019-9958(67)91165-5}, \href
  {http://dx.doi.org/10.1016/S0019-9958(67)91165-5}
  {\path{doi:10.1016/S0019-9958(67)91165-5}}.

\bibitem[\protect\citeauthoryear{Greenberger, Horne, Shimony \&
  Zeilinger}{Greenberger et~al.}{1990}]{ghsz}
Greenberger, D.~M., Horne, M.~A., Shimony, A., \& Zeilinger, A. (1990).
\newblock {B}ell's theorem without inequalities.
\newblock {\em American Journal of Physics}, {\em 58}, 1131--1143.
\newblock Available from: \url{http://dx.doi.org/10.1119/1.16243}, \href
  {http://dx.doi.org/10.1119/1.16243} {\path{doi:10.1119/1.16243}}.

\bibitem[\protect\citeauthoryear{Greenberger, Horne \& Zeilinger}{Greenberger
  et~al.}{1993}]{green-horn-zei}
Greenberger, D.~M., Horne, M.~A., \& Zeilinger, A. (1993).
\newblock Multiparticle interferometry and the superposition principle.
\newblock {\em Physics Today}, {\em 46}, 22--29.

\bibitem[\protect\citeauthoryear{Greenberger \& YaSin}{Greenberger \&
  YaSin}{1989}]{greenberger2}
Greenberger, D.~M. \& YaSin, A. (1989).
\newblock ``{H}aunted'' measurements in quantum theory.
\newblock {\em Foundation of Physics}, {\em 19\/}(6), 679--704.
\newblock Available from: \url{http://dx.doi.org/10.1007/BF00731905}, \href
  {http://dx.doi.org/10.1007/BF00731905} {\path{doi:10.1007/BF00731905}}.

\bibitem[\protect\citeauthoryear{Gr{\"{u}}nwald \&
  Vit{\'{a}}nyi}{Gr{\"{u}}nwald \& Vit{\'{a}}nyi}{1987}]{gruenwald-vitanyi}
Gr{\"{u}}nwald, P.~D. \& Vit{\'{a}}nyi, P. M.~B. (1987).
\newblock Algorithmic information theory.
\newblock In P.~Adriaans \& J.~F. van Benthem (Eds.), {\em Handbook of the
  Philosophy of Information}  (pp.\ 281--320). New York, NY: Freeman.
\newblock a volume in the Handbook of the Philosophy of Science, ed. by Dov
  Gabbay, Paul Thagard, and John Wood.
\newblock Available from:
  \url{http://homepages.cwi.nl/~paulv/papers/handbooklogic07.pdf}, \href
  {http://arxiv.org/abs/arXiv:0809.2754} {\path{arXiv:arXiv:0809.2754}}.

\bibitem[\protect\citeauthoryear{Hai-Qiang, Su-Mei, Da, Jun-Tao, Ling-Ling,
  Yan-Xue \& Ling-An}{Hai-Qiang et~al.}{2004}]{0256-307X-21-10-027}
Hai-Qiang, M., Su-Mei, W., Da, Z., Jun-Tao, C., Ling-Ling, J., Yan-Xue, H., \&
  Ling-An, W. (2004).
\newblock A random number generator based on quantum entangled photon pairs.
\newblock {\em Chinese Physics Letters}, {\em 21\/}(10), 1961--1964.
\newblock Available from: \url{http://dx.doi.org/10.1088/0256-307X/21/10/027},
  \href {http://dx.doi.org/10.1088/0256-307X/21/10/027}
  {\path{doi:10.1088/0256-307X/21/10/027}}.

\bibitem[\protect\citeauthoryear{Halmos}{Halmos}{1974}]{halmos-vs}
Halmos, P.~R. (1974).
\newblock {\em Finite-dimensional Vector Spaces}.
\newblock New York, Heidelberg, Berlin: Springer.

\bibitem[\protect\citeauthoryear{Hertz}{Hertz}{1894}]{hertz-94}
Hertz, H. (1894).
\newblock {\em {P}rinzipien der {M}echanik}.
\newblock Leipzig: Barth.

\bibitem[\protect\citeauthoryear{Herzog, Kwiat, Weinfurter \& Zeilinger}{Herzog
  et~al.}{1995}]{hkwz}
Herzog, T.~J., Kwiat, P.~G., Weinfurter, H., \& Zeilinger, A. (1995).
\newblock Complementarity and the quantum eraser.
\newblock {\em Physical Review Letters}, {\em 75\/}(17), 3034--3037.
\newblock Available from: \url{http://dx.doi.org/10.1103/PhysRevLett.75.3034},
  \href {http://dx.doi.org/10.1103/PhysRevLett.75.3034}
  {\path{doi:10.1103/PhysRevLett.75.3034}}.

\bibitem[\protect\citeauthoryear{Heywood \& Redhead}{Heywood \&
  Redhead}{1983}]{hey-red}
Heywood, P. \& Redhead, M. L.~G. (1983).
\newblock Nonlocality and the {K}ochen-{S}pecker paradox.
\newblock {\em Foundations of Physics}, {\em 13\/}(5), 481--499.
\newblock Available from: \url{http://dx.doi.org/10.1007/BF00729511}, \href
  {http://dx.doi.org/10.1007/BF00729511} {\path{doi:10.1007/BF00729511}}.

\bibitem[\protect\citeauthoryear{Hilbert}{Hilbert}{1902}]{hilbert-1900e}
Hilbert, D. (1902).
\newblock Mathematical problems.
\newblock {\em Bulletin of the American Mathematical Society}, {\em 8\/}(10),
  437--479.
\newblock Available from:
  \url{http://dx.doi.org/10.1090/S0002-9904-1902-00923-3}, \href
  {http://dx.doi.org/10.1090/S0002-9904-1902-00923-3}
  {\path{doi:10.1090/S0002-9904-1902-00923-3}}.

\bibitem[\protect\citeauthoryear{Hilbert}{Hilbert}{1926}]{hilbert-26}
Hilbert, D. (1926).
\newblock {\"{U}}ber das {U}nendliche.
\newblock {\em Mathematische Annalen}, {\em 95\/}(1), 161--190.
\newblock Available from: \url{http://dx.doi.org/10.1007/BF01206605}, \href
  {http://dx.doi.org/10.1007/BF01206605} {\path{doi:10.1007/BF01206605}}.

\bibitem[\protect\citeauthoryear{Hirsch}{Hirsch}{1985}]{1985cfd..book.....F}
Hirsch, M.~W. (1985).
\newblock The chaos of dynamical systems.
\newblock In {\em Chaos, fractals, and dynamics ({G}uelph, {O}nt., 1981/1983)},
  volume~98 of {\em The chaos of dynamical systems. {L}ecture notes in pure and
  applied mathematics}  (pp.\ 189--196). New York: Dekker.

\bibitem[\protect\citeauthoryear{Hole}{Hole}{1994}]{1994IJTP...33.1085H}
Hole, A. (1994).
\newblock Predictability in deterministic theories.
\newblock {\em International Journal of Theoretical Physics}, {\em 33},
  1085--1111.
\newblock Available from: \url{http://dx.doi.org/10.1007/BF01882755}, \href
  {http://dx.doi.org/10.1007/BF01882755} {\path{doi:10.1007/BF01882755}}.

\bibitem[\protect\citeauthoryear{Hooker}{Hooker}{1975}]{hooker}
Hooker, C.~A. (1975).
\newblock {\em The Logico-Algebraic Approach to Quantum Mechanics. {V}olume
  {I}: Historical Evolution}.
\newblock Dordrecht: Reidel.

\bibitem[\protect\citeauthoryear{Hume}{Hume}{1748}]{hume-thu}
Hume, D. (1748).
\newblock {\em An Enquiry Concerning Human Understanding}.
\newblock Available from: \url{http://www.gutenberg.org/etext/9662}.

\bibitem[\protect\citeauthoryear{Jammer}{Jammer}{1966}]{jammer:66}
Jammer, M. (1966).
\newblock {\em The Conceptual Development of Quantum Mechanics}.
\newblock New York: McGraw-Hill Book Company.

\bibitem[\protect\citeauthoryear{Jammer}{Jammer}{1974}]{jammer1}
Jammer, M. (1974).
\newblock {\em The Philosophy of Quantum Mechanics}.
\newblock New York: John Wiley \& Sons.

\bibitem[\protect\citeauthoryear{Jammer}{Jammer}{1989}]{jammer:89}
Jammer, M. (1989).
\newblock {\em The Conceptual Development of Quantum Mechanics. {T}he History
  of Modern Physics, 1800-1950; v. 12\/} (second ed.).
\newblock New York: American Institute of Physics.

\bibitem[\protect\citeauthoryear{Jammer}{Jammer}{1992}]{jammer-92}
Jammer, M. (1992).
\newblock John {S}teward {B}ell and the debate on the significance of his
  contributions to the foundations of quantum mechanics.
\newblock In A.~van~der Merwe, F.~Selleri, \& G.~Tarozzi (Eds.), {\em Bell's
  Theorem and the Foundations of Modern Physics}  (pp.\ 1--23). Singapore:
  World Scientific.

\bibitem[\protect\citeauthoryear{Jennewein, Achleitner, Weihs, Weinfurter \&
  Zeilinger}{Jennewein et~al.}{2000}]{zeilinger:qct}
Jennewein, T., Achleitner, U., Weihs, G., Weinfurter, H., \& Zeilinger, A.
  (2000).
\newblock A fast and compact quantum random number generator.
\newblock {\em Review of Scientific Instruments}, {\em 71}, 1675--1680.
\newblock Available from: \url{http://dx.doi.org/10.1063/1.1150518}, \href
  {http://arxiv.org/abs/quant-ph/9912118} {\path{arXiv:quant-ph/9912118}},
  \href {http://dx.doi.org/10.1063/1.1150518} {\path{doi:10.1063/1.1150518}}.

\bibitem[\protect\citeauthoryear{Kalmbach}{Kalmbach}{1983}]{kalmbach-83}
Kalmbach, G. (1983).
\newblock {\em Orthomodular Lattices}.
\newblock New York: Academic Press.

\bibitem[\protect\citeauthoryear{Kalmbach}{Kalmbach}{1986}]{kalmbach-86}
Kalmbach, G. (1986).
\newblock {\em Measures and Hilbert Lattices}.
\newblock Singapore: World Scientific.

\bibitem[\protect\citeauthoryear{Kamber}{Kamber}{1964}]{kamber64}
Kamber, F. (1964).
\newblock Die {S}truktur des {A}ussagenkalk{\"{u}}ls in einer physikalischen
  {T}heorie.
\newblock {\em {N}achrichten der {A}kademie der {W}issenschaften in
  {G}{\"{o}}ttingen, {M}athematisch-{P}hysikalische {K}lasse}, {\em 10},
  103--124.

\bibitem[\protect\citeauthoryear{Kamber}{Kamber}{1965}]{kamber65}
Kamber, F. (1965).
\newblock Zweiwertige {W}ahrscheinlichkeitsfunktionen auf
  orthokomplement{\"{a}}ren {V}erb{\"{a}}nden.
\newblock {\em Mathematische Annalen}, {\em 158\/}(3), 158--196.
\newblock Available from: \url{http://dx.doi.org/10.1007/BF01359975}, \href
  {http://dx.doi.org/10.1007/BF01359975} {\path{doi:10.1007/BF01359975}}.

\bibitem[\protect\citeauthoryear{Kanter}{Kanter}{1990}]{kanter}
Kanter, I. (1990).
\newblock Undecidability principle and the uncertainty principle even for
  classical systems.
\newblock {\em Physical Review Letters}, {\em 64}, 332--335.
\newblock Available from: \url{http://link.aps.org/abstract/PRL/v64/p332},
  \href {http://dx.doi.org/10.1103/PhysRevLett.64.332}
  {\path{doi:10.1103/PhysRevLett.64.332}}.

\bibitem[\protect\citeauthoryear{Kauffman}{Kauffman}{1987}]{Kauffman198753}
Kauffman, L.~H. (1987).
\newblock Self-reference and recursive forms.
\newblock {\em Journal of Social and Biological Structures}, {\em 10\/}(1),
  53--72.
\newblock Available from: \url{http://dx.doi.org/10.1016/0140-1750(87)90034-0},
  \href {http://dx.doi.org/10.1016/0140-1750(87)90034-0}
  {\path{doi:10.1016/0140-1750(87)90034-0}}.

\bibitem[\protect\citeauthoryear{Kochen \& Specker}{Kochen \&
  Specker}{1965}]{kochen3}
Kochen, S. \& Specker, E.~P. (1965).
\newblock The calculus of partial propositional functions.
\newblock In {\em Proceedings of the 1964 International Congress for Logic,
  Methodology and Philosophy of Science, Jerusalem}, (pp.\ 45--57)., Amsterdam.
  North Holland.
\newblock Reprinted in Ref.~\cite[pp. 222--234]{specker-ges}.

\bibitem[\protect\citeauthoryear{Kochen \& Specker}{Kochen \&
  Specker}{1967}]{kochen1}
Kochen, S. \& Specker, E.~P. (1967).
\newblock The problem of hidden variables in quantum mechanics.
\newblock {\em Journal of Mathematics and Mechanics (now Indiana University
  Mathematics Journal)}, {\em 17\/}(1), 59--87.
\newblock Reprinted in Ref.~\cite[pp. 235--263]{specker-ges}.
\newblock Available from: \url{http://dx.doi.org/10.1512/iumj.1968.17.17004},
  \href {http://dx.doi.org/10.1512/iumj.1968.17.17004}
  {\path{doi:10.1512/iumj.1968.17.17004}}.

\bibitem[\protect\citeauthoryear{Kragh}{Kragh}{1997}]{Kragh-1997AHESradioact}
Kragh, H. (1997).
\newblock The origin of radioactivity: from solvable problem to unsolved
  non-problem.
\newblock {\em Archive for History of Exact Sciences}, {\em 50}, 331--358.
\newblock Available from: \url{http://dx.doi.org/10.1007/BF00374597}, \href
  {http://dx.doi.org/10.1007/BF00374597} {\path{doi:10.1007/BF00374597}}.

\bibitem[\protect\citeauthoryear{Kragh}{Kragh}{1999}]{Kragh-qg}
Kragh, H. (1999).
\newblock {\em Quantum Generations: A History of Physics in the Twentieth
  Century}.
\newblock Princeton, NJ: Princeton University Press.

\bibitem[\protect\citeauthoryear{Kragh}{Kragh}{2009}]{Kragh-2009_RePoss5}
Kragh, H. (2009).
\newblock Subatomic determinism and causal models of radioactive decay,
  1903-1923.
\newblock RePoSS: Research Publications on Science Studies 5. Department of
  Science Studies, University of Aarhus.

\bibitem[\protect\citeauthoryear{Kreisel}{Kreisel}{1974}]{kreisel}
Kreisel, G. (1974).
\newblock A notion of mechanistic theory.
\newblock {\em Synthese}, {\em 29}, 11--26.
\newblock Available from: \url{http://dx.doi.org/10.1007/BF00484949}, \href
  {http://dx.doi.org/10.1007/BF00484949} {\path{doi:10.1007/BF00484949}}.

\bibitem[\protect\citeauthoryear{Krenn \& Svozil}{Krenn \&
  Svozil}{1998}]{svozil-krenn}
Krenn, G. \& Svozil, K. (1998).
\newblock Stronger-than-quantum correlations.
\newblock {\em Foundations of Physics}, {\em 28\/}(6), 971--984.
\newblock Available from: \url{http://dx.doi.org/10.1023/A:1018821314465},
  \href {http://dx.doi.org/10.1023/A:1018821314465}
  {\path{doi:10.1023/A:1018821314465}}.

\bibitem[\protect\citeauthoryear{Kwiat, Steinberg \& Chiao}{Kwiat
  et~al.}{1992}]{PhysRevA.45.7729}
Kwiat, P.~G., Steinberg, A.~M., \& Chiao, R.~Y. (1992).
\newblock Observation of a ``quantum eraser:'' a revival of coherence in a
  two-photon interference experiment.
\newblock {\em Physical Review A}, {\em 45\/}(11), 7729--7739.
\newblock Available from: \url{http://dx.doi.org/10.1103/PhysRevA.45.7729},
  \href {http://dx.doi.org/10.1103/PhysRevA.45.7729}
  {\path{doi:10.1103/PhysRevA.45.7729}}.

\bibitem[\protect\citeauthoryear{Lakatos}{Lakatos}{1978}]{lakatosch}
Lakatos, I. (1978).
\newblock {\em Philosophical Papers. 1.~The Methodology of Scientific Research
  Programmes}.
\newblock Cambridge: Cambridge University Press.

\bibitem[\protect\citeauthoryear{Landauer}{Landauer}{1986}]{landauer:86}
Landauer, R. (1986).
\newblock Computation and physics: {W}heeler's meaning circuit?
\newblock {\em Foundations of Physics}, {\em 16}, 551--564.
\newblock Available from: \url{http://dx.doi.org/10.1007/BF01886520}, \href
  {http://dx.doi.org/10.1007/BF01886520} {\path{doi:10.1007/BF01886520}}.

\bibitem[\protect\citeauthoryear{Landauer}{Landauer}{1991}]{landauer}
Landauer, R. (1991).
\newblock Information is physical.
\newblock {\em Physics Today}, {\em 44\/}(5), 23--29.
\newblock Available from: \url{http://dx.doi.org/10.1063/1.881299}, \href
  {http://dx.doi.org/10.1063/1.881299} {\path{doi:10.1063/1.881299}}.

\bibitem[\protect\citeauthoryear{Landauer}{Landauer}{1994}]{landauer-95}
Landauer, R. (1994).
\newblock Advertisement for a paper {I} like.
\newblock In J.~L. Casti \& J.~F. Traub (Eds.), {\em On Limits}  (pp.\~39).
  Santa Fe, NM: Santa Fe Institute.
\newblock Santa Fe Institute Report 94-10-056.
\newblock Available from:
  \url{http://www.santafe.edu/research/publications/workingpapers/94-10-056.pdf}.

\bibitem[\protect\citeauthoryear{Laplace}{Laplace}{1998}]{laplace-prob}
Laplace, P.-S. (1995,1998).
\newblock {\em Philosophical Essay on Probabilities. {T}ranslated from the
  fifth {F}rench edition of 1825}.
\newblock Berlin, New York: Springer.
\newblock Available from:
  \url{http://www.archive.org/details/philosophicaless00lapliala}.

\bibitem[\protect\citeauthoryear{L'Ecuyer \& Simard}{L'Ecuyer \&
  Simard}{2007}]{1268777}
L'Ecuyer, P. \& Simard, R. (2007).
\newblock {TestU01}: A {C} library for empirical testing of random number
  generators.
\newblock {\em ACM Transactions on Mathematical Software (TOMS)}, {\em
  33\/}(4), Article~22, 1--40.
\newblock Available from: \url{http://dx.doi.org/10.1145/1268776.1268777},
  \href {http://dx.doi.org/10.1145/1268776.1268777}
  {\path{doi:10.1145/1268776.1268777}}.

\bibitem[\protect\citeauthoryear{Leff \& Rex}{Leff \&
  Rex}{1990a}]{maxwell-demon}
Leff, H.~S. \& Rex, A.~F. (1990a).
\newblock {\em Maxwell's Demon}.
\newblock Princeton, NJ: Princeton University Press.

\bibitem[\protect\citeauthoryear{Leff \& Rex}{Leff \&
  Rex}{1990b}]{maxwell-demon2}
Leff, H.~S. \& Rex, A.~F. (1990b).
\newblock {\em Maxwell's Demon 2. Entropy, Classical and Quantum Information,
  Computing}.
\newblock Bristol and Philadelphia: Institute of Physics Publishing.

\bibitem[\protect\citeauthoryear{Li \& Vit{\'{a}}nyi}{Li \&
  Vit{\'{a}}nyi}{1992}]{li:92}
Li, M. \& Vit{\'{a}}nyi, P. M.~B. (1992).
\newblock Inductive reasoning and {K}olmogorov complexity.
\newblock {\em Journal of Computer and System Science}, {\em 44}, 343--384.
\newblock Available from: \url{http://dx.doi.org/10.1016/0022-0000(92)90026-F},
  \href {http://dx.doi.org/10.1016/0022-0000(92)90026-F}
  {\path{doi:10.1016/0022-0000(92)90026-F}}.

\bibitem[\protect\citeauthoryear{Lichtenberg \& Lieberman}{Lichtenberg \&
  Lieberman}{1983}]{li-83}
Lichtenberg, A.~J. \& Lieberman, M.~A. (1983).
\newblock {\em Regular and Stochastic Motion}.
\newblock New York: Springer.

\bibitem[\protect\citeauthoryear{Locke}{Locke}{1690}]{lock-thu}
Locke, J. (1690).
\newblock {\em An Essay Concerning Human Understanding}.
\newblock Available from: \url{http://www.gutenberg.org/etext/10615 and
  http://www.gutenberg.org/etext/10616}.

\bibitem[\protect\citeauthoryear{Lyapunov}{Lyapunov}{1992}]{Lyapunov-92}
Lyapunov, A.~M. (1992).
\newblock The general problem of the stability of motion.
\newblock {\em International Journal of Control}, {\em 55\/}(3), 531--534.
\newblock Available from: \url{http://dx.doi.org/10.1080/00207179208934253},
  \href {http://dx.doi.org/10.1080/00207179208934253}
  {\path{doi:10.1080/00207179208934253}}.

\bibitem[\protect\citeauthoryear{Margolus}{Margolus}{2002}]{margolus-02}
Margolus, N. (2002).
\newblock Universal cellular automata based on the collisions of soft spheres.
\newblock In A.~Adamatzky (Ed.), {\em Collision-based computing}  (pp.\
  107--134). London: Springer.
\newblock Available from: \url{http://people.csail.mit.edu/nhm/cca.pdf}.

\bibitem[\protect\citeauthoryear{Marsaglia}{Marsaglia}{1995}]{diehard}
Marsaglia, G. (1995).
\newblock The {M}arsaglia random number {CDROM} including the diehard battery
  of tests of randomness.
\newblock available from URL www.stat.fsu.edu/pub/diehard/.
\newblock Available from: \url{http://www.stat.fsu.edu/pub/diehard/}.

\bibitem[\protect\citeauthoryear{Martin-L{\"{o}}f}{Martin-L{\"{o}}f}{1966}]{MartinLöf1966602}
Martin-L{\"{o}}f, P. (1966).
\newblock The definition of random sequences.
\newblock {\em Information and Control}, {\em 9\/}(6), 602--619.
\newblock Available from: \url{http://dx.doi.org/10.1016/0030-4018(87)90271-9},
  \href {http://dx.doi.org/10.1016/S0019-9958(66)80018-9}
  {\path{doi:10.1016/S0019-9958(66)80018-9}}.

\bibitem[\protect\citeauthoryear{Mermin}{Mermin}{1993}]{mermin-93}
Mermin, N.~D. (1993).
\newblock Hidden variables and the two theorems of {J}ohn {B}ell.
\newblock {\em Reviews of Modern Physics}, {\em 65}, 803--815.
\newblock Available from: \url{http://dx.doi.org/10.1103/RevModPhys.65.803},
  \href {http://dx.doi.org/10.1103/RevModPhys.65.803}
  {\path{doi:10.1103/RevModPhys.65.803}}.

\bibitem[\protect\citeauthoryear{Moore}{Moore}{1990}]{moore}
Moore, C.~D. (1990).
\newblock Unpredictability and undecidability in dynamical systems.
\newblock {\em Physical Review Letters}, {\em 64}, 2354--2357.
\newblock Cf. Ch. Bennett, {\sl Nature}, {\bf 346}, 606 (1990).
\newblock Available from: \url{http://dx.doi.org/10.1103/PhysRevLett.64.2354},
  \href {http://dx.doi.org/10.1103/PhysRevLett.64.2354}
  {\path{doi:10.1103/PhysRevLett.64.2354}}.

\bibitem[\protect\citeauthoryear{Moore}{Moore}{1956}]{e-f-moore}
Moore, E.~F. (1956).
\newblock Gedanken-experiments on sequential machines.
\newblock In C.~E. Shannon \& J.~McCarthy (Eds.), {\em Automata Studies}  (pp.\
  129--153). Princeton, NJ: Princeton University Press.

\bibitem[\protect\citeauthoryear{Nagel}{Nagel}{1986}]{nagel-ViewFromNowhere}
Nagel, T. (1986).
\newblock {\em The View from Nowhere}.
\newblock New York and Oxford: Oxford University Press.

\bibitem[\protect\citeauthoryear{Navara \& Rogalewicz}{Navara \&
  Rogalewicz}{1991}]{nav:91}
Navara, M. \& Rogalewicz, V. (1991).
\newblock The pasting constructions for orthomodular posets.
\newblock {\em Mathematische Nachrichten}, {\em 154}, 157--168.
\newblock Available from: \url{http://dx.doi.org/10.1002/mana.19911540113},
  \href {http://dx.doi.org/10.1002/mana.19911540113}
  {\path{doi:10.1002/mana.19911540113}}.

\bibitem[\protect\citeauthoryear{Odifreddi}{Odifreddi}{1989}]{odi:89}
Odifreddi, P. (1989).
\newblock {\em Classical Recursion Theory, Vol. 1}.
\newblock Amsterdam: North-Holland.

\bibitem[\protect\citeauthoryear{Ou, Hong \& Mandel}{Ou
  et~al.}{1987}]{Mandel-Ou1987118}
Ou, Z., Hong, C., \& Mandel, L. (1987).
\newblock Relation between input and output states for a beam splitter.
\newblock {\em Optics Communications}, {\em 63\/}(2), 118--122.
\newblock Available from: \url{http://dx.doi.org/10.1016/0030-4018(87)90271-9},
  \href {http://dx.doi.org/10.1016/0030-4018(87)90271-9}
  {\path{doi:10.1016/0030-4018(87)90271-9}}.

\bibitem[\protect\citeauthoryear{Paterek, Kofler, Prevedel, Klimek, Aspelmeyer,
  Zeilinger \& Brukner}{Paterek et~al.}{2010}]{1367-2630-12-1-013019}
Paterek, T., Kofler, J., Prevedel, R., Klimek, P., Aspelmeyer, M., Zeilinger,
  A., \& Brukner, {\v{C}}. (2010).
\newblock Logical independence and quantum randomness.
\newblock {\em New Journal of Physics}, {\em 12\/}(1), 013019.
\newblock Available from:
  \url{http://dx.doi.org/10.1088/1367-2630/12/1/013019}, \href
  {http://dx.doi.org/10.1088/1367-2630/12/1/013019}
  {\path{doi:10.1088/1367-2630/12/1/013019}}.

\bibitem[\protect\citeauthoryear{Pauli}{Pauli}{1958}]{pauli:58}
Pauli, W. (1958).
\newblock {D}ie allgemeinen {P}rinzipien der {W}ellenmechanik.
\newblock In S.~Fl{\"{u}}gge (Ed.), {\em {H}andbuch der {P}hysik. {B}and {V},
  {T}eil 1. {P}rinzipien der {Q}uantentheorie {I}}  (pp.\ 1--168). Berlin,
  G{\"{o}}ttingen and Heidelberg: Springer.

\bibitem[\protect\citeauthoryear{Peres}{Peres}{1978}]{peres222}
Peres, A. (1978).
\newblock Unperformed experiments have no results.
\newblock {\em American Journal of Physics}, {\em 46}, 745--747.
\newblock Available from: \url{http://dx.doi.org/10.1119/1.11393}, \href
  {http://dx.doi.org/10.1119/1.11393} {\path{doi:10.1119/1.11393}}.

\bibitem[\protect\citeauthoryear{Peres}{Peres}{1980}]{PhysRevD.22.879}
Peres, A. (1980).
\newblock Can we undo quantum measurements?
\newblock {\em Physical Review D}, {\em 22\/}(4), 879--883.
\newblock Available from: \url{http://dx.doi.org/10.1103/PhysRevD.22.879},
  \href {http://dx.doi.org/10.1103/PhysRevD.22.879}
  {\path{doi:10.1103/PhysRevD.22.879}}.

\bibitem[\protect\citeauthoryear{Peterson}{Peterson}{1993}]{peterson-NC}
Peterson, I. (1993).
\newblock {\em Chaos in the Solar System}.
\newblock New York: W. H. Freeman and Company.

\bibitem[\protect\citeauthoryear{Pfau, Sp\"alter, Kurtsiefer, Ekstrom \&
  Mlynek}{Pfau et~al.}{1994}]{PhysRevLett.73.1223}
Pfau, T., Sp\"alter, S., Kurtsiefer, C., Ekstrom, C.~R., \& Mlynek, J. (1994).
\newblock Loss of spatial coherence by a single spontaneous emission.
\newblock {\em Physical Review Letters}, {\em 73\/}(9), 1223--1226.
\newblock Available from: \url{http://dx.doi.org/10.1103/PhysRevLett.73.1223},
  \href {http://dx.doi.org/10.1103/PhysRevLett.73.1223}
  {\path{doi:10.1103/PhysRevLett.73.1223}}.

\bibitem[\protect\citeauthoryear{Pironio, Ac{\'i}n, Massar, {Boyer de la
  Giroday}, Matsukevich, Maunz, Olmschenk, Hayes, Luo, Manning \&
  Monroe}{Pironio et~al.}{2010}]{10.1038/nature09008}
Pironio, S., Ac{\'i}n, A., Massar, S., {Boyer de la Giroday}, A., Matsukevich,
  D.~N., Maunz, P., Olmschenk, S., Hayes, D., Luo, L., Manning, T.~A., \&
  Monroe, C. (2010).
\newblock Random numbers certified by {B}ell's theorem.
\newblock {\em Nature}, {\em 464}, 1021--1024.
\newblock Available from: \url{http://dx.doi.org/10.1038/nature09008}, \href
  {http://dx.doi.org/10.1038/nature09008} {\path{doi:10.1038/nature09008}}.

\bibitem[\protect\citeauthoryear{Pitowsky}{Pitowsky}{1989a}]{pitowsky-89a}
Pitowsky, I. (1989a).
\newblock From {G}eorge {B}oole to {J}ohn {B}ell: The origin of {B}ell's
  inequality.
\newblock In M.~Kafatos (Ed.), {\em {B}ell's Theorem, Quantum Theory and the
  Conceptions of the Universe}  (pp.\ 37--49). Dordrecht: Kluwer.

\bibitem[\protect\citeauthoryear{Pitowsky}{Pitowsky}{1989b}]{pitowsky}
Pitowsky, I. (1989b).
\newblock {\em Quantum Probability---Quantum Logic}.
\newblock Berlin: Springer.

\bibitem[\protect\citeauthoryear{Pitowsky}{Pitowsky}{1994}]{Pit-94}
Pitowsky, I. (1994).
\newblock {G}eorge {B}oole's `conditions of possible experience' and the
  quantum puzzle.
\newblock {\em The British Journal for the Philosophy of Science}, {\em 45},
  95--125.
\newblock Available from: \url{http://dx.doi.org/10.1093/bjps/45.1.95}, \href
  {http://dx.doi.org/10.1093/bjps/45.1.95} {\path{doi:10.1093/bjps/45.1.95}}.

\bibitem[\protect\citeauthoryear{Pitowsky \& Svozil}{Pitowsky \&
  Svozil}{2001}]{2000-poly}
Pitowsky, I. \& Svozil, K. (2001).
\newblock New optimal tests of quantum nonlocality.
\newblock {\em Physical Review A}, {\em 64}, 014102.
\newblock Available from: \url{http://dx.doi.org/10.1103/PhysRevA.64.014102},
  \href {http://arxiv.org/abs/quant-ph/0011060}
  {\path{arXiv:quant-ph/0011060}}, \href
  {http://dx.doi.org/10.1103/PhysRevA.64.014102}
  {\path{doi:10.1103/PhysRevA.64.014102}}.

\bibitem[\protect\citeauthoryear{Poincar{\'{e}}}{Poincar{\'{e}}}{1890}]{poincare-1890}
Poincar{\'{e}}, H. (1890).
\newblock Sur le probl{\'{e}}me des trois corps et les {\'{e}}quations de la
  dynamique.
\newblock {\em Acta Mathematica}, {\em 13\/}(1), A3--A270.
\newblock Available from: \url{http://dx.doi.org/10.1007/BF02392506}, \href
  {http://dx.doi.org/10.1007/BF02392506} {\path{doi:10.1007/BF02392506}}.

\bibitem[\protect\citeauthoryear{Poincar{\'{e}}}{Poincar{\'{e}}}{1914}]{poincare14}
Poincar{\'{e}}, H. (1914).
\newblock {\em Wissenschaft und Hypothese}.
\newblock Leipzig: Teubner.

\bibitem[\protect\citeauthoryear{Popescu \& Rohrlich}{Popescu \&
  Rohrlich}{1994}]{pop-rohr}
Popescu, S. \& Rohrlich, D. (1994).
\newblock Quantum nonlocality as an axiom.
\newblock {\em Foundations of Physics}, {\em 24\/}(3), 379--358.
\newblock Available from: \url{http://dx.doi.org/10.1007/BF02058098}, \href
  {http://dx.doi.org/10.1007/BF02058098} {\path{doi:10.1007/BF02058098}}.

\bibitem[\protect\citeauthoryear{Popescu \& Rohrlich}{Popescu \&
  Rohrlich}{1997}]{popescu-97b}
Popescu, S. \& Rohrlich, D. (1997).
\newblock Action and passion at a distance: an essay in honor of professor
  {A}bner {S}himony.
\newblock {\em Boston Studies in the Philosophy of Science}, {\em 194},
  197--206.
\newblock Available from: \url{http://arxiv.org/abs/quant-ph/9605004}, \href
  {http://arxiv.org/abs/quant-ph/9605004} {\path{arXiv:quant-ph/9605004}}.

\bibitem[\protect\citeauthoryear{Popper}{Popper}{1950a}]{popper-50i}
Popper, K.~R. (1950a).
\newblock Indeterminism in quantum physics and in classical physics {I}.
\newblock {\em The British Journal for the Philosophy of Science}, {\em 1},
  117--133.
\newblock Available from: \url{http://dx.doi.org/10.1093/bjps/I.2.117}, \href
  {http://dx.doi.org/10.1093/bjps/I.2.117} {\path{doi:10.1093/bjps/I.2.117}}.

\bibitem[\protect\citeauthoryear{Popper}{Popper}{1950b}]{popper-50ii}
Popper, K.~R. (1950b).
\newblock Indeterminism in quantum physics and in classical physics {II}.
\newblock {\em The British Journal for the Philosophy of Science}, {\em 1},
  173--195.
\newblock Available from: \url{http://dx.doi.org/10.1093/bjps/I.3.173}, \href
  {http://dx.doi.org/10.1093/bjps/I.3.173} {\path{doi:10.1093/bjps/I.3.173}}.

\bibitem[\protect\citeauthoryear{Popper}{Popper}{1959}]{popper-en}
Popper, K.~R. (1959).
\newblock {\em The Logic of Scientific Discovery}.
\newblock New York: Basic Books.

\bibitem[\protect\citeauthoryear{Popper \& Eccles}{Popper \&
  Eccles}{1977}]{popper-eccles}
Popper, K.~R. \& Eccles, J.~C. (1977).
\newblock {\em The Self and Its Brain}.
\newblock Berlin, Heidelberg, London, New York: Springer.

\bibitem[\protect\citeauthoryear{Pour-El \& Richards}{Pour-El \&
  Richards}{1989}]{pr1}
Pour-El, M.~B. \& Richards, J.~I. (1989).
\newblock {\em Computability in Analysis and Physics}.
\newblock Berlin: Springer.

\bibitem[\protect\citeauthoryear{Pt{\'{a}}k \& Pulmannov{\'{a}}}{Pt{\'{a}}k \&
  Pulmannov{\'{a}}}{1991}]{pulmannova-91}
Pt{\'{a}}k, P. \& Pulmannov{\'{a}}, S. (1991).
\newblock {\em Orthomodular Structures as Quantum Logics}.
\newblock Dordrecht: Kluwer Academic Publishers.

\bibitem[\protect\citeauthoryear{Purrington}{Purrington}{1997}]{purrington}
Purrington, R.~D. (1997).
\newblock {\em Physics in the Nineteenth Century}.
\newblock New Brunswick, NJ: Rutgers University Press.

\bibitem[\protect\citeauthoryear{Putnam}{Putnam}{1981}]{putnam:81}
Putnam, H. (1981).
\newblock {\em Reason, Truth and History}.
\newblock Cambridge: Cambridge University Press.

\bibitem[\protect\citeauthoryear{Rado}{Rado}{1962}]{rado}
Rado, T. (1962).
\newblock On non-computable functions.
\newblock {\em The Bell System Technical Journal}, {\em XLI(41)\/}(3),
  877--884.

\bibitem[\protect\citeauthoryear{Rarity, Owens \& Tapster}{Rarity
  et~al.}{1994}]{rarity-94}
Rarity, J.~G., Owens, M. P.~C., \& Tapster, P.~R. (1994).
\newblock Quantum random-number generation and key sharing.
\newblock {\em Journal of Modern Optics}, {\em 41}, 2435--2444.
\newblock Available from: \url{http://dx.doi.org/10.1080/09500349414552281},
  \href {http://dx.doi.org/10.1080/09500349414552281}
  {\path{doi:10.1080/09500349414552281}}.

\bibitem[\protect\citeauthoryear{Redhead}{Redhead}{1990}]{redhead}
Redhead, M. (1990).
\newblock {\em Incompleteness, Nonlocality, and Realism: A Prolegomenon to the
  Philosophy of Quantum Mechanics}.
\newblock Oxford: Clarendon Press.

\bibitem[\protect\citeauthoryear{Richardson}{Richardson}{1968}]{richardson68}
Richardson, D. (1968).
\newblock Some undecidable problems involving elementary functions of a real
  variable.
\newblock {\em Journal of Symbolic Logic}, {\em 33\/}(4), 514--520.
\newblock Available from: \url{http://www.jstor.org/stable/2271358}.

\bibitem[\protect\citeauthoryear{{Rogers, Jr.}}{{Rogers, Jr.}}{1967}]{rogers1}
{Rogers, Jr.}, H. (1967).
\newblock {\em Theory of Recursive Functions and Effective Computability}.
\newblock New York: MacGraw-Hill.

\bibitem[\protect\citeauthoryear{R{\"{o}}ssler}{R{\"{o}}ssler}{1998}]{roessler-98}
R{\"{o}}ssler, O.~E. (1998).
\newblock {\em Endophysics. {T}he World as an Interface}.
\newblock Singapore: World Scientific.
\newblock With a foreword by Peter Weibel.

\bibitem[\protect\citeauthoryear{Rousseau}{Rousseau}{2004}]{rousseau-2004}
Rousseau, C. (2004).
\newblock Divergent series: Past, present, future~$\ldots$.
\newblock preprint.
\newblock Available from:
  \url{http://www.dms.umontreal.ca/~rousseac/divergent.pdf}.

\bibitem[\protect\citeauthoryear{Rucker}{Rucker}{1982}]{rucker}
Rucker, R. (1982).
\newblock {\em Infinity and the Mind}.
\newblock Boston: Birkh{\"{a}}user.

\bibitem[\protect\citeauthoryear{Rukhin, Soto, Nechvatal, Smid, Barker, Leigh,
  Levenson, Vangel, Banks, Hekert, Dray \& Vo}{Rukhin
  et~al.}{2001}]{Rukhin-nist}
Rukhin, A., Soto, J., Nechvatal, J., Smid, M., Barker, E., Leigh, S., Levenson,
  M., Vangel, M., Banks, D., Hekert, A., Dray, J., \& Vo, S. (2001).
\newblock {\em A Statistical Test Suite for Random and Pseudorandom Number
  Generators for Cryptographic Applications. NIST Special Publication 800-22}.
\newblock Gaithersburg, MD: National Institute of Standards and Technology
  (NIST).
\newblock Available from:
  \url{http://csrc.nist.gov/groups/ST/toolkit/rng/documents/SP800-22b.pdf}.

\bibitem[\protect\citeauthoryear{Samuelson}{Samuelson}{1968}]{Samuelson-1968}
Samuelson, P.~A. (1968).
\newblock Constructing an unbiased random sequence.
\newblock {\em Journal of the American Statistical Association}, {\em
  63\/}(324), 1526--1527.
\newblock Available from: \url{http://www.jstor.org/stable/2285902}.

\bibitem[\protect\citeauthoryear{Scarpellini}{Scarpellini}{1963}]{Scarpellini-63}
Scarpellini, B. (1963).
\newblock Zwei unentscheidbare {P}robleme der {A}nalysis.
\newblock {\em Zeitschrift f{\"{u}}r Mathematische Logik und Grundlagen der
  Mathematik}, {\em 9}, 265--289.
\newblock Available from: \url{http://dx.doi.org/10.1002/malq.19630091802},
  \href {http://dx.doi.org/10.1002/malq.19630091802}
  {\path{doi:10.1002/malq.19630091802}}.

\bibitem[\protect\citeauthoryear{Schaller \& Svozil}{Schaller \&
  Svozil}{1996}]{schaller-96}
Schaller, M. \& Svozil, K. (1996).
\newblock Automaton logic.
\newblock {\em International Journal of Theoretical Physics}, {\em 35\/}(5),
  911--940.
\newblock Available from: \url{http://dx.doi.org/10.1007/BF02302381}, \href
  {http://dx.doi.org/10.1007/BF02302381} {\path{doi:10.1007/BF02302381}}.

\bibitem[\protect\citeauthoryear{Schlick}{Schlick}{1932}]{schlick}
Schlick, M. (1932).
\newblock Causality in everday life and in recent science.
\newblock {\em University of California Publications in Philosophy}, {\em 15},
  99--125.
\newblock reprinted in \cite[pp.~515-533]{Schlick1} and
  \cite[pp.~415-445]{schlick-ges-v6}.
\newblock Available from: \url{http://dx.doi.org/10.1007/978-3-211-33116-3_24},
  \href {http://dx.doi.org/10.1007/978-3-211-33116-3\_24}
  {\path{doi:10.1007/978-3-211-33116-3\_24}}.

\bibitem[\protect\citeauthoryear{Schlick}{Schlick}{1935}]{schlick-35}
Schlick, M. (1935).
\newblock Unanswerable questions?
\newblock {\em The Philosopher}, {\em 13}, 98--104.
\newblock reprinted in
  \cite[pp.~621-634]{springerlink:10.1007/978-3-211-33116-3_33}.
\newblock Available from: \url{http://dx.doi.org/10.1007/978-3-211-33116-3_33},
  \href {http://dx.doi.org/10.1007/978-3-211-33116-3\_33}
  {\path{doi:10.1007/978-3-211-33116-3\_33}}.

\bibitem[\protect\citeauthoryear{Schlick}{Schlick}{1949}]{Schlick1}
Schlick, M. (1949).
\newblock Causality in everday life and in recent science.
\newblock In H.~Feigl \& W.~Sellars (Eds.), {\em Readings in Philosophical
  Analysis}  (pp.\ 515--533). New York: Appleton-Century-Crofts.
\newblock previously published in \cite{schlick}.

\bibitem[\protect\citeauthoryear{Schlick}{Schlick}{2008a}]{schlick-ges-v6}
Schlick, M. (2008a).
\newblock Causality in everday life and in recent science.
\newblock In F.~Stadler \& H.~J. Wendel (Eds.), {\em {G}esamtausgabe. {D}ie
  Wiener {Z}eit. {A}ufs{\"{a}}tze, {B}eitr{\"{a}}ge, {R}ezensionen 1926--1936},
  volume~6  (pp.\ 415--445). Wien and New York: Springer.
\newblock Available from: \url{http://dx.doi.org/10.1007/978-3-211-33116-3_24},
  \href {http://dx.doi.org/10.1007/978-3-211-33116-3_24}
  {\path{doi:10.1007/978-3-211-33116-3_24}}.

\bibitem[\protect\citeauthoryear{Schlick}{Schlick}{2008b}]{springerlink:10.1007/978-3-211-33116-3_33}
Schlick, M. (2008b).
\newblock Unanswerable questions?
\newblock In F.~Stadler \& H.~J. Wendel (Eds.), {\em {G}esamtausgabe. {D}ie
  Wiener {Z}eit. {A}ufs{\"{a}}tze, {B}eitr{\"{a}}ge, {R}ezensionen 1926--1936},
  volume~6  (pp.\ 621--634). Wien and New York: Springer.
\newblock Available from: \url{http://dx.doi.org/10.1007/978-3-211-33116-3_33},
  \href {http://dx.doi.org/10.1007/978-3-211-33116-3_33}
  {\path{doi:10.1007/978-3-211-33116-3_33}}.

\bibitem[\protect\citeauthoryear{Schr{\"{o}}dinger}{Schr{\"{o}}dinger}{1935a}]{schrodinger}
Schr{\"{o}}dinger, E. (1935a).
\newblock Die gegenw{\"{a}}rtige {S}ituation in der {Q}uantenmechanik.
\newblock {\em Naturwissenschaften}, {\em 23}, 807--812, 823--828, 844--849.
\newblock {E}nglish translation in Ref.~\cite{trimmer} and in Ref.~\cite[pp.
  152-167]{wheeler-Zurek:83}.
\newblock Available from: \url{http://dx.doi.org/10.1007/BF01491891,
  http://dx.doi.org/10.1007/BF01491914, http://dx.doi.org/10.1007/BF01491987},
  \href {http://dx.doi.org/10.1007/BF01491891, 10.1007/BF01491914,
  10.1007/BF01491987} {\path{doi:10.1007/BF01491891, 10.1007/BF01491914,
  10.1007/BF01491987}}.

\bibitem[\protect\citeauthoryear{Schr{\"{o}}dinger}{Schr{\"{o}}dinger}{1935b}]{CambridgeJournals:1737068}
Schr{\"{o}}dinger, E. (1935b).
\newblock Discussion of probability relations between separated systems.
\newblock {\em Mathematical Proceedings of the Cambridge Philosophical
  Society}, {\em 31\/}(04), 555--563.
\newblock Available from: \url{http://dx.doi.org/10.1017/S0305004100013554},
  \href {http://dx.doi.org/10.1017/S0305004100013554}
  {\path{doi:10.1017/S0305004100013554}}.

\bibitem[\protect\citeauthoryear{Schr{\"{o}}dinger}{Schr{\"{o}}dinger}{1936}]{CambridgeJournals:2027212}
Schr{\"{o}}dinger, E. (1936).
\newblock Probability relations between separated systems.
\newblock {\em Mathematical Proceedings of the Cambridge Philosophical
  Society}, {\em 32\/}(03), 446--452.
\newblock Available from: \url{http://dx.doi.org/10.1017/S0305004100019137},
  \href {http://dx.doi.org/10.1017/S0305004100019137}
  {\path{doi:10.1017/S0305004100019137}}.

\bibitem[\protect\citeauthoryear{Schr{\"{o}}dinger}{Schr{\"{o}}dinger}{1995}]{schroedinger-interpretation}
Schr{\"{o}}dinger, E. (1995).
\newblock {\em The Interpretation of Quantum Mechanics. {D}ublin Seminars
  (1949-1955) and Other Unpublished Essays}.
\newblock Woodbridge, Connecticut: Ox Bow Press.

\bibitem[\protect\citeauthoryear{Scully \& Dr\"uhl}{Scully \&
  Dr\"uhl}{1982}]{PhysRevA.25.2208}
Scully, M.~O. \& Dr\"uhl, K. (1982).
\newblock Quantum eraser: A proposed photon correlation experiment concerning
  observation and ``delayed choice'' in quantum mechanics.
\newblock {\em Physical Review A}, {\em 25\/}(4), 2208--2213.
\newblock Available from: \url{http://dx.doi.org/10.1103/PhysRevA.25.2208},
  \href {http://dx.doi.org/10.1103/PhysRevA.25.2208}
  {\path{doi:10.1103/PhysRevA.25.2208}}.

\bibitem[\protect\citeauthoryear{Scully, Englert \& Walther}{Scully
  et~al.}{1991}]{Nature351}
Scully, M.~O., Englert, B.-G., \& Walther, H. (1991).
\newblock Quantum optical tests of complementarity.
\newblock {\em Nature}, {\em 351}, 111--116.
\newblock Available from: \url{http://dx.doi.org/10.1038/351111a0}, \href
  {http://dx.doi.org/10.1038/351111a0} {\path{doi:10.1038/351111a0}}.

\bibitem[\protect\citeauthoryear{Shaw}{Shaw}{1981}]{shaw}
Shaw, R.~S. (1981).
\newblock Strange attractors, chaotic behavior, and information flow.
\newblock {\em Zeitschrift f{\"{u}}r Naturforschung A}, {\em 36}, 80--112.

\bibitem[\protect\citeauthoryear{Sherr, Bainbridge \& Anderson}{Sherr
  et~al.}{1941}]{PhysRev.60.473}
Sherr, R., Bainbridge, K.~T., \& Anderson, H.~H. (1941).
\newblock Transmutation of mercury by fast neutrons.
\newblock {\em Physical Review}, {\em 60\/}(7), 473--479.
\newblock Available from: \url{http://dx.doi.org/10.1103/PhysRev.60.473}, \href
  {http://dx.doi.org/10.1103/PhysRev.60.473}
  {\path{doi:10.1103/PhysRev.60.473}}.

\bibitem[\protect\citeauthoryear{Shimony}{Shimony}{1984}]{shimony2}
Shimony, A. (1984).
\newblock Controllable and uncontrollable non-locality.
\newblock In Kamefuchi, S. \& Gakkai, N.~B. (Eds.), {\em Proceedings of the
  International Symposium... Proceedings of the International Symposium
  Foundations of Quantum Mechanics in the Light of New Technology}, (pp.\
  225--230)., Tokyo. Physical Society of Japan.
\newblock See also J. Jarrett, {\sl Bell's Theorem, Quantum Mechanics and Local
  Realism}, Ph. D. thesis, Univ. of Chicago, 1983; {\sl Nous}, {\bf 18}, 569
  (1984).

\bibitem[\protect\citeauthoryear{Smale}{Smale}{1967}]{smale-hm}
Smale, S. (1967).
\newblock Differentiable dynamical systems.
\newblock {\em Bulletin of the American Mathematical Society}, {\em 73},
  747--817.
\newblock Available from:
  \url{http://dx.doi.org/10.1090/S0002-9904-1967-11798-1}, \href
  {http://dx.doi.org/10.1090/S0002-9904-1967-11798-1}
  {\path{doi:10.1090/S0002-9904-1967-11798-1}}.

\bibitem[\protect\citeauthoryear{Smullyan}{Smullyan}{1992a}]{smullyan-92}
Smullyan, R.~M. (1992a).
\newblock {\em {G}{\"{o}}del's Incompleteness Theorems}.
\newblock New York, New York: Oxford University Press.

\bibitem[\protect\citeauthoryear{Smullyan}{Smullyan}{1992b}]{smullyan-78}
Smullyan, R.~M. (1992b).
\newblock {\em What is the Name of This Book?}
\newblock Englewood Cliffs, NJ: Prentice-Hall, Inc.

\bibitem[\protect\citeauthoryear{Sosa}{Sosa}{2009}]{sosa-rk2}
Sosa, E. (2009).
\newblock {\em Reflective Knowledge. {A}pt Belief and Reflective Knowledge,
  Volume {II}}.
\newblock Oxford: Clarendon Press.

\bibitem[\protect\citeauthoryear{Specker}{Specker}{1949}]{Specker49}
Specker, E. (1949).
\newblock Nicht konstruktiv beweisbare {S}{\"{a}}tze der {A}nalysis.
\newblock {\em The Journal of Smbolic Logic}, {\em 14}, 145--158.
\newblock Reprinted in Ref.~\cite[pp. 35--48]{specker-ges}; {E}nglish
  translation: {\it Theorems of Analysis which cannot be proven
  constructively}.
\newblock Available from: \url{http://www.jstor.org/stable/2267043}.

\bibitem[\protect\citeauthoryear{Specker}{Specker}{1959}]{Specker57}
Specker, E. (1959).
\newblock Der {S}atz vom {M}aximum in der rekursiven {A}nalysis.
\newblock In Heyting, A. (Ed.), {\em Constructivity in mathematics :
  proceedings of the colloquium held at Amsterdam, 1957}, (pp.\ 254--265).,
  Amsterdam. North-Holland Publishing Company.
\newblock Reprinted in Ref.~\cite[pp. 148--159]{specker-ges}; {E}nglish
  translation: {\it Theorems of Analysis which cannot be proven
  constructively}.

\bibitem[\protect\citeauthoryear{Specker}{Specker}{1960}]{specker-60}
Specker, E. (1960).
\newblock {D}ie {L}ogik nicht gleichzeitig entscheidbarer {A}ussagen.
\newblock {\em Dialectica}, {\em 14\/}(2-3), 239--246.
\newblock Reprinted in Ref.~\cite[pp. 175--182]{specker-ges}; {E}nglish
  translation: {\it The logic of propositions which are not simultaneously
  decidable}, Reprinted in Ref.~\cite[pp. 135-140]{hooker} and as an
  eprint~arXiv:1103.4537.
\newblock Available from:
  \url{http://dx.doi.org/10.1111/j.1746-8361.1960.tb00422.x}, \href
  {http://arxiv.org/abs/http://arxiv.org/abs/1103.4537}
  {\path{arXiv:http://arxiv.org/abs/1103.4537}}, \href
  {http://dx.doi.org/10.1111/j.1746-8361.1960.tb00422.x}
  {\path{doi:10.1111/j.1746-8361.1960.tb00422.x}}.

\bibitem[\protect\citeauthoryear{Specker}{Specker}{1990}]{specker-ges}
Specker, E. (1990).
\newblock {\em Selecta}.
\newblock Basel: Birkh{\"{a}}user Verlag.

\bibitem[\protect\citeauthoryear{Stace}{Stace}{1934}]{stace}
Stace, W.~T. (1934).
\newblock The refutation of realism.
\newblock {\em Mind}, {\em 43\/}(170), 145--155.
\newblock reprinted in \cite[pp.~364-372]{stace1}.
\newblock Available from: \url{http://www.jstor.org/stable/2250077}.

\bibitem[\protect\citeauthoryear{Stace}{Stace}{1949}]{stace1}
Stace, W.~T. (1949).
\newblock The refutation of realism.
\newblock In H.~Feigl \& W.~Sellars (Eds.), {\em Readings in Philosophical
  Analysis}  (pp.\ 364--372). New York: Appleton-Century-Crofts.
\newblock previously published in {\em Mind} {\bf 53}, 349-353 (1934).

\bibitem[\protect\citeauthoryear{Stefanov, Gisin, Guinnard, Guinnard \&
  Zbinden}{Stefanov et~al.}{2000}]{stefanov-2000}
Stefanov, A., Gisin, N., Guinnard, O., Guinnard, L., \& Zbinden, H. (2000).
\newblock Optical quantum random number generator.
\newblock {\em Journal of Modern Optics}, {\em 47}, 595--598.
\newblock Available from: \url{http://dx.doi.org/10.1080/095003400147908},
  \href {http://dx.doi.org/10.1080/095003400147908}
  {\path{doi:10.1080/095003400147908}}.

\bibitem[\protect\citeauthoryear{Sterne}{Sterne}{1767}]{sterne}
Sterne, L. (1760-1767).
\newblock {\em The Life and Opinions of Tristram Shandy, Gentleman}.
\newblock London.
\newblock Available from: \url{http://www.gutenberg.org/etext/1079}.

\bibitem[\protect\citeauthoryear{Stewart}{Stewart}{1991}]{Stewart-91}
Stewart, I. (1991).
\newblock Deciding the undecidable.
\newblock {\em Nature}, {\em 352}, 664--665.
\newblock Available from: \url{http://dx.doi.org/10.1038/352664a0}, \href
  {http://dx.doi.org/10.1038/352664a0} {\path{doi:10.1038/352664a0}}.

\bibitem[\protect\citeauthoryear{Sundman}{Sundman}{1912}]{Sundman12}
Sundman, K.~F. (1912).
\newblock Memoire sur le probl{\`{e}}me de trois corps.
\newblock {\em Acta Mathematica}, {\em 36}, 105--179.
\newblock Available from: \url{http://dx.doi.org/10.1007/BF02422379}, \href
  {http://dx.doi.org/10.1007/BF02422379} {\path{doi:10.1007/BF02422379}}.

\bibitem[\protect\citeauthoryear{Suppes}{Suppes}{1993}]{suppes-1993}
Suppes, P. (1993).
\newblock The transcendental character of determinism.
\newblock {\em Midwest Studies In Philosophy}, {\em 18\/}(1), 242--257.
\newblock Available from:
  \url{http://dx.doi.org/10.1111/j.1475-4975.1993.tb00266.x}, \href
  {http://dx.doi.org/10.1111/j.1475-4975.1993.tb00266.x}
  {\path{doi:10.1111/j.1475-4975.1993.tb00266.x}}.

\bibitem[\protect\citeauthoryear{Svozil}{Svozil}{1990}]{svozil-qct}
Svozil, K. (1990).
\newblock The quantum coin toss---testing microphysical undecidability.
\newblock {\em Physics Letters A}, {\em 143}, 433--437.
\newblock Available from: \url{http://dx.doi.org/10.1016/0375-9601(90)90408-G},
  \href {http://dx.doi.org/10.1016/0375-9601(90)90408-G}
  {\path{doi:10.1016/0375-9601(90)90408-G}}.

\bibitem[\protect\citeauthoryear{Svozil}{Svozil}{1993}]{svozil-93}
Svozil, K. (1993).
\newblock {\em Randomness \& Undecidability in Physics}.
\newblock Singapore: World Scientific.

\bibitem[\protect\citeauthoryear{Svozil}{Svozil}{1994}]{svozil-94}
Svozil, K. (1994).
\newblock Extrinsic-intrinsic concept and complementarity.
\newblock In H.~Atmanspacher \& G.~J. Dalenoort (Eds.), {\em Inside versus
  Outside}  (pp.\ 273--288). Heidelberg: Springer.
\newblock Available from:
  \url{http://tph.tuwien.ac.at/~svozil/publ/1994-exin.pdf}.

\bibitem[\protect\citeauthoryear{Svozil}{Svozil}{1995a}]{svozil-nat-acad}
Svozil, K. (1995a).
\newblock A constructivist manifesto for the physical sciences ---
  {C}onstructive re-interpretation of physical undecidability.
\newblock In Schimanovich, W.~D., K{\"{o}}hler, E., \& Stadler, F. (Eds.), {\em
  The Foundational Debate, Complexity and Constructivity in Mathematics and
  Physics}, (pp.\ 65--88)., Dordrecht, Boston, London. Kluwer.
\newblock Available from:
  \url{http://tph.tuwien.ac.at/~svozil/publ/1994-interface.pdf}.

\bibitem[\protect\citeauthoryear{Svozil}{Svozil}{1995b}]{svozil-set}
Svozil, K. (1995b).
\newblock Set theory and physics.
\newblock {\em Foundations of Physics}, {\em 25}, 1541--1560.
\newblock Available from: \url{http://dx.doi.org/10.1007/BF02055507}, \href
  {http://dx.doi.org/10.1007/BF02055507} {\path{doi:10.1007/BF02055507}}.

\bibitem[\protect\citeauthoryear{Svozil}{Svozil}{1998}]{svozil-ql}
Svozil, K. (1998).
\newblock {\em Quantum Logic}.
\newblock Singapore: Springer.

\bibitem[\protect\citeauthoryear{Svozil}{Svozil}{2002a}]{svozil-2001-convention}
Svozil, K. (2002a).
\newblock Conventions in relativity theory and quantum mechanics.
\newblock {\em Foundations of Physics}, {\em 32}, 479--502.
\newblock Available from: \url{http://dx.doi.org/10.1023/A:1015017831247},
  \href {http://arxiv.org/abs/quant-ph/0110054}
  {\path{arXiv:quant-ph/0110054}}, \href
  {http://dx.doi.org/10.1023/A:1015017831247}
  {\path{doi:10.1023/A:1015017831247}}.

\bibitem[\protect\citeauthoryear{Svozil}{Svozil}{2002b}]{svozil-2002-statepart-prl}
Svozil, K. (2002b).
\newblock Quantum information in base $n$ defined by state partitions.
\newblock {\em Physical Review A}, {\em 66}, 044306.
\newblock Available from: \url{http://dx.doi.org/10.1103/PhysRevA.66.044306},
  \href {http://arxiv.org/abs/quant-ph/0205031}
  {\path{arXiv:quant-ph/0205031}}, \href
  {http://dx.doi.org/10.1103/PhysRevA.66.044306}
  {\path{doi:10.1103/PhysRevA.66.044306}}.

\bibitem[\protect\citeauthoryear{Svozil}{Svozil}{2004}]{svozil-2003-garda}
Svozil, K. (2004).
\newblock Quantum information via state partitions and the context translation
  principle.
\newblock {\em Journal of Modern Optics}, {\em 51}, 811--819.
\newblock Available from: \url{http://dx.doi.org/10.1080/09500340410001664179},
  \href {http://arxiv.org/abs/quant-ph/0308110}
  {\path{arXiv:quant-ph/0308110}}, \href
  {http://dx.doi.org/10.1080/09500340410001664179}
  {\path{doi:10.1080/09500340410001664179}}.

\bibitem[\protect\citeauthoryear{Svozil}{Svozil}{2005a}]{svozil-2004-brainteaser}
Svozil, K. (2005a).
\newblock Communication cost of breaking the {B}ell barrier.
\newblock {\em Physical Review A}, {\em 72\/}(9), 050302(R).
\newblock Available from: \url{http://dx.doi.org/10.1103/PhysRevA.72.050302},
  \href {http://arxiv.org/abs/physics/0510050} {\path{arXiv:physics/0510050}},
  \href {http://dx.doi.org/10.1103/PhysRevA.72.050302}
  {\path{doi:10.1103/PhysRevA.72.050302}}.

\bibitem[\protect\citeauthoryear{Svozil}{Svozil}{2005b}]{svozil-2001-eua}
Svozil, K. (2005b).
\newblock Logical equivalence between generalized urn models and finite
  automata.
\newblock {\em International Journal of Theoretical Physics}, {\em 44},
  745--754.
\newblock Available from: \url{http://dx.doi.org/10.1007/s10773-005-7052-0},
  \href {http://arxiv.org/abs/quant-ph/0209136}
  {\path{arXiv:quant-ph/0209136}}, \href
  {http://dx.doi.org/10.1007/s10773-005-7052-0}
  {\path{doi:10.1007/s10773-005-7052-0}}.

\bibitem[\protect\citeauthoryear{Svozil}{Svozil}{2005c}]{svozil-2004-analog}
Svozil, K. (2005c).
\newblock Noncontextuality in multipartite entanglement.
\newblock {\em J. Phys. A: Math. Gen.}, {\em 38}, 5781--5798.
\newblock Available from: \url{http://dx.doi.org/10.1088/0305-4470/38/25/013},
  \href {http://arxiv.org/abs/quant-ph/0401113}
  {\path{arXiv:quant-ph/0401113}}, \href
  {http://dx.doi.org/10.1088/0305-4470/38/25/013}
  {\path{doi:10.1088/0305-4470/38/25/013}}.

\bibitem[\protect\citeauthoryear{Svozil}{Svozil}{2006a}]{svozil-2006-uniquenessprinciple}
Svozil, K. (2006a).
\newblock Are simultaneous {B}ell measurements possible?
\newblock {\em New Journal of Physics}, {\em 8}, 39, 1--8.
\newblock Available from: \url{http://dx.doi.org/10.1088/1367-2630/8/3/039},
  \href {http://arxiv.org/abs/quant-ph/0401113}
  {\path{arXiv:quant-ph/0401113}}, \href
  {http://dx.doi.org/10.1088/1367-2630/8/3/039}
  {\path{doi:10.1088/1367-2630/8/3/039}}.

\bibitem[\protect\citeauthoryear{Svozil}{Svozil}{2006b}]{svozil-2005-cu}
Svozil, K. (2006b).
\newblock Computational universes.
\newblock {\em Chaos, Solitons \& Fractals}, {\em 25\/}(4), 845--859.
\newblock Available from: \url{http://dx.doi.org/10.1016/j.chaos.2004.11.055},
  \href {http://arxiv.org/abs/physics/0305048} {\path{arXiv:physics/0305048}},
  \href {http://dx.doi.org/10.1016/j.chaos.2004.11.055}
  {\path{doi:10.1016/j.chaos.2004.11.055}}.

\bibitem[\protect\citeauthoryear{Svozil}{Svozil}{2006c}]{svozil-2005-ln1e}
Svozil, K. (2006c).
\newblock Staging quantum cryptography with chocolate balls.
\newblock {\em American Journal of Physics}, {\em 74\/}(9), 800--803.
\newblock Available from: \url{http://dx.doi.org/10.1119/1.2205879}, \href
  {http://arxiv.org/abs/physics/0510050} {\path{arXiv:physics/0510050}}, \href
  {http://dx.doi.org/10.1119/1.2205879} {\path{doi:10.1119/1.2205879}}.

\bibitem[\protect\citeauthoryear{Svozil}{Svozil}{2007}]{svozil-2007-cestial}
Svozil, K. (2007).
\newblock Omega and the time evolution of the n-body problem.
\newblock In Calude, C.~S. (Ed.), {\em Randomness and Complexity, from
  {L}eibniz to {C}haitin}, (pp.\ 231--236)., Singapore. World Scientific.
\newblock eprint arXiv:physics/0703031.
\newblock Available from: \url{http://arxiv.org/abs/physics/0703031}, \href
  {http://arxiv.org/abs/arXiv:physics/0703031}
  {\path{arXiv:arXiv:physics/0703031}}.

\bibitem[\protect\citeauthoryear{Svozil}{Svozil}{2009a}]{svozil-2008-ql}
Svozil, K. (2009a).
\newblock Contexts in quantum, classical and partition logic.
\newblock In K.~Engesser, D.~M. Gabbay, \& D.~Lehmann (Eds.), {\em Handbook of
  Quantum Logic and Quantum Structures}  (pp.\ 551--586). Amsterdam: Elsevier.
\newblock Available from: \url{http://arxiv.org/abs/quant-ph/0609209}, \href
  {http://arxiv.org/abs/arXiv:quant-ph/0609209}
  {\path{arXiv:arXiv:quant-ph/0609209}}.

\bibitem[\protect\citeauthoryear{Svozil}{Svozil}{2009b}]{1612095}
Svozil, K. (2009b).
\newblock On the brightness of the {T}homson lamp: A prolegomenon to quantum
  recursion theory.
\newblock In Calude, C.~S., Costa, J.~F., Dershowitz, N., Freire, E., \&
  Rozenberg, G. (Eds.), {\em UC '09: Proceedings of the 8th International
  Conference on Unconventional Computation}, (pp.\ 236--246)., Berlin,
  Heidelberg. Springer.
\newblock Available from: \url{http://dx.doi.org/10.1007/978-3-642-03745-0_26},
  \href {http://dx.doi.org/10.1007/978-3-642-03745-0_26}
  {\path{doi:10.1007/978-3-642-03745-0_26}}.

\bibitem[\protect\citeauthoryear{Svozil}{Svozil}{2009c}]{svozil:040102}
Svozil, K. (2009c).
\newblock Proposed direct test of a certain type of noncontextuality in quantum
  mechanics.
\newblock {\em Physical Review A}, {\em 80\/}(4), 040102.
\newblock Available from: \url{http://dx.doi.org/10.1103/PhysRevA.80.040102},
  \href {http://dx.doi.org/10.1103/PhysRevA.80.040102}
  {\path{doi:10.1103/PhysRevA.80.040102}}.

\bibitem[\protect\citeauthoryear{Svozil}{Svozil}{2009d}]{svozil-2006-omni}
Svozil, K. (2009d).
\newblock Quantum scholasticism: On quantum contexts, counterfactuals, and the
  absurdities of quantum omniscience.
\newblock {\em Information Sciences}, {\em 179}, 535--541.
\newblock Available from: \url{http://dx.doi.org/10.1016/j.ins.2008.06.012},
  \href {http://dx.doi.org/10.1016/j.ins.2008.06.012}
  {\path{doi:10.1016/j.ins.2008.06.012}}.

\bibitem[\protect\citeauthoryear{Svozil}{Svozil}{2009e}]{svozil-2009-howto}
Svozil, K. (2009e).
\newblock Three criteria for quantum random-number generators based on beam
  splitters.
\newblock {\em Physical Review A}, {\em 79\/}(5), 054306.
\newblock Available from: \url{http://dx.doi.org/10.1103/PhysRevA.79.054306},
  \href {http://arxiv.org/abs/arXiv:quant-ph/0903.2744}
  {\path{arXiv:arXiv:quant-ph/0903.2744}}, \href
  {http://dx.doi.org/10.1103/PhysRevA.79.054306}
  {\path{doi:10.1103/PhysRevA.79.054306}}.

\bibitem[\protect\citeauthoryear{Svozil}{Svozil}{2010a}]{2010-qchocolate}
Svozil, K. (2010a).
\newblock Chocolate cryptography.
\newblock eprint {arXiv:0903.0231}.
\newblock Available from: \url{http://http://arxiv.org/abs/0903.0231}, \href
  {http://arxiv.org/abs/arXiv:0903.0231} {\path{arXiv:arXiv:0903.0231}}.

\bibitem[\protect\citeauthoryear{Svozil}{Svozil}{2010b}]{svozil_2010-pc09}
Svozil, K. (2010b).
\newblock Quantum value indefiniteness.
\newblock {\em Natural Computing}, {\em online first}, 1--12.
\newblock Available from: \url{http://dx.doi.org/10.1007/s11047-010-9241-x},
  \href {http://arxiv.org/abs/arXiv:1001.1436} {\path{arXiv:arXiv:1001.1436}},
  \href {http://dx.doi.org/10.1007/s11047-010-9241-x}
  {\path{doi:10.1007/s11047-010-9241-x}}.

\bibitem[\protect\citeauthoryear{Svozil \& Tkadlec}{Svozil \&
  Tkadlec}{1996}]{svozil-tkadlec}
Svozil, K. \& Tkadlec, J. (1996).
\newblock Greechie diagrams, nonexistence of measures in quantum logics and
  {K}ochen--{S}pecker type constructions.
\newblock {\em Journal of Mathematical Physics}, {\em 37\/}(11), 5380--5401.
\newblock Available from: \url{http://dx.doi.org/10.1063/1.531710}, \href
  {http://dx.doi.org/10.1063/1.531710} {\path{doi:10.1063/1.531710}}.

\bibitem[\protect\citeauthoryear{Tarski}{Tarski}{1932}]{tarski:32}
Tarski, A. (1932).
\newblock {D}er {W}ahrheitsbegriff in den {S}prachen der deduktiven
  {D}isziplinen.
\newblock {\em Akademie der Wissenschaften in Wien.
  Mathematisch-naturwissenschaftliche Klasse, Akademischer Anzeiger}, {\em 69},
  9--12.

\bibitem[\protect\citeauthoryear{Tarski}{Tarski}{1956}]{tarski:56}
Tarski, A. (1956).
\newblock {\em Logic, Semantics and Metamathematics}.
\newblock Oxford: Oxford University Press.

\bibitem[\protect\citeauthoryear{{The RAND Corporation}}{{The RAND
  Corporation}}{1955}]{rand-55}
{The RAND Corporation} (1955).
\newblock {\em A Million Random Digits with 100,000 Normal Deviates Free Press
  Publishers}.
\newblock Glencoe, Illinois: Knolls Atomic Power Lab. Report KAPL-3147.
\newblock Available from:
  \url{http://www.rand.org/pubs/monograph\_reports/MR1418/}.

\bibitem[\protect\citeauthoryear{Toffoli}{Toffoli}{1978}]{toffoli:79}
Toffoli, T. (1978).
\newblock The role of the observer in uniform systems.
\newblock In G.~J. Klir (Ed.), {\em Applied General Systems Research, Recent
  Developments and Trends}  (pp.\ 395--400). New York, London: Plenum Press.

\bibitem[\protect\citeauthoryear{Trimmer}{Trimmer}{1980}]{trimmer}
Trimmer, J.~D. (1980).
\newblock The present situation in quantum mechanics: a translation of
  {S}chr{\"{o}}dinger's ``cat paradox''.
\newblock {\em Proceedings of the American Philosophical Society}, {\em 124},
  323--338.
\newblock Reprinted in Ref.~\cite[pp. 152-167]{wheeler-Zurek:83}.
\newblock Available from:
  \url{http://www.tu-harburg.de/rzt/rzt/it/QM/cat.html}.

\bibitem[\protect\citeauthoryear{{T}uring}{{T}uring}{1937}]{turing-36}
{T}uring, A.~M. (1936-7 and 1937).
\newblock On computable numbers, with an application to the
  {E}ntscheidungsproblem.
\newblock {\em Proceedings of the London Mathematical Society, Series 2}, {\em
  42, 43}, 230--265, 544--546.
\newblock Reprinted in Ref.~\cite{davis}.
\newblock Available from: \url{http://dx.doi.org/10.1112/plms/s2-42.1.230,
  http://dx.doi.org/10.1112/plms/s2-43.6.544}, \href
  {http://dx.doi.org/10.1112/plms/s2-42.1.230, 10.1112/plms/s2-43.6.544}
  {\path{doi:10.1112/plms/s2-42.1.230, 10.1112/plms/s2-43.6.544}}.

\bibitem[\protect\citeauthoryear{{T}uring}{{T}uring}{1968}]{Turing-Intelligent_Machinery}
{T}uring, A.~M. (1968).
\newblock Intelligent machinery.
\newblock In C.~R. Evans \& A.~D.~J. Robertson (Eds.), {\em {C}ybernetics: Key
  Papers}  (pp.\ 27--52). London: Butterworths.
\newblock Available from: \url{http://dx.doi.org/10.1007/978-3-540-70626-7_40},
  \href {http://dx.doi.org/10.1007/978-3-540-70626-7_40}
  {\path{doi:10.1007/978-3-540-70626-7_40}}.

\bibitem[\protect\citeauthoryear{Vaidman}{Vaidman}{2007}]{vaidman:2009}
Vaidman, L. (2007).
\newblock Counterfactuals in quantum mechanics.
\newblock In D.~Greenberger, K.~Hentschel, \& F.~Weinert (Eds.), {\em
  Compendium of Quantum Physics}  (pp.\ 132--136). Berlin, Heidelberg:
  Springer.
\newblock Available from: \url{http://dx.doi.org/10.1007/978-3-540-70626-7_40},
  \href {http://arxiv.org/abs/arXiv:0709.0340} {\path{arXiv:arXiv:0709.0340}},
  \href {http://dx.doi.org/10.1007/978-3-540-70626-7_40}
  {\path{doi:10.1007/978-3-540-70626-7_40}}.

\bibitem[\protect\citeauthoryear{{von Neumann}}{{von
  Neumann}}{1951}]{von-neumann1}
{von Neumann}, J. (1951).
\newblock Various techniques used in connection with random digits.
\newblock {\em National Bureau of Standards Applied Math Series}, {\em 12},
  36--38.
\newblock Reprinted in {\sl John {von Neumann}, Collected Works, (Vol. V)}, A.
  H. Traub, editor, MacMillan, New York, 1963, p. 768--770.

\bibitem[\protect\citeauthoryear{{von Neumann}}{{von
  Neumann}}{1966}]{v-neumann-66}
{von Neumann}, J. (1966).
\newblock {\em Theory of Self-Reproducing Automata}.
\newblock Urbana: University of Illinois Press.
\newblock A. W. Burks, editor.

\bibitem[\protect\citeauthoryear{Wagon}{Wagon}{1986}]{wagon1}
Wagon, S. (1986).
\newblock {\em The {B}anach-{T}arski Paradox}.
\newblock Cambridge: Cambridge University Press.

\bibitem[\protect\citeauthoryear{Wang}{Wang}{1974}]{wang}
Wang, P.~S. (1974).
\newblock The undecidability of the existence of zeros of real elementary
  functions.
\newblock {\em Journal of the Association for Computing Machinery (JACM)}, {\em
  21}, 586--589.
\newblock Available from: \url{http://dx.doi.org/10.1145/321850.321856}, \href
  {http://dx.doi.org/10.1145/321850.321856} {\path{doi:10.1145/321850.321856}}.

\bibitem[\protect\citeauthoryear{Wang, Long \& Li}{Wang
  et~al.}{2006}]{wang:056107}
Wang, P.~X., Long, G.~L., \& Li, Y.~S. (2006).
\newblock Scheme for a quantum random number generator.
\newblock {\em Journal of Applied Physics}, {\em 100\/}(5), 056107.
\newblock Available from: \url{http://dx.doi.org/10.1063/1.2338830}, \href
  {http://dx.doi.org/10.1063/1.2338830} {\path{doi:10.1063/1.2338830}}.

\bibitem[\protect\citeauthoryear{Wang}{Wang}{1991}]{Wang91}
Wang, Q.~D. (1991).
\newblock The global solution of the $n$-body problem.
\newblock {\em Celestial Mechanics}, {\em 50}, 73--88.
\newblock Available from: \url{http://dx.doi.org/10.1007/BF00048987}, \href
  {http://dx.doi.org/10.1007/BF00048987} {\path{doi:10.1007/BF00048987}}.

\bibitem[\protect\citeauthoryear{Wang}{Wang}{2001}]{Wang01}
Wang, Q.~D. (2001).
\newblock Power series solutions and integral manifold of the n-body problem.
\newblock {\em Regular \& Chaotic Dynamics}, {\em 6\/}(4), 433--442.
\newblock Available from:
  \url{http://dx.doi.org/10.1070/RD2001v006n04ABEH000187}, \href
  {http://dx.doi.org/10.1070/RD2001v006n04ABEH000187}
  {\path{doi:10.1070/RD2001v006n04ABEH000187}}.

\bibitem[\protect\citeauthoryear{Weierstrass, Hermite \&
  Mittag-Leffler}{Weierstrass et~al.}{1885}]{weierstrass-1885}
Weierstrass, C., Hermite, C., \& Mittag-Leffler, G. (1885).
\newblock {M}ittheilung, einen von {K}{\"o}nig {O}scar {II} gestifteten
  mathematischen {P}reis betreffend. {C}ommunication sur un prix de
  math{\'e}matiques fond{\'e} par le roi {O}scar {II}.
\newblock {\em Acta Mathematica}, {\em 7\/}(1), I--VI.
\newblock Available from: \url{http://dx.doi.org/10.1007/BF02402191}, \href
  {http://dx.doi.org/10.1007/BF02402191} {\path{doi:10.1007/BF02402191}}.

\bibitem[\protect\citeauthoryear{Weihrauch}{Weihrauch}{2000}]{Weihrauch}
Weihrauch, K. (2000).
\newblock {\em Computable Analysis. An Introduction}.
\newblock Berlin, Heidelberg: Springer.

\bibitem[\protect\citeauthoryear{Weihs, Jennewein, Simon, Weinfurter \&
  Zeilinger}{Weihs et~al.}{1998a}]{wjswz-98}
Weihs, G., Jennewein, T., Simon, C., Weinfurter, H., \& Zeilinger, A. (1998a).
\newblock Violation of {B}ell's inequality under strict {E}instein locality
  conditions.
\newblock {\em Physical Review Letters}, {\em 81}, 5039--5043.
\newblock Available from: \url{http://dx.doi.org/10.1103/PhysRevLett.81.5039},
  \href {http://dx.doi.org/10.1103/PhysRevLett.81.5039}
  {\path{doi:10.1103/PhysRevLett.81.5039}}.

\bibitem[\protect\citeauthoryear{Weihs, Jennewein, Simon, Weinfurter \&
  Zeilinger}{Weihs et~al.}{1998b}]{zeilinger-epr-98}
Weihs, G., Jennewein, T., Simon, C., Weinfurter, H., \& Zeilinger, A. (1998b).
\newblock Violation of {B}ell's inequality under strict {E}instein locality
  conditions.
\newblock {\em Physical Review Letters}, {\em 81}, 5039--5043.
\newblock Available from: \url{http://dx.doi.org/10.1103/PhysRevLett.81.5039},
  \href {http://dx.doi.org/10.1103/PhysRevLett.81.5039}
  {\path{doi:10.1103/PhysRevLett.81.5039}}.

\bibitem[\protect\citeauthoryear{Wheeler \& Zurek}{Wheeler \&
  Zurek}{1983}]{wheeler-Zurek:83}
Wheeler, J.~A. \& Zurek, W.~H. (1983).
\newblock {\em Quantum Theory and Measurement}.
\newblock Princeton, NJ: Princeton University Press.

\bibitem[\protect\citeauthoryear{Wigner}{Wigner}{1960}]{wigner}
Wigner, E.~P. (1960).
\newblock The unreasonable effectiveness of mathematics in the natural
  sciences. {R}ichard {C}ourant {L}ecture delivered at {N}ew {Y}ork
  {U}niversity, {M}ay 11, 1959.
\newblock {\em Communications on Pure and Applied Mathematics}, {\em 13},
  1--14.
\newblock Available from: \url{http://dx.doi.org/10.1002/cpa.3160130102}, \href
  {http://dx.doi.org/10.1002/cpa.3160130102}
  {\path{doi:10.1002/cpa.3160130102}}.

\bibitem[\protect\citeauthoryear{Wigner}{Wigner}{1961}]{wigner:mb}
Wigner, E.~P. (1961).
\newblock Remarks on the mind-body question.
\newblock In I.~J. Good (Ed.), {\em The Scientist Speculates}  (pp.\ 284--302).
  London and New York: Heinemann and Basic Books.
\newblock Reprinted in Ref.~\cite[pp. 168-181]{wheeler-Zurek:83}.
\newblock Available from:
  \url{http://www.phys.uu.nl/igg/jos/foundQM/wigner.pdf}.

\bibitem[\protect\citeauthoryear{Wolfram}{Wolfram}{1984}]{wolfram84}
Wolfram, S. (1984).
\newblock Computation theory of cellular automata.
\newblock {\em Communications in Mathematical Physics}, {\em 96\/}(1), 15--57.
\newblock Available from: \url{http://dx.doi.org/10.1007/BF01217347}, \href
  {http://dx.doi.org/10.1007/BF01217347} {\path{doi:10.1007/BF01217347}}.

\bibitem[\protect\citeauthoryear{Wolfram}{Wolfram}{1985}]{wolfram85b}
Wolfram, S. (1985).
\newblock Undecidability and intractability in theoretical physics.
\newblock {\em Physical Review Letters}, {\em 54\/}(8), 735--738.
\newblock Available from: \url{http://dx.doi.org/10.1103/PhysRevLett.54.7357},
  \href {http://dx.doi.org/10.1103/PhysRevLett.54.735}
  {\path{doi:10.1103/PhysRevLett.54.735}}.

\bibitem[\protect\citeauthoryear{Wolfram}{Wolfram}{2002}]{wolfram-2002}
Wolfram, S. (2002).
\newblock {\em A New Kind of Science}.
\newblock Champaign, IL: Wolfram Media, Inc.

\bibitem[\protect\citeauthoryear{Wolpert}{Wolpert}{2001}]{PhysRevE.65.016128}
Wolpert, D.~H. (2001).
\newblock Computational capabilities of physical systems.
\newblock {\em Physical Review E}, {\em 65\/}(1), 016128.
\newblock Available from: \url{http://dx.doi.org/10.1103/PhysRevE.65.016128},
  \href {http://dx.doi.org/10.1103/PhysRevE.65.016128}
  {\path{doi:10.1103/PhysRevE.65.016128}}.

\bibitem[\protect\citeauthoryear{Wright}{Wright}{1978}]{wright:pent}
Wright, R. (1978).
\newblock The state of the pentagon. {A} nonclassical example.
\newblock In A.~R. Marlow (Ed.), {\em Mathematical Foundations of Quantum
  Theory}  (pp.\ 255--274). New York: Academic Press.

\bibitem[\protect\citeauthoryear{Wright}{Wright}{1990}]{wright}
Wright, R. (1990).
\newblock Generalized urn models.
\newblock {\em Foundations of Physics}, {\em 20\/}(7), 881--903.
\newblock Available from: \url{http://dx.doi.org/10.1007/BF01889696}, \href
  {http://dx.doi.org/10.1007/BF01889696} {\path{doi:10.1007/BF01889696}}.

\bibitem[\protect\citeauthoryear{Zajonc, Wang, Zou \& Mandel}{Zajonc
  et~al.}{1991}]{Zajonc-91}
Zajonc, A.~G., Wang, L.~J., Zou, X.~Y., \& Mandel, L. (1991).
\newblock Quantum eraser.
\newblock {\em Nature}, {\em 353}, 507--508.
\newblock Available from: \url{http://dx.doi.org/10.1038/353507b0}, \href
  {http://dx.doi.org/10.1038/353507b0} {\path{doi:10.1038/353507b0}}.

\bibitem[\protect\citeauthoryear{Zeilinger}{Zeilinger}{1981}]{zeilinger:882}
Zeilinger, A. (1981).
\newblock General properties of lossless beam splitters in interferometry.
\newblock {\em American Journal of Physics}, {\em 49\/}(9), 882--883.
\newblock Available from: \url{http://dx.doi.org/10.1119/1.12387}, \href
  {http://dx.doi.org/10.1119/1.12387} {\path{doi:10.1119/1.12387}}.

\bibitem[\protect\citeauthoryear{Zeilinger}{Zeilinger}{1999}]{zeil-99}
Zeilinger, A. (1999).
\newblock A foundational principle for quantum mechanics.
\newblock {\em Foundations of Physics}, {\em 29\/}(4), 631--643.
\newblock Available from: \url{http://dx.doi.org/10.1023/A:1018820410908},
  \href {http://dx.doi.org/10.1023/A:1018820410908}
  {\path{doi:10.1023/A:1018820410908}}.

\bibitem[\protect\citeauthoryear{Zeilinger}{Zeilinger}{2005}]{zeil-05_nature_ofQuantum}
Zeilinger, A. (2005).
\newblock The message of the quantum.
\newblock {\em Nature}, {\em 438}, 743.
\newblock Available from: \url{http://dx.doi.org/10.1038/438743a}, \href
  {http://dx.doi.org/10.1038/438743a} {\path{doi:10.1038/438743a}}.

\bibitem[\protect\citeauthoryear{Zierler \& Schlessinger}{Zierler \&
  Schlessinger}{1965}]{ZirlSchl-65}
Zierler, N. \& Schlessinger, M. (1965).
\newblock Boolean embeddings of orthomodular sets and quantum logic.
\newblock {\em Duke Mathematical Journal}, {\em 32}, 251--262.

\bibitem[\protect\citeauthoryear{Zurek}{Zurek}{2003}]{RevModPhys.75.715}
Zurek, W.~H. (2003).
\newblock Decoherence, einselection, and the quantum origins of the classical.
\newblock {\em Reviews of Modern Physics}, {\em 75\/}(3), 715--775.
\newblock Available from: \url{http://dx.doi.org/10.1103/RevModPhys.75.715},
  \href {http://dx.doi.org/10.1103/RevModPhys.75.715}
  {\path{doi:10.1103/RevModPhys.75.715}}.

\bibitem[\protect\citeauthoryear{Zuse}{Zuse}{1970}]{zuse-70}
Zuse, K. (1970).
\newblock {\em Calculating Space. MIT Technical Translation AZT-70-164-GEMIT}.
\newblock Cambridge, MA: MIT (Proj. MAC).

\end{thebibliography}

\end{document}